\documentclass{elsart}

\usepackage{amssymb,amsmath,graphicx,cite,xcolor}

\DeclareSymbolFont{bbold}{U}{bbold}{m}{n}
\DeclareSymbolFontAlphabet{\mathbbold}{bbold}

\def\undertilde#1{\mathord{\vtop{\ialign{##\crcr
$\hfil\displaystyle{#1}\hfil$\crcr\noalign{\kern1.5pt\nointerlineskip}
$\hfil\tilde{}\hfil$\crcr\noalign{\kern1.5pt}}}}}

\begin{document}

\begin{frontmatter}

\maketitle

\title{Towards a high-supersaturation theory of crystal growth: Nonlinear one-step flow model\\ in 1+1 dimensions}

\author[a]{Joshua P. Schneider\corauthref{cor}}\ead{jschneid@math.umd.edu},
\author[b]{Paul N. Patrone}\ead{paul.patrone@nist.gov},
\author[a,c,d]{Dionisios Margetis}\ead{dio@math.umd.edu}

\corauth[cor]{Corresponding author. Tel.: (301) 405 5455; fax: (301) 314 0827.}

\address[a]{Department of Mathematics, University of Maryland,
        College Park, MD 20742}
        
 \address[b]{National Institute of Standards and Technology, Gaithersburg, MD 20899}

\address[c]{Institute for Physical Science and Technology, University of Maryland, College Park,
MD 20742}

\address[d]{Center for Scientific Computation and Mathematical Modeling, University of Maryland, College Park, MD 20742}

\begin{abstract}
Starting with a many-atom master equation of a kinetic, restricted solid-on-solid (KRSOS) model with external material deposition, we investigate nonlinear aspects of the passage to a mesoscale description for a crystal surface in 1+1 dimensions. This latter description focuses on the motion of an atomic line defect (i.e. a step), which is defined via appropriate statistical average over KRSOS microstates. Near thermodynamic equilibrium and for low enough supersaturation, we show that this mesoscale picture is reasonably faithful to the Burton-Cabrera-Frank (BCF) step-flow model. More specifically, we invoke a maximum principle in conjunction with asymptotic error estimates to derive the elements of the BCF model: (i) a diffusion equation for the density of adsorbed adatoms; (ii) a step velocity law; and (iii) a linear relation for the mass flux of adatoms at the step. In this vein, we also provide a criterion by which the adatom flux remains linear in supersaturation, suggesting a range of non-equilibrium conditions under which the BCF model remains valid. Lastly, we make use of kinetic Monte Carlo simulations to numerically study effects that drive the system to higher supersaturations -- e.g. deposition of material onto the surface from above. Using these results, we describe empirical corrections to the BCF model that amount to a nonlinear relation for the adatom flux at the step edge. 
\end{abstract}

\begin{keyword}Crystal surface; Burton-Cabrera-Frank (BCF) model; Line defects; Atom hopping; kinetic Monte Carlo; Mesoscale Limit

\PACS 81.10.Aj; 68.55.-a; 68.35.Md; 64.60.De

\end{keyword}

\end{frontmatter}

\section{Introduction}
\label{sec:intro}

A critical task in materials science  has been to understand how atomic defects on crystal surfaces form and evolve, so that one can assess the stability of novel nanostructures and small devices. Below the roughening transition, the growth of nanostructures is mediated by the motion of atomic line defects which resemble steps \cite{BCF51, PimpinelliVillain98, JeongWilliams99}. These steps act as sources and sinks for adsorbed atoms (adatoms). A goal is to understand the dynamics of steps, which can in turn elucidate the morphological evolution of crystal surfaces at large scales~\cite{Misbah10}. Studies of the Burton-Cabrera-Frank (BCF) model of step flow~\cite{BCF51} and its variants~\cite{JeongWilliams99} have long played a key role in achieving this goal. 

The BCF model \cite{BCF51} was originally formulated as a phenomenological mesoscale theory for {\em near-equilibrium} processes that cause steps to move via mass conservation. The principal ingredients are: (i) a step velocity law; (ii) the adatom diffusion on terraces, i.e. nanoscale domains bounded by steps; and (iii) a {\em linear} kinetic law for the density of flux at steps. This model has been enriched to the point where it is now used to describe step flow under an apparently wide range of kinetics on a crystal surface. Nonetheless, basic aspects of step flow remain poorly understood. Recently, it was indicated that BCF theory in 1+1 dimensions can be interpreted as a dilute limit of the adatom gas~\cite{PatroneEinsteinMargetis14,LuLiuMargetis14}.  In this context, the relative deviation (supersaturation) of the adatom density from its equilibrium value is a variable of importance. An emerging question is the following: How can {\em corrections} to the BCF theory be described consistently with atomistic dynamics on a crystal lattice?

The goals of this paper are twofold. First, we seek to demonstrate the validity of the BCF model via appropriate selection of atomistic parameters. Second, we wish to describe empirical corrections to a BCF-like model of one-step flow in correspondence to atomistic dynamics in 1+1 dimensions. We carry out (i) a formal analysis based on a master equation for a kinetic restricted solid-on-solid (KRSOS) model, in the spirit of~\cite{PatroneEinsteinMargetis14}; and (ii) numerical computations via kinetic Monte Carlo (KMC) simulations of the KRSOS model.
Our main results in this study are summarized as follows.

\begin{itemize}

\item Assuming that an arbitrary yet finite number of particle states contribute to the system evolution, we derive a formal power-series expansion for the steady-state solution to a master equation with external material deposition.

\item We prove a ``maximum principle'' for a master equation including material deposition onto the surface from above.

\item By a microscopic averaging in the time-dependent setting, we derive exact expressions for the adatom flux at the step edge. Our formulas separate the linear
kinetic law of the BCF model, which is exact in the dilute limit, from higher-order corrections that result from many-particle states.

\item Heuristically, we show how the BCF-type model results as the mesoscale limit of the master equation at low enough supersaturations.

\item We find bounds for possible deviations, which are expressed in terms of discrete averages over special microscale configurations, from the linear kinetic law for adatom fluxes at the step and from the diffusion equation on the terrace. Our derivation makes use of estimates resulting from our ``maximum principle''. 

\item By using KMC simulations,
we empirically determine these supersaturation-driven corrections to the BCF theory, particularly a nonlinear (quadratic) relation for the adatom mass flux at the 
step edge. Furthermore, we empirically determine a condition by which the linear kinetic law for adatom flux at the step is reasonably accurate (see Remark 7 in Section \ref{sec:Corrections}).
\end{itemize}

We assume that the reader is familiar with the basic concepts of epitaxial growth. For broad reviews on the subject, see, e.g., \cite{PimpinelliVillain98, JeongWilliams99,Misbah10}.

%%%%%%%%%%%%%%%%%%%%%%%%
\subsection{Approach and key outcomes}
\label{subsec:Intro-approach}
%%%%%%%%%%%%%%%%%%%%%%%

Our starting point is a master equation for the probability density of atomistic configurations. This description, which includes external material deposition with rate (atoms per unit time) $F$, is an extension of the deposition-free master equation for adatoms invoked in~\cite{PatroneMargetis14,PatroneEinsteinMargetis14}. Our equation embodies generic microscopic rules of the KRSOS model for: adatom hopping on terraces (or, the microscopic process for surface diffusion); and detachment and attachment of atoms at the step edge according to detailed balance. The corresponding mathematical formalism also accounts for
events that do not conserve the total mass, or adatom number, as a result of the nonzero $F$.  
A simplifying assumption inherent to our formalism is that adatoms do not form bonds away from the step edge and do not interact with each other electrostatically or elastically or by indirect (through-substrate) mechanisms; hence, adatom-adatom correlations originate solely from kinetics.

Our atomistic model allows for the analytical derivation of an exact, closed-form expression
for the equilibrium concentration of adatoms on the terrace if $F=0$. 
%This result follows from general considerations of statistical mechanics, without any approximations due to the truncation of the governing hierachy of particle equations~\cite{PatroneMargetis14}.
Furthermore, for arbitrary $F$, we derive formulas for the adatom fluxes at step edges via formal analysis of the master equation. In this vein, kinetic contributions to the adatom mass flux are manifestly connected to atomistic variables. 

For low enough supersaturation, we formally show that the adatom motion gives rise  
to the familiar linear kinetic relation characteristic of BCF-type models. On the other hand, at high supersaturations, nonlinear corrections to this relation must be taken into account. 
By use of KMC simulations, we demonstrate this nonlinearity and numerically estimate the dependence of the coefficients of the
emergent nonlinear kinetic relation on the attachment-detachment and deposition rates of the atomistic processes. 

Specifically, within the mesoscale picture of a crystal surface, the adatom fluxes, $J_\pm$, toward the step edge are empirically
described by an expansion of the form
\begin{equation} \label{nlkr_bcf}
J_\pm \approx c^{eq}\sum_{n=1}^{N_*} \kappa_\pm^{(n)} \,\bar{\sigma}_{\pm}^n~.
\end{equation}
In this relation, $c^{eq}$ is the equilibrium adatom concentration at the step edge (for $F=0$); $\mathcal C_\pm$ is the adatom concentration at the upper ($-$) or lower ($+$) terrace; $\bar{\sigma}_{\pm}=\mathcal C_{\pm}/c^{eq}-1$ is the corresponding supersaturation; and $\kappa_{\pm}^{(n)}$ are kinetic coefficients that in principle depend on atomistic rates.
Note that only the $n=1$ term in~\eqref{nlkr_bcf} is present in BCF-type models~\cite{JeongWilliams99}.
In the present work, we provide evidence for the origin of nonlinear terms in~\eqref{nlkr_bcf}; and numerically compute the coefficients $\kappa_{\pm}^{(n)}$ for $n=1,\,2$ in certain typical cases of atomistic dynamics. A related outcome of our numerical computations is an empirical condition for the validity of the linear kinetic law. 

%%%%%%%%%%%%%%%%%
\subsection{Past works}
\label{subsec:Intro-Past}
%%%%%%%%%%%%%%%%%

It is worthwhile placing our work in the appropriate context of past literature, e.g.~\cite{PatroneEinsteinMargetis14, PatroneMargetis14,LuLiuMargetis14,Zhao05,AckermanEvans11,Zhaoetal15,Saum09,Zangwill92}. Our goal here is to explore the connection of an atomistic model 
to mesoscale descriptions of crystals. This theme bears resemblance to the main objectives of \cite{PatroneEinsteinMargetis14, PatroneMargetis14,LuLiuMargetis14,Zhao05}. However, in these works \cite{PatroneEinsteinMargetis14, PatroneMargetis14,LuLiuMargetis14,Zhao05} the authors' attention focuses on near-equilibrium processes, whereas in the present paper we study the kinetic regime in which external material deposition tends to drive the system away from equilibrium. A similar task is undertaken in~\cite{PatroneCaflischDM_12}, albeit via a (coarse-grained) ``terrace-step-kink'' model. We herein avoid {\em a priori} approximations associated with the diluteness of the adatom system which is a key, explicit assumption in \cite{PatroneEinsteinMargetis14, PatroneMargetis14, LuLiuMargetis14}. 

Our study here has a perspective distinct from that of~\cite{AckermanEvans11,Zhaoetal15,Saum09} in which extensive computations are carried out in 2+1 dimensions.
In particular, in~\cite{AckermanEvans11,Zhaoetal15} the authors derive a set of refined boundary conditions at the step edge that depend on the local environment on the basis of a discrete diffusion equation with a {\em fixed} step position.
In~\cite{Saum09} only numerical comparisons of KMC simulations to aspects of the BCF model are shown. A different view is adopted in~\cite{Zangwill92} where a high-dimensional master equation is reduced to a Langevin-type equation for height columns on the crystal lattice. We should also mention the probabilistic approach in~\cite{MarzuolaWeare13}, which addresses the passage from an atomistic description within a solid-on-solid model to a {\em fully} continuum picture.

A discussion that the boundary condition involving the mass flux toward the step edge may exhibit a nonlinear behavior as a function of the adatom density can be found in \cite{VoigtBalykov2006a,CermelliJabbour2005}. Specifically, in \cite{VoigtBalykov2006a} the authors carry out numerical simulations of a  ``terrace-step-kink'' model that reveal a nonlinear dependence of adatom fluxes on the supersaturation, $\bar{\sigma}_\pm$; and relate this behavior to the step-continuum thermodynamic approach of \cite{CermelliJabbour2005}. 

In the present treatment, we point out such a nonlinearity at the mesoscale from a kinetic atomistic perspective, in an effort to avoid continuum thermodynamic principles. 
By recourse to atomistic mechanisms, we argue that nonlinear terms in the boundary conditions for the mass flux naturally emerge as the system is driven farther from equilibrium; cf.~\eqref{eq:cond-micro}. Building on our results, our long-term goal is to address phenomena in 2D settings, for which more complicated atomistic models are necessary~\cite{Caflischetal_99}.

%%%%%%%%%%%%%%%%%%
\subsection{Limitations }
\label{subsec:Intro-Limitations}
%%%%%%%%%%%%%%%%%%
 
Our work has several limitations. To start with, our atomistic model is one dimensional (1D). This simplification has some unphysical consequences. For example, it leaves out surface features intimately connected to the effect of step stiffness~\cite{BCF51}. Step meandering, an important 2D effect, is completely absent in our approach. Furthermore, we do not account for nucleation, which at low enough temperatures is known to cause deviations from the usual kinetic law for the step velocity~\cite{Shitara93}. In a related vein, we are unable to adequately model advection at the atomistic scale and derive corresponding terms in the continuum limit. In our 1D setting, island formation dictates that equilibrium cannot be established; thus we exclude bonding between adatoms~\cite{PatroneMargetis14}. Furthermore, we focus on the motion of a single step.
This consideration leaves out elastic and entropic step-step interactions~\cite{JeongWilliams99, PimpinelliVillain98}. Admittedly, our version of a microscopic master equation is simplified since it forms a direct extension, by addition of external deposition, of the description in~\cite{PatroneMargetis14}. This model is deemed suitable for low and moderate adatom densities. Thus, our formalism may not entirely capture the full range of effects arising at high supersaturation. Our analysis, which focuses on averages of  microscopic variables such as the number of adatoms per lattice site, leaves unexplored the issue of stochastic fluctuations in step motion~\cite{LuLiuMargetis14}.

%%%%%%%%%%%%%%%%%%%%%
\subsection{Outline of paper}
\label{subsec:Intro-Outline}
%%%%%%%%%%%%%%%%%%%%%

The remainder of the paper is organized as follows. Section \ref{sec:ReviewModels} provides a review of the main models: the BCF-type model of step flow in 1D (Section~\ref{subsec:BCFmodel}), which offers the usual mesoscale picture phenomenologically; and the atomistic KRSOS model (Section~\ref{subsec:SOSmodel}), which is the starting point and basis of our analysis. In Section \ref{sec:Max-Steady_SOS}, we introduce a maximum principle and calculate the steady-state solution for a master equation of the KRSOS model with material deposition onto the surface from above. A discrete version of the BCF
model is developed in Section \ref{sec:DiscreteBCF}. The BCF-type model corresponding to our atomistic dynamics is formally derived in Section \ref{sec:BCFregime}. In Section \ref{sec:Corrections}, we characterize high-supersaturation corrections for the mass flux at the step edge; and propose a more general kinetic relation for mesoscale step flow models. Finally, Section \ref{sec:Discussion} contains a summary and discussion of our results.

{\bf Notation.} We write $f = \mathcal O(g)$ ($f =o(g)$) to imply that $|f/g|$ is bounded by a nonzero constant (approaches zero) in a prescribed limit. For ease in notation, we write $f=\mathcal O(h,g)$ to imply a relation of the form $f=\mathcal O(h)+\mathcal O(g)$. We use the symbol $\lesssim$ to denote boundedness up to the constant factor. The symbol $\mathbbold{R}$ denotes the set of reals.

%%%%%%%%%%%%%%%%%%%%%%%%%%%%%%%%%%%%%%%%%%%%%%%%%%%%%%%%%%%%
%%%%%%%%%%%%%%%%%%%%%%%%%%%%%%%%%%%%%%%%%%%%%%%%%%%%%%%%%%%%
\section{Background: Mesoscale and atomistic models} 
\label{sec:ReviewModels}
%%%%%%%%%%%%%%%%%%%%%%%%%%%%%%%%%%%%%%%%%%%%%%%%%%%%%%%%%%%%
%%%%%%%%%%%%%%%%%%%%%%%%%%%%%%%%%%%%%%%%%%%%%%%%%%%%%%%%%%%%

In this section, we describe ingredients of the mesoscale and atomistic models, which form the core of our paper. In particular, we review basics of the BCF model, and introduce the master equation of atomistic dynamics with material deposition onto the surface from above.

%%%%%%%%%%%%%%%%%%%%%%%%%%%%%%%%%%%%%%%%%%%%%%%%%%%%%%%
\subsection{Mesoscale: BCF model} 
\label{subsec:BCFmodel}
%%%%%%%%%%%%%%%%%%%%%%%%%%%%%%%%%%%%%%%%%%%%%%%%%%%%%%%

By phenomenological principles, the BCF model treats adatoms and the crystal surface in a continuum fashion in the lateral direction, yet retains the atomistic detail of the crystal in the vertical direction~\cite{BCF51,JeongWilliams99}. 
This approach makes use of an adatom concentration, $\rho$, on each terrace in the laboratory frame. Accordingly, the BCF theory is comprised of the following major elements: (a) A step velocity law, which expresses mass conservation for adatoms; (b) a diffusion equation for $\rho$; and (c) a linear kinetic relation for the adatom flux normal to the step edge. To simplify the model without losing sight of the essential physics, we assume that the desorption and evaporation of atoms is negligible. 
(It should be noted, however, that desorption is treated by BCF~\cite{BCF51}.)

\begin{figure}[!h]
$\begin{array}{c}
    \includegraphics[width=\textwidth]{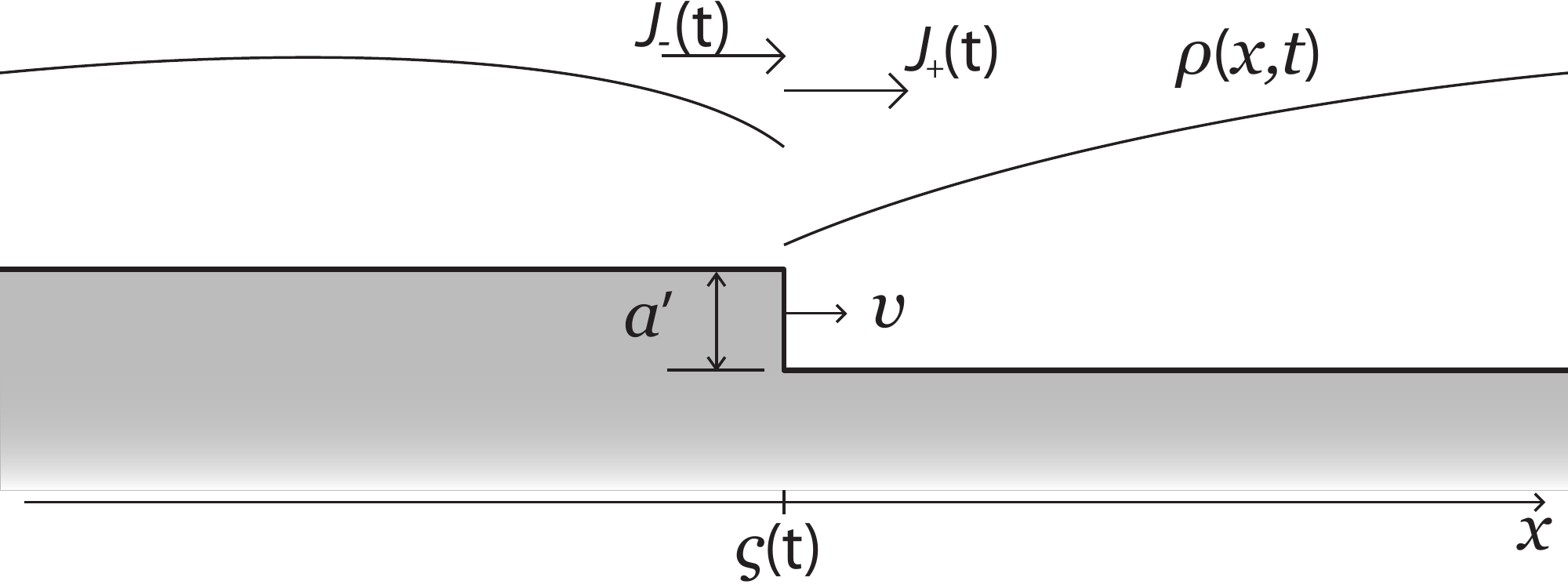}
\end{array}$
\caption{Mesoscale view: Schematic of a step with atomic height $a'$ ajoining two terraces. The dependent variable  $\rho(x,t)$ is the adatom concentration field on each terrace in the laboratory frame. The step velocity, $v=\dot\varsigma (t)$, is determined by the adatom fluxes $J_{\pm}$ at the step edge on the lower ($+$) or upper ($-$) terrace via mass conservation; cf.~\eqref{eq:step_velocity_bcf}. }
\label{fig:bcf_picture}
\end{figure}

The geometry of a step ajoining two terraces is depicted in Fig. \ref{fig:bcf_picture}. The upper ($-$) terrace, on the left of step edge, and lower ($+$) terrace, on the right of step edge, differ in height by $a'$, an atomic length. 
In this view, the adatoms are represented by the concentration field $\rho(x,t)$.
Let $\varsigma(t)$ be the position of the step edge. We apply screw-periodic boundary conditions in the spatial coordinate, $x$.

Now consider the motion of the step. The step velocity, $v(t)=\dot \varsigma(t)$, is determined by mass conservation:  
\begin{equation}\label{eq:step_velocity_bcf}
 v = \frac{\Omega}{a'} \left( J_- - J_+ \right)~,
\end{equation}
where $J_\pm$ denotes the $x$-directed mass flux at the step edge on
the upper ($-$) or lower ($+$) terrace, $\Omega = aa'$ is the atomic area, and $a$ is the lattice spacing in the lateral ($x$-) direction.

For later algebraic convenience, we define $\hat{x}:=x-\varsigma(t)$ which is the coordinate relative to the step edge. On each terrace, the variable $\mathcal C(\hat{x},t)=\rho(x,t)$ satisfies the diffusion equation~\cite{BCF51}
\begin{equation}\label{eq:diffusion_eq_bcf}
 \frac{\partial \mathcal{C}}{\partial t} = \mathcal{D} \frac{\partial^2 \mathcal{C}}{\partial {\hat{x}}^2} + v\frac{\partial \mathcal{C}}{\partial \hat{x}} + \mathcal{F}~,
\end{equation}
where $\mathcal{D}$ is the macroscopic adatom diffusivity and $\mathcal{F}$ is the mesoscopic external deposition flux  \cite{PimpinelliVillain98, JeongWilliams99}. Note the presence of the advection term, $v (\partial\mathcal C/\partial \hat{x})$, on the right-hand side of~\eqref{eq:diffusion_eq_bcf}; this term originates from $\partial\rho/\partial t$ in the corresponding diffusion equation for $\rho(x,t)$, viz., $\partial\rho/\partial t=\mathcal D (\partial^2 \mathcal \rho/\partial x^2)+\mathcal F$. Thus, the flux at the step edge consistent with Fick's law  is
\begin{equation}\label{eq:J-Fick_BCF}
 J_\pm=-\mathcal D (\partial \mathcal C/\partial \hat{x})_\pm -v \mathcal C_\pm=-\mathcal D (\partial \rho/\partial x)_\pm -v \rho_\pm~.
 \end{equation}

The remaining ingredient of the BCF model is a set of boundary conditions for $\mathcal C$, or $\rho$, at the step edge through the mass flux, $J(x,t)$.
BCF originally introduced Dirichlet boundary conditions, by which the restriction $\mathcal C_\pm$  of $\mathcal C$ at the step edge is set equal to an equilibrium value, $c^{eq}$~\cite{BCF51}. Later on, a Robin boundary condition 
was imposed~\cite{Chernov61} via the linear version of condition~\eqref{nlkr_bcf}. Note that the Robin boundary condition is typically a linear relation between the solution of a partial differential equation and its normal derivative at a free boundary. This condition was later improved by incorporation of the Ehrlich-Schwoebel barrier~\cite{EhrlichHudda66, SchwoebelShipsey66}; see~\cite{PimpinelliVillain98}. 
The linear kinetic relation for the mass flux at the step is
\begin{equation}\label{eq:lkr_bcf}
 J_\pm =\mp \kappa_\pm \left(\mathcal{C}_\pm - c^{eq}\right)~, 
\end{equation}
where $\kappa_\pm$ describes the rate of attachment/detachment of atoms at the step in the presence of an Ehrlich-Schwoebel barrier.

In \cite{VoigtBalykov2006a}, numerical simulations based on a ``terrace-step-kink'' model suggest a nonlinear dependence of  $J_\pm$ on $\mathcal{C}_\pm - c^{eq}$. The authors argue that this can be explained by the thermodynamic approach of \cite{CermelliJabbour2005}. 
The observation of such a nonlinear effect motivates us to conjecture a generalized relation of form~\eqref{nlkr_bcf}, where the terms corresponding to $n\ge 2$ account for far-from-equilibrium, high-supersaturation corrections to the traditional linear kinetic law~\eqref{eq:lkr_bcf}. 
In Section~\ref{sec:Corrections}, we provide evidence for~\eqref{nlkr_bcf} that emerges from kinetic aspects of our simplified 1D atomistic model.

%%%%%%%%%%%%%%%%%%%%%%%%%%%%%%%%%%%%%%%%%%%%%%%%%%%%%%%%%%%%
\subsection{Microscale: KRSOS model} 
\label{subsec:SOSmodel}
%%%%%%%%%%%%%%%%%%%%%%%%%%%%%%%%%%%%%%%%%%%%%%%%%%%%%%%%%%%%

At the microscale, we consider a simple-cubic crystal surface with a single step~\cite{PatroneMargetis14}. The surface consists of distinct height columns on an 1D lattice of lateral spacing $a$, with total length $L=Na$; see Fig.~\ref{fig:sos_rates}. We consider $L=\mathcal O(1)$ as $a\to 0$, e.g., by setting $L=1$. Screw-periodic boundary conditions are applied in the $x$-direction. 

Atoms of the top layer that have two in-plane nearest neighbors are {\em step atoms}~\cite{PatroneMargetis14}; these atoms are immobile in our model. In contrast, the atom of the step edge, which lies at one end of the top layer, has a single in-plane nearest neighbor and is referred to as an {\em edge atom};
it may detach from the step and move to one of the adjacent terraces. By this picture, the adatoms are movable atoms that are neither edge atoms nor step atoms~\cite{PatroneMargetis14}. 

In our model, we do not allow  islands to form; thus, if any two adatoms become nearest neighbors on a terrace, they do not form a bond with each other. Adatoms are free to diffuse across the surface until they reach the step, which acts as a sink or source of them. Externally deposited atoms are assumed to become adatoms on the terrace instantly, and may not attach to the step directly~\cite{LuLiuMargetis14}. 

%%%%%%%%%%%%%%%%%%%%%%%%%%%%%%%%%%%%%%%%%%%%%%%%%%%%%%%%%%%%
\subsubsection{On atomistic processes and system representation}
\label{sssec:tr-multi}
%%%%%%%%%%%%%%%%%%%%%%%%%%%%%%%%%%%%%%%%%%%%%%%%%%%%%%%%%%%%

Our model is characterized by transitions between discrete configurations of adatoms. The total mass of these configurations is not conserved if $F\neq 0$. The transitions are controlled by $F$ as well as by Arrhenius rates which have the form $\nu\exp[-E/(k_BT)]$, where $\nu$ is an attempt frequency, 
$T$ is the absolute temperature, $k_B$ is Boltzmann's constant, and $E$ is an appropriate activation energy; see~\cite{PatroneMargetis14}. These kinetic rates correspond to atomistic processes of surface diffusion and attachment/detachment at the step.

\begin{figure}[!h]
$\begin{array}{c}
    \includegraphics[width=\textwidth]{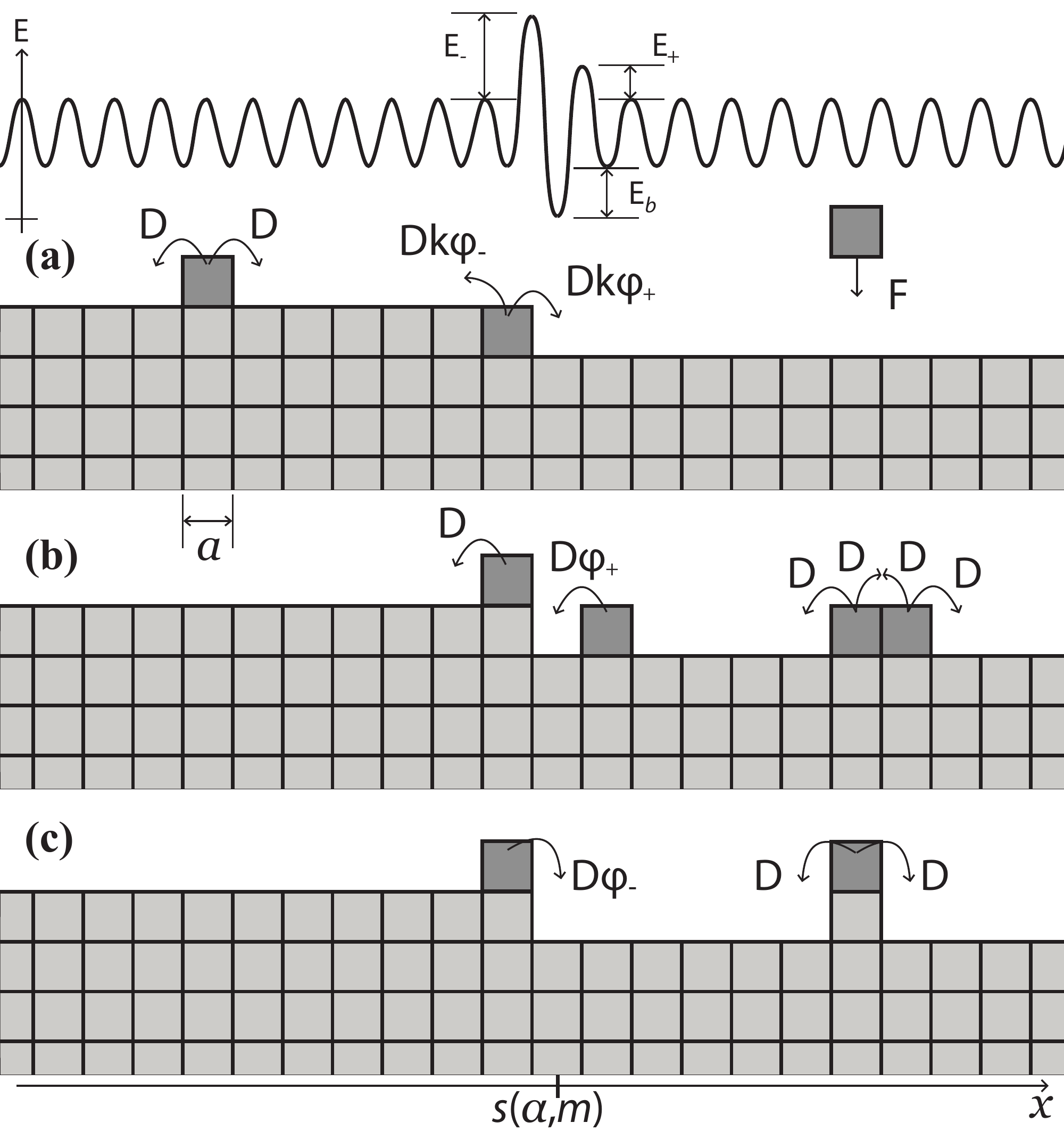}
\end{array}$
\caption{Microscopic view: Schematic of atomistic processes allowed by the 1D atomistic model on a lattice with lateral spacing $a$. Movable atoms (adatoms) are shown in dark grey. (a) Top panel: Hopping of adatoms on each terrace with rate $D$, detachment of edge atom from step to upper ($-$) or lower ($+$) terrace with rate $Dk\phi_{\pm}$, and deposition of atoms from above with rate $F$. The deposited atom becomes adatom instantly. (b) Middle panel: Hopping of adatom at step to same terrace with rate $D$, attachment of adatom from lower terrace to step edge with rate $D\phi_+$, and hopping of (unbonded) adatoms forming a pair with rate $D$. Adatoms cannot form islands. (c) Bottom panel: Attachment of adatom from upper terrace to step edge, and hopping of adatom lying on top of another adatom to same terrace. The site to the right of the step is labeled $s(\boldsymbol{\alpha},m)$; see Definition 2.}
\label{fig:sos_rates}
\end{figure}

The basic processes allowed by our 1D atomistic model are shown in Fig. \ref{fig:sos_rates}. The requisite atomistic rates can be described as follows~\cite{PatroneMargetis14}. First, the rate $D=\nu \exp[-E_h/(k_B T)]$ accounts for unbiased adatom hopping on terraces, sufficiently away from a step edge. The extra factor $\phi_\pm = \exp[-E_\pm/(k_BT)]$ expresses additional energy barriers, $E_{\pm}$, corresponding to adatom attachment to the step edge from the lower ($+$) or upper ($-$) terrace. Lastly, the factor $k=\exp[-E_b/(k_BT)]$ accounts for the extra energy, $E_b$, that is necessary for the breaking of the edge-atom bonds with step atoms so that the atom detaches from the step edge. 

In principle, our 1D system can have an (countably) infinite number of adatom configurations under the influence of a nonzero deposition, $F$. Following \cite{PatroneMargetis14}, we are compelled to represent such atomistic configurations by {\em multisets}. A multiset, $\boldsymbol{\alpha}$, is an unordered list whose entries correspond to the positions of adatoms on the 1D lattice; in particular, $\boldsymbol{\alpha} = \{\}$ expresses a configuration that is void of adatoms. Departing slightly from \cite{PatroneMargetis14}, we use multisets $\boldsymbol{\alpha}$ containing the {\em Lagrangian coordinates} of adatoms. For notational clarity, the indices $i$ and $j$ reference lattice sites in a fixed (Eulerian) coordinate system, whereas $\hat{\imath}$ and $\hat{\jmath}$ label sites in a (Lagrangian) coordinate frame relative to the step. Accordingly, repeated entries in $\boldsymbol{\alpha}$ indicate multiple adatoms occupying the same lattice site. For example, the system configuration represented by $\boldsymbol{\alpha} = \{\hat{\imath},\hat{\jmath},\hat{\jmath}\}$ has one adatom at site $\hat{\imath}$ and two adatoms at site $\hat{\jmath}$. The number of adatoms correponding to 
$\boldsymbol{\alpha}$ is simply $|\boldsymbol{\alpha}|$, the cardinality of the multiset.

In connecting the atomistic model to one-step flow with $F=0$ in 1D, our approach relies on explicitly determining the position of the step edge at time $t>0$ from the number of adatoms, $|\boldsymbol{\alpha}|$, and the initial adatom configuration~\cite{PatroneMargetis14}. This is a consequence of mass conservation. If $F\neq 0$, however, more information is needed in order to track the step edge: At every atomistic transition, one must account for the atoms deposited from above. 

For our purposes, a system representation that allows this bookeeping results from using an integer, $m$, in addition to using $\boldsymbol{\alpha}$. This $m$ is the total {\em mass}, or number of atoms, of the system. Thus, if $m_0$ is the initial mass then $m-m_0$ measures the overall mass increase because of external deposition. Finally, given the initial position of the site to the right of the step edge, $s_0$, in a fixed (Eulerian) coordinate system, we can explicitly track the step for all time $t>0$.
\medskip

{\bf Definition 1.} (Representation of atomistic system.) The pair $(\boldsymbol{\alpha}, m)$ defines the {\em state} of the atomistic system: the multiset $\boldsymbol\alpha$ expresses the adatom configuration and the index $m$ is an integer that counts the overall mass of the system. Thus, if $m_0$ is the initial mass then $m-m_0$ counts how many adatoms are deposited on the surface from above.

{\bf Definition 2.} (Discrete step position.) For each state $(\boldsymbol{\alpha}, m)$, the discrete {\em step position} in Eulerian coordinates is $s(\boldsymbol{\alpha}, m) = s_0-|\boldsymbol{\alpha}|+m-m_0$. For fixed mass $m$, the step position is uniquely determined from the number of adatoms $|\boldsymbol{\alpha}|$ and the initial position of the site to immediately to the right of the step edge, $s_0$. Accordingly, $s(\boldsymbol{\alpha}, m)$ also references the site to the right of the step edge; see Figure \ref{fig:sos_rates}.

%%%%%%%%%%%%%%%%%%%%%%%%%%%%%%%%%%%%%%%%%%%%%%%%%%%%%%%
\subsubsection{Master equation} 
\label{sssec:MasterEqn}
%%%%%%%%%%%%%%%%%%%%%%%%%%%%%%%%%%%%%%%%%%%%%%%%%%%%%%%

The KRSOS model is characterized by a time-dependent probability density function (PDF), $p_{\boldsymbol{\alpha},m}(t)$, defined over the domain of discrete states $(\boldsymbol{\alpha}, m)$. 
Accordingly, the time evolution of the system is described by the master equation
\begin{equation}\label{eq:master_eq}
\dot{p}_{\boldsymbol{\alpha},m}(t) = \sum_{\boldsymbol{\alpha}',m'} T_{(\boldsymbol{\alpha},m),(\boldsymbol{\alpha}',m')} p_{\boldsymbol{\alpha}',m'}(t)~,
\end{equation}
under given initial data, $p_{\boldsymbol\alpha, m}(0)$.
In the above, $T_{(\boldsymbol{\alpha},m),(\boldsymbol{\alpha}',m')}$ expresses the overall transition of the system from state $(\boldsymbol{\alpha}',m')$ to state $(\boldsymbol{\alpha},m)$. Evidently, master equation (\ref{eq:master_eq}) governs a Markov process with countably infinite states. 

Next, we describe the rates $T_{(\boldsymbol{\alpha},m),(\boldsymbol{\alpha}',m')}$. The nonzero transition rates obey the following rules:
\vspace{-1mm}
\begin{subequations}\label{eq:transition_rates}
\begin{align}
T_{(\boldsymbol{\alpha},m),(\boldsymbol{\alpha}',m')} &= D, & \mbox{  if  } m=m' \mbox{  and  } |\boldsymbol{\alpha}| = |\boldsymbol{\alpha}'| \mbox{  and  } \left|\boldsymbol{\alpha}\setminus\boldsymbol{\alpha}'\right| = 1 \nonumber\\
    & & \mbox{  and  } \Big| ||\boldsymbol{\alpha}\setminus\boldsymbol{\alpha}'|| - ||\boldsymbol{\alpha}'\setminus\boldsymbol{\alpha}||\Big| = 1; \\
T_{(\boldsymbol{\alpha},m),(\boldsymbol{\alpha}',m')} &= D\phi_\pm, & \mbox{  if  } m=m' \mbox{  and  } |\boldsymbol{\alpha}| = |\boldsymbol{\alpha}'| - 1 \nonumber\\
    & & \mbox{  and  } \boldsymbol{\alpha}'\setminus\boldsymbol{\tilde{\alpha}} = \{\pm 1\}; \\
T_{(\boldsymbol{\alpha},m),(\boldsymbol{\alpha}',m')} &= Dk\phi_\pm, & \mbox{  if  } m=m' \mbox{  and  } |\boldsymbol{\alpha}| = |\boldsymbol{\alpha}'| + 1  \nonumber\\
    & & \mbox{  and  } \boldsymbol{\alpha}\setminus\boldsymbol{\tilde{\alpha}'} = \{\pm 1\}; \\
T_{(\boldsymbol{\alpha},m),(\boldsymbol{\alpha}',m')} &= \frac{F}{N-1}, & \mbox{  if  } m=m'+1 \mbox{  and  } |\boldsymbol{\alpha}| = |\boldsymbol{\alpha}'| + 1  \nonumber\\
    & & \mbox{  and  } |\boldsymbol{\alpha}\setminus\boldsymbol{\alpha}'| = 1;
\end{align}
\vspace{-4mm}
and, so that probability is conserved,
\begin{equation}\label{eq:transitions_probability_conservation}
T_{(\boldsymbol{\alpha}',m'),(\boldsymbol{\alpha}',m')} =\quad -\sum\limits_{(\boldsymbol{\alpha},m)\atop (\boldsymbol{\alpha},m)\neq (\boldsymbol{\alpha}',m')} T_{(\boldsymbol{\alpha},m),(\boldsymbol{\alpha}',m')} \qquad \mbox{for all  } (\boldsymbol{\alpha}',m'). \qquad
\end{equation}
\end{subequations}
All transition rates not listed in (\ref{eq:transition_rates}) are zero. Here, we introduce the multiset difference $\boldsymbol{\alpha}\setminus\boldsymbol{\alpha}'$, which itself is a multiset containing the elements in $\boldsymbol{\alpha}$ that are not in $\boldsymbol{\alpha}'$, counting multiplicity. For example, $\{\hat{\imath},\hat{\jmath},\hat{\jmath}\}\setminus\{\hat{\jmath}\} = \{\hat{\imath},\hat{\jmath}\}$. Additionally, the symbol $||\cdot||$ indicates the $\ell^p$-norm with $p\geq 1$, and we define the ``multiset increment operation'' as $\boldsymbol{\tilde{\alpha}} = \{\hat{\imath}+1|\mbox{ for all } \hat{\imath}\in \boldsymbol{\alpha}\}$, i.e. the set $\boldsymbol{\tilde{\alpha}}$ is just $\boldsymbol{\alpha}$ after each element has been incremented by one. Setting $F=0$ in (\ref{eq:transition_rates}) reduces~\eqref{eq:master_eq} to the master equation governing surface relaxation~\cite{PatroneMargetis14}.

Among the transition rates that are zero, notable examples include
\begin{subequations}\label{eq:transition_zero_rates}
\begin{align}
T_{(\boldsymbol{\alpha},m),(\boldsymbol{\alpha}',m')} &= 0, & \mbox{  if  } m=m' \mbox{  and  } |\boldsymbol{\alpha}| < |\boldsymbol{\alpha}'| - 1 \mbox{  or  } |\boldsymbol{\alpha}| > |\boldsymbol{\alpha}'| + 1; \\
T_{(\boldsymbol{\alpha},m),(\boldsymbol{\alpha}',m')} &= 0, & \mbox{  if  } m=m' \mbox{  and  } |\boldsymbol{\alpha}| = |\boldsymbol{\alpha}'|+1 \nonumber\\
    & & \mbox{  and  } -1 \in \boldsymbol{\alpha}'; \\
T_{(\boldsymbol{\alpha},m),(\boldsymbol{\alpha}',m')} &= 0, & \mbox{  if  } m=m'+1 \mbox{  and  } |\boldsymbol{\alpha}| \leq |\boldsymbol{\alpha}'|.
\end{align}
\end{subequations}
Equation (\ref{eq:transition_zero_rates}a) indicates that no more than one atom may attach to or detach from the step in a single transition. Equation (\ref{eq:transition_zero_rates}b) asserts that no atoms may detach if the site directly above the edge atom is occupied, and (\ref{eq:transition_zero_rates}c) prevents atoms from being deposited at $s(\boldsymbol{\alpha},m)$. Note that the transitions described in (\ref{eq:transition_rates}a)-(\ref{eq:transition_rates}c), along with (\ref{eq:transition_zero_rates}a) and (\ref{eq:transition_zero_rates}b) are subject to detailed balance~\cite{LuLiuMargetis14, PatroneMargetis14}.

The master equation (\ref{eq:master_eq}) along with transition rates (\ref{eq:transition_rates}) and (\ref{eq:transition_zero_rates}) completely govern the full mass-dependent microscale model. For some of our purposes, notably the maximum principle of Section \ref{sec:Max-Steady_SOS} and its application, we require an alternate version of equation (\ref{eq:master_eq}) that describes the evolution of a {\em marginalized} PDF, $p_{\boldsymbol{\alpha}}(t)$, where the mass variable has been summed.
\medskip

{\bf Definition 3.} (Marginal probability density function.) The marginal probability density is
\begin{equation}\label{eq:marginal_pdf}
p_{\boldsymbol{\alpha}}(t) = \sum_m p_{\boldsymbol{\alpha},m}(t),
\end{equation}
where $p_{\boldsymbol{\alpha},m}(t)$ satisfies (\ref{eq:master_eq}) with transition rates (\ref{eq:transition_rates}) and (\ref{eq:transition_zero_rates}).
\medskip

The marginal PDF in Definition 3 satisfies what will be referred to as the marginalized master equation, found by summing over the mass variable $m$ on both sides of (\ref{eq:master_eq}). It is important to note that a sum over $m$ on the right-hand side of master equation (\ref{eq:master_eq}) involves careful consideration of rules (\ref{eq:transition_rates}) since the transition rates $T_{(\boldsymbol{\alpha},m),(\boldsymbol{\alpha}',m')}$ depend on $m$ in addition to the PDF $p_{\boldsymbol{\alpha},m}(t)$.
 
Next, we give the marginalized master equation and rules for the associated transition rates. The master equation for the marginalized PDF of Definition 3 is
\begin{align}\label{eq:master_eq_marginal}
\dot{p}_{\boldsymbol{\alpha}}(t) 
    &= \sum_{\boldsymbol{\alpha}'} T_{\boldsymbol{\alpha},\boldsymbol{\alpha}'} p_{\boldsymbol{\alpha}'}(t) \nonumber\\
    &= D\sum_{\boldsymbol{\alpha}'} \left[ A_{\boldsymbol{\alpha},\boldsymbol{\alpha}'} + \epsilon B_{\boldsymbol{\alpha},\boldsymbol{\alpha}'} \right] p_{\boldsymbol{\alpha}'}(t)~.
\end{align}
Here, $A_{\boldsymbol{\alpha},\boldsymbol{\alpha}'}$ accounts for the atomistic processes of attachment and detachment at the step edge, and atom hopping on each terrace, as described in~\cite{PatroneMargetis14}; and $B_{\boldsymbol{\alpha},\boldsymbol{\alpha}'}$ together with the non-dimensional parameter $\epsilon \equiv F/D$ account for material deposition onto the surface from above. In (\ref{eq:master_eq_marginal}), $\epsilon$ plays the role of a P\' eclet number, measuring the deposition rate relative to terrace diffusion. Note that the symbol $R$ has previously been used for the inverse ratio, $R=D/F$, in part of the physics literature~\cite{AmarFamily95}.

The scaled, non-zero transition rates $A_{\boldsymbol{\alpha},\boldsymbol{\alpha}'}$ and $B_{\boldsymbol{\alpha},\boldsymbol{\alpha}'}$ can be described by rules similar to those in (\ref{eq:transition_rates}), viz.

\begin{align}\label{eq:transition_rates_marginal}
A_{\boldsymbol{\alpha},\boldsymbol{\alpha}'}=& 1~, \qquad\ \mbox{if  } |\boldsymbol{\alpha}| = |\boldsymbol{\alpha}'| \mbox{  and  } \left|\boldsymbol{\alpha}\setminus\boldsymbol{\alpha}'\right| = 1 \mbox{  and  } \Big| ||\boldsymbol{\alpha}\setminus\boldsymbol{\alpha}'|| - ||\boldsymbol{\alpha}'\setminus\boldsymbol{\alpha}||\Big| = 1~; \nonumber \\
A_{\boldsymbol{\alpha},\boldsymbol{\alpha}'}=& \phi_\pm~, \quad\ \  \mbox{if  }\ |\boldsymbol{\alpha}| = |\boldsymbol{\alpha}'| - 1 \mbox{  and  } \boldsymbol{\alpha}'\setminus\boldsymbol{\tilde{\alpha}} = \{\pm 1\}~;  \nonumber\\ 
A_{\boldsymbol{\alpha},\boldsymbol{\alpha}'} =& k\phi_\pm~,\quad\,  \mbox{if  }\ |\boldsymbol{\alpha}| = |\boldsymbol{\alpha}'| + 1 \mbox{  and  } \boldsymbol{\alpha}\setminus\boldsymbol{\tilde{\alpha}}' = \{\pm 1\}~; \nonumber\\
B_{\boldsymbol{\alpha},\boldsymbol{\alpha}'}=& \frac{1}{N-1}~,\quad  \mbox{if  } |\boldsymbol{\alpha}| = |\boldsymbol{\alpha}'| + 1 \mbox{  and  } \left|\boldsymbol{\alpha}\setminus\boldsymbol{\alpha}'\right| = 1;  \nonumber\\
A_{\boldsymbol{\alpha}',\boldsymbol{\alpha}'}=& -\sum\limits_{\boldsymbol{\alpha}\atop \boldsymbol{\alpha}\neq \boldsymbol{\alpha}'} A_{\boldsymbol{\alpha},\boldsymbol{\alpha}'}~,\quad
B_{\boldsymbol{\alpha}',\boldsymbol{\alpha}'}= -\sum\limits_{\boldsymbol{\alpha}\atop \boldsymbol{\alpha}\neq \boldsymbol{\alpha}'} B_{\boldsymbol{\alpha},\boldsymbol{\alpha}'} \quad  \mbox{for all  } \boldsymbol{\alpha}'~.  
\end{align}

In the spirit of~\cite{PatroneMargetis14}, one may view master equation~\eqref{eq:master_eq_marginal} as a kinetic hierarchy of coupled particle equations for adatoms.
For fixed number of adatoms, $|\boldsymbol{\alpha}|=n$, a combinatorial argument entails that there are 
\begin{equation}\label{eq:num_states}
\omega(n)=\left( \begin{array}{c} n+N-2 \\ n \end{array} \right)
\end{equation}
distinct atomistic configurations on the 1D lattice of size $N$ ($N\ge 2$). By~\eqref{eq:num_states}, $\ln\omega(n)$ grows as $\mathcal O(n)$ for $n\gg 1$ with $n=\mathcal O(N)$.  Evidently, situations with high enough supersaturation imply the contribution to $p_{\boldsymbol{\alpha}}$ from states that have many adatoms, i.e., large $|\boldsymbol{\alpha}|$. In~\cite{PatroneMargetis14,PatroneEinsteinMargetis14} this complication is avoided by restricting attention to the dilute regime, in which the dynamics of \eqref{eq:master_eq_marginal} are dominated primarily by $0$- and $1$-particle states. In the present work, we aim to account for higher particle states. Unless we state otherwise, we assume that the number of adatoms cannot exceed a certain bound, $M$: $|\boldsymbol{\alpha}|\le M$, where $M$ is a fixed yet arbitrary positive integer.

%%%%%%%%%%%%%%%%%%%%%%%%%%%%%%%%%%%%%%%%%%%%%%%%%%%%%%%%%%%%%%%%%%%
%%%%%%%%%%%%%%%%%%%%%%%%%%%%%%%%%%%%%%%%%%%%%%%%%%%%%%%%%%%%%%%%%%%
\section{KRSOS model: Maximum principle and steady state solution}
\label{sec:Max-Steady_SOS}
%%%%%%%%%%%%%%%%%%%%%%%%%%%%%%%%%%%%%%%%%%%%%%%%%%%%%%%%%%%%%%%%%%%
%%%%%%%%%%%%%%%%%%%%%%%%%%%%%%%%%%%%%%%%%%%%%%%%%%%%%%%%%%%%%%%%%%%
In this section, we point out a significant property of the KRSOS model. First, we state and prove a maximum principle for marginalized master equation~\eqref{eq:master_eq_marginal}, which forms an extension of the maximum principle in~\cite{PatroneMargetis14} (Section~\ref{subsec:max}). This principle relies on the presumed existence of a steady-state solution. For $F=0$, the steady-state solution is known to correspond to the equilibrium distribution of adatoms~\cite{PatroneMargetis14}. Here, we derive an exact, closed-form expression for this equilibrium solution
via the canonical ensemble~\cite{Huang} (Section~\ref{subsec:EquilibriumNoDep}).  We also discuss the steady-state solution of~\eqref{eq:master_eq_marginal} for sufficiently small $F$, $F\neq 0$, by assuming that a finite number of particle states contribute to the governing kinetic hierarchy (Section~\ref{subsec:SteadyStateSOS}).

%%%%%%%%%%%%%%%%%%%%%%%%%%%%%%%%%%%%%%%%%%%%%%%%%%%%%%
\subsection{Maximum principle for master equation}
\label{subsec:max}
%%%%%%%%%%%%%%%%%%%%%%%%%%%%%%%%%%%%%%%%%%%%%%%%%%%%%%
Now consider marginalized master equation~\eqref{eq:master_eq_marginal} with $\epsilon\ge 0$.

{\bf Proposition 1.} {\em If a non-trivial steady-state solution, $p_{\boldsymbol\alpha}^{ss}$, of~\eqref{eq:master_eq_marginal} exists, then any solution $p_{\boldsymbol\alpha}(t)$ satisfies}
\begin{equation}\label{eq:maximum_principle}
\max_{\boldsymbol{\alpha}} \frac{p_{\boldsymbol{\alpha}}(t)}{p_{\boldsymbol{\alpha}}^{ss}} \leq \max_{\boldsymbol{\alpha}} \frac{p_{\boldsymbol{\alpha}}(0)}{p_{\boldsymbol{\alpha}}^{ss}}~,\quad  t>0~.
\end{equation}

We proceed to prove Proposition~1 by invoking the identity
\begin{equation}\label{eq:ss_condition}
\sum_{\boldsymbol{\alpha}'} T_{\boldsymbol{\alpha},\boldsymbol{\alpha}'} p_{\boldsymbol{\alpha}'}^{ss} = 0~.
\end{equation}

{\em Proof of Proposition 1.} Equation~\eqref{eq:master_eq_marginal} can be written as
\begin{align}\label{eq:mp_algebra}
\dot{p}_{\boldsymbol{\alpha}}(t) 
    &= \sum_{\boldsymbol{\alpha}'} T_{\boldsymbol{\alpha},\boldsymbol{\alpha}'} p_{\boldsymbol{\alpha}',m'}(t) \notag \\
    &= T_{\boldsymbol{\alpha},\boldsymbol{\alpha}} p_{\boldsymbol{\alpha}}(t) + \sum_{\boldsymbol{\alpha}' \neq \boldsymbol{\alpha}} T_{\boldsymbol{\alpha},\boldsymbol{\alpha}'} p_{\boldsymbol{\alpha}'}(t) \notag \\
    &= T_{\boldsymbol{\alpha},\boldsymbol{\alpha}} p_{\boldsymbol{\alpha}}^{ss} \frac{p_{\boldsymbol{\alpha}}(t)}{p_{\boldsymbol{\alpha}}^{ss}} + \sum_{\boldsymbol{\alpha}' \neq \boldsymbol{\alpha}} T_{\boldsymbol{\alpha},\boldsymbol{\alpha}'} p_{\boldsymbol{\alpha}'}^{ss} \frac{p_{\boldsymbol{\alpha}'}(t)}{p_{\boldsymbol{\alpha}'}^{ss}} \notag \\
    &= \sum_{\boldsymbol{\alpha}' \neq \boldsymbol{\alpha}} T_{\boldsymbol{\alpha},\boldsymbol{\alpha}'} p_{\boldsymbol{\alpha}'}^{ss} \left\{ \frac{p_{\boldsymbol{\alpha}'}(t)}{p_{\boldsymbol{\alpha}'}^{ss}} - 
    \frac{p_{\boldsymbol{\alpha}}(t)}{p_{\boldsymbol{\alpha}}^{ss}} \right\}~.
\end{align}
Note that $T_{\boldsymbol{\alpha},\boldsymbol{\alpha}'} p_{\boldsymbol{\alpha}'}^{ss} \geq 0$ for all $\boldsymbol{\alpha}' \neq \boldsymbol{\alpha}$. Thus, the sign of $\dot{p}_{\boldsymbol{\alpha}}(t)$ is determined by the quantity in brackets. In particular, if $\boldsymbol{\alpha}$ maximizes (minimizes) $p_{\boldsymbol{\alpha}'}(t)/p_{\boldsymbol{\alpha}'}^{ss}$ over all $\boldsymbol{\alpha}'$, then $\dot{p}_{\boldsymbol{\alpha}}(t) \leq 0$ ($\dot{p}_{\boldsymbol{\alpha}}(t) \geq 0$). 
This assertion implies the desired maximum principle (and corresponding minimum principle), thus concluding the proof. $\square$
\medskip

Proposition 1 states that, relative to the steady state, the solution $p_\alpha(t)$ of (\ref{eq:master_eq_marginal}) will never deviate more than the initial data. In other words, the system cannot be driven away from the steady state distribution.
\medskip

{\bf Remark 1.} If the initial data $p_{\boldsymbol{\alpha}}(0)$ satisfies
\begin{equation}\label{eq:inital_data_near_eq}
 \max_{\boldsymbol{\alpha}} \frac{p_{\boldsymbol{\alpha}}(0)}{p_{\boldsymbol{\alpha}}^{ss}} \leq C~, 
\end{equation}
for a parameter-independent constant $C$, then Proposition 1 implies $p_{\boldsymbol{\alpha}}(t) \lesssim p_{\boldsymbol{\alpha}}^{ss}$ for all $t>0$. This property will enable us to estimate certain averages in Section~\ref{sec:DiscreteBCF}.

{\bf Remark 2.} In general, Proposition 1 cannot be applied to the full mass-dependent master equation (\ref{eq:master_eq}) since the assumption of existence of a steady-state, and therefore equation (\ref{eq:ss_condition}), is violated when $\epsilon>0$. To see this, consider (\ref{eq:master_eq}) after marginalizing in $\boldsymbol{\alpha}$. The resulting equation,
\begin{equation}\label{eq:mass_master_eq}
 \dot{p}_{m}(t) = F\left[p_{m-1}(t) - p_{m}(t)\right],
\end{equation}
subject to $p_0(0) = 1$, is satisfied by the Poisson distribution $p_m(t) = \frac{(Ft)^m e^{-Ft}}{m!}$, for which $p_m(t)=0$ for any $m=\mathcal O(1)$ in the limit as $t\to \infty$. Thus, no non-trivial steady state exists.

%%%%%%%%%%%%%%%%%%%%%%%%%%%%%%%%%%%%%%%%%%%%%%%%%%%%%%%%%%%%%%%%%%%
\subsection{Equilibrium distribution, $F=0$} 
\label{subsec:EquilibriumNoDep}
%%%%%%%%%%%%%%%%%%%%%%%%%%%%%%%%%%%%%%%%%%%%%%%%%%%%%%%%%%%%%%%%%%%

Next, we discuss the case without external deposition, $F=0$ ($\epsilon=0$). For this purpose we will assume that the system is initially in a state whose mass is $m_0$ with probability one. Then, the use of mass index $m$ for states of the KRSOS model is unnecessary since $m=m_0$ for all time $t>0$. In this situation, the full master equation (\ref{eq:master_eq}) is equivalent to the marginalized master equation (\ref{eq:master_eq_marginal}). With this in mind, we will refer to the latter equation in the case $\epsilon=0$.

When external deposition is absent, master equation~\eqref{eq:master_eq_marginal} has an equilibrium solution, $p_{\boldsymbol{\alpha}}^{eq}$, by Kolmogorov's criterion~\cite{PatroneMargetis14}. 
This equilibrium solution was given in~\cite{PatroneMargetis14} by recourse to the canonical ensemble of statistical mechanics. We will follow the same approach and improve the result of \cite{PatroneMargetis14} by representing $p_{\boldsymbol{\alpha}}^{eq}$ in simple closed form.

Recall that $Dk\phi_\pm$ is the detachment rate, where $k = \exp[-E_b/(k_BT)]$ and $E_b$ measures the energy of the adatom resulting from detachment of an edge atom.  

We apply the formalism of the canonical ensemble to particle states of our system~\cite{Huang}. In the KRSOS model, the energy of each adatom configuration, $\boldsymbol{\alpha}$, is simply $|\boldsymbol{\alpha}|E_b$. By applying the Boltzmann-Gibbs distribution at equilibrium, we can assert that the probability of having $n$ adatoms is ${\rm Prob}\{|\boldsymbol{\alpha}|=n\}=c\, \exp[-nE_b/(k_BT)]$ where $c$ is a normalization constant. Consequently, the partition function, $Z$, for adatoms is computed as
\begin{align}\label{eq:partition_function}
Z =& \sum\limits_{\boldsymbol{\alpha}} e^{-|\boldsymbol{\alpha}|E_b/(k_BT)}\nonumber\\
  =& \sum_{n=0}^\infty \omega(n) \,e^{-nE_b/(k_BT)}=\frac{1}{(1-k)^{N-1}}~,
\end{align}
by using~\eqref{eq:num_states} and the binomial expansion. Thus, the equilibrium solution is
\begin{equation}\label{eq:p_eq}
p_{\boldsymbol{\alpha}}^{eq} = \frac{1}{Z} e^{-|\boldsymbol{\alpha}|E_b/T} 
  = (1-k)^{N-1} k^{|\boldsymbol{\alpha}|}~.
\end{equation}
Notice that $p_{\boldsymbol{\alpha}}^{eq} \propto k^{|\boldsymbol{\alpha}|}$ if $k\ll 1$; cf. \cite{PatroneMargetis14}.

%%%%%%%%%%%%%%%%%%%%%%%%%%%%%%%%%%%%%%%%%%%%%%%%%%%%%%
\subsection{Steady state of atomistic model with $F>0$} 
\label{subsec:SteadyStateSOS}
%%%%%%%%%%%%%%%%%%%%%%%%%%%%%%%%%%%%%%%%%%%%%%%%%%%%%%
In this subsection, we discuss a plausible steady-state solution of our KRSOS model for sufficiently small $\epsilon$ ($\epsilon=F/D$). To this end, we employ a formal argument based on the assumption that only a finite number of particle (adatom) states contribute to the system evolution; 
$|\boldsymbol\alpha|\le M$ for arbitrary yet fixed $M$. 

In contrast to the case with $F=0$, for which the Kolmogorov criterion holds and entails the existence of the equilibrium solution ~\cite{PatroneMargetis14}, microscopic reversibility is lost for nonzero deposition flux, $F>0$. In fact, by use of a known kinetic model, in 
Appendix~\ref{app:BirthDeath} we show that no steady-state solution of master equation~\eqref{eq:master_eq_marginal} exists in the kinetic regime with $\epsilon > \phi_+ + \phi_-$.
This latter condition implies that the external deposition rate is larger than the rate of attachment; consequently, from a physical viewpoint, a steady accumulation of adatoms on the terrace can occur for long enough times. 

In the remainder of this section, we make the conjecture (but do not prove) that a finite number of particle states contribute to the system evolution. Accordingly, we restrict attention to the kinetic regime with sufficiently small $\epsilon$ ($\epsilon\ll 1$). Our conjecture
is favored by KMC simulations, a sample of which is shown in Fig.~\ref{fig:simulation_mp_states}.
\medskip

{\renewcommand*{\arraystretch}{0.5}
\begin{figure}[!h]
$\begin{array}{ccc}
    \includegraphics[width=0.34\textwidth]{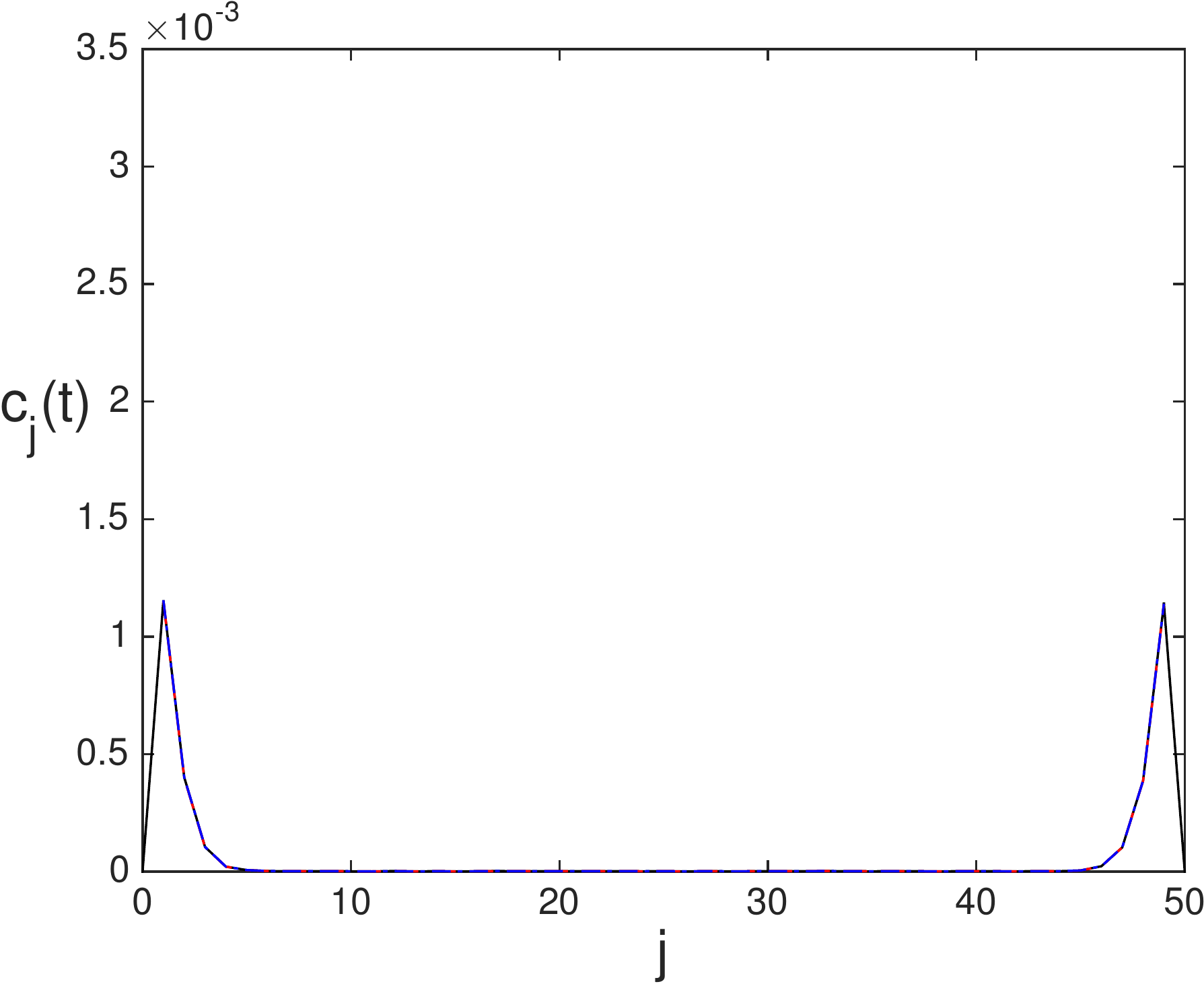} &
    \includegraphics[width=0.34\textwidth]{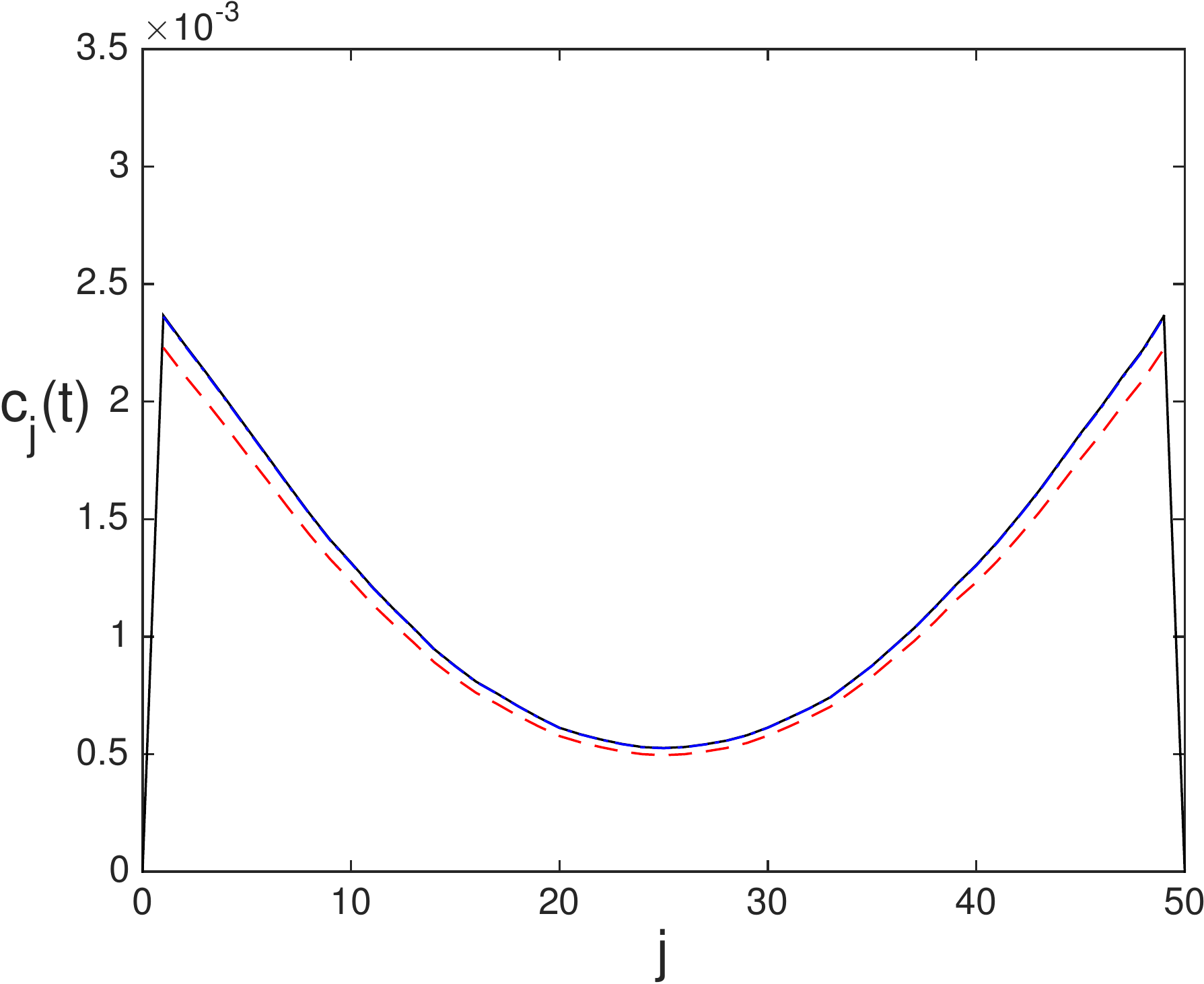} &
    \includegraphics[width=0.34\textwidth]{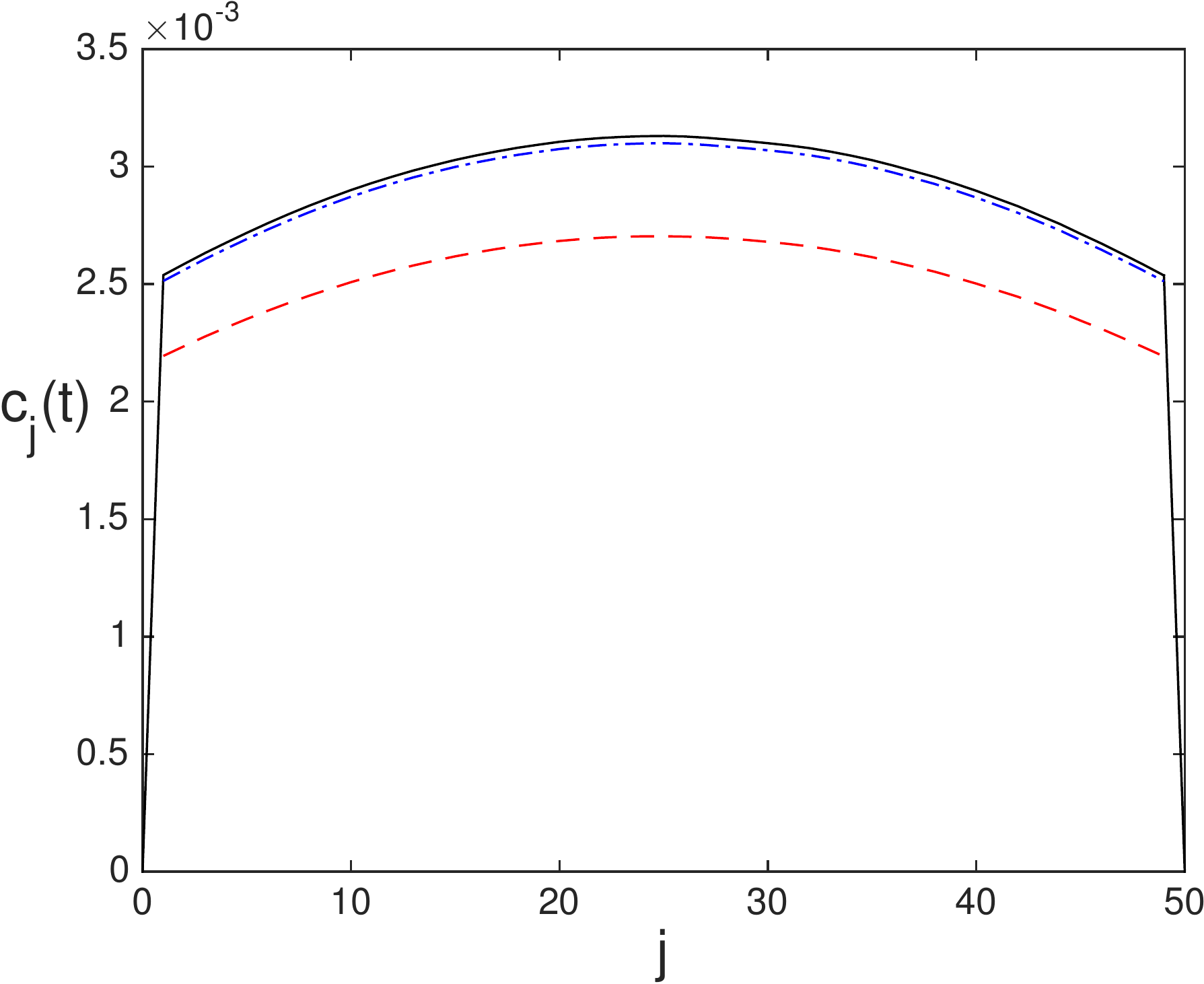}\\
    \mathbf{(a)} & \mathbf{(b)} & \mathbf{(c)}
\end{array}$
\caption{(Color online) Snapshots of the number, $c_j(t)$, of adatoms per lattice site, $j$, versus $j$ (Lagrangian coordinate) for fixed time, $t$, by KMC simulations, with: $N=50$, $D=10^{10}$ sec$^{-1}$, $k=2.5\times 10^{-3}$, $\phi_\pm=1$, and $F=10^7$ sec$^{-1}$ ($\epsilon=10^{-3}$). 
The time of simulation increases from left to right: (a) $t=10^{-10}$; (b) $t=10^{-8}$; and (c) $t=10^{-6}$. 
In each plot, the red dashed curve shows the contribution to $c_j(t)$ from $1$-particle states; the blue dash-dot curve includes both $1$- and $2$-particle states; and the black solid curve is the overall adatom number density. 
The standard deviation of densities in $10$ ensembles of $10^6$ simulations estimate errors smaller than the thickness of the plotted curves.}
\label{fig:simulation_mp_states} 
\end{figure}
} 

By enforcement of restriction $|\boldsymbol{\alpha}| \leq M$, master equation~\eqref{eq:master_eq_marginal} reduces to 
\begin{align}\label{eq:truncated_master_equation}
\dot{\mathbf{p}}^\epsilon(t) &= \mathfrak T\mathbf{p}^\epsilon(t) \notag \\
           &= D(\mathfrak A+\epsilon \mathfrak B)\mathbf{p}^\epsilon(t)~,
\end{align}
where $\mathbf{p}^\epsilon$ is the $\epsilon$-dependent vector of dimension $\Omega(M):=\sum_{n=0}^M \omega(n)=(M+N-1)!/[M!\,(N-1)!]$ formed by $p_{\boldsymbol{\alpha}}$, and the $\mathfrak T$ matrix is split into the (finite-dimensional) attachment/detachment matrix, $\mathfrak A$, and deposition matrix, $\epsilon\,\mathfrak B$, in correspondence to the
$A_{\boldsymbol\alpha,\boldsymbol\alpha'}$ and $\epsilon B_{\boldsymbol\alpha,\boldsymbol\alpha'}$ of~\eqref{eq:transition_rates}.  We impose the $\epsilon$-independent initial data $\mathbf p(0)=:\mathbf p_0$. 

The approximation of the full microscopic model by~\eqref{eq:truncated_master_equation} offers two obvious advantages. First, the time-dependent solution of~\eqref{eq:truncated_master_equation} can be expressed conveniently in terms of a matrix exponential. Second, the unique steady-state distribution of (\ref{eq:truncated_master_equation}) always exists; it is the normalized eigenvector of the $\mathfrak T$ matrix with zero eigenvalue\cite{VanKampen2007}.

We proceed to formally express the steady-state solution, $\mathbf p^{ss,\epsilon}$, of~\eqref{eq:truncated_master_equation} as an appropriate series expansion in $\epsilon$. This task can be carried out in several ways; for example, through the conversion of~\eqref{eq:truncated_master_equation} to a Volterra integral equation, an approach that we choose to apply here.
By treating $\epsilon D \mathfrak B \mathbf p^\epsilon(t)$ as a forcing term in~\eqref{eq:truncated_master_equation}, the method of variation of parameters yields
\begin{equation}\label{eq:integral_eq}
 \mathbf{p}^\epsilon(t) = \Phi(t)\, \left[ \mathbf{p}_0 + \epsilon D\int_0^t \Phi^{-1}(t') \mathfrak B \mathbf{p}^\epsilon(t') dt' \right]~,
\end{equation}
where $\Phi(t):=\exp(D\mathfrak A\,t)$. We mention in passing that, by the usual theory of Volterra equations, \eqref{eq:integral_eq} has a unique solution locally in time \cite{Tricomi1985}.

The matrix $\mathfrak A$ is diagonalizable because it corresponds to the transition matrix of a Markov process satisfying detailed balance \cite{VanKampen2007}. Thus, we apply the decomposition $\mathfrak A = V\Lambda V^{-1}$ where $\Lambda = \mbox{diag}\{\lambda_j\}_{j=1}^{\Omega(M)}$, $\{\lambda_j\}$ are the (non-dimensional) eigenvalues of $\mathfrak A$, and $V$ is a matrix whose column vectors are the respective eigenvectors. Let $\{\lambda_j\}_{j=1}^{\Omega(M)}$ be ordered, $0=\lambda_1>\lambda_2\ge \cdots\ge \lambda_{\Omega(M)}$ \cite{VanKampen2007}.
By $\Phi(t) = Ve^{Dt\Lambda}V^{-1}$, we have $\Phi^{-1}(t) = Ve^{-Dt\Lambda}V^{-1}$. Hence, \eqref{eq:integral_eq} is recast to
\begin{equation}\label{eq:integral_eq_conv}
 \mathbf{p}^\epsilon(t) =  Ve^{Dt\Lambda}V^{-1}  \mathbf{p}_0 + \epsilon D\int_0^t Ve^{D(t-t')\Lambda} V^{-1} \mathfrak B \mathbf{p}^\epsilon(t') dt'~.
\end{equation}
At this stage, a formula for $\mathbf p^\epsilon(t)$ ensues by standard methods. We resort to the Laplace transform, $\widehat{\mathbf{p}}^\epsilon(s)=\int_0^\infty e^{-st}\,\mathbf p^\epsilon(t)\,dt$, of $\mathbf p^\epsilon(t)$ with ${\rm Re} s> c >0$ for some suitable positive number $c$; by~\eqref{eq:integral_eq_conv}, we directly obtain
\begin{equation}\label{eq:laplace_trans_p}
 \widehat{\mathbf{p}}^\epsilon(s) = \left[I - \epsilon D \,V\mathfrak D(s)V^{-1}\mathfrak B\right]^{-1} V\mathfrak D(s)V^{-1}\mathbf{p}_0~,
\end{equation}
where $\mathfrak D(s) := \mbox{diag}\{(s-D\lambda_j)^{-1}\}_{j=1}^{\Omega(M)}$ and $I$ is the unit matrix. In~\eqref{eq:laplace_trans_p}, we assume that $\epsilon$ is small enough so that the requisite inverse matrix makes sense.

The next step in this approach is to compute the inverse transform of~\eqref{eq:laplace_trans_p}. However, in principle, this choice requires carrying out in the right half of the $s$-plane
the inversion of the matrix $I-\epsilon D\, V\mathfrak D(s) V^{-1}\mathfrak B$ for arbitrary $M$. This task is considerably simplified for the steady-state solution, $\mathbf p^{ss,\epsilon}$, in the limit $t\to\infty$, as shown in Appendix~\ref{app:Asymptotics}. The resulting formula reads
\begin{equation}\label{eq:p_ss}
 \mathbf{p}^{ss,\epsilon} = \mathbf{p}^{0} + \sum_{l=1}^\infty \left( -\epsilon \mathfrak A^\dagger \mathfrak B \right)^l \mathbf{p}^0~.
\end{equation}
In the above, $\mathbf{p}^0$ corresponds to the equilibrium solution in the absence of external deposition ($\epsilon=0$), and $\mathfrak A^\dagger$ denotes the Moore-Penrose pseudoinverse of $\mathfrak A$. Equation (\ref{eq:p_ss}) indicates the relative contributions of external deposition and diffusion/attachment/detachment processes to the steady state of the hypothetical $M$-particle KRSOS model underlying this calculation. 

Our heuristics leave several open questions regarding the meaning of~\eqref{eq:p_ss} for large particle number $M$. For instance, the behavior with $M$ of the bound for $\epsilon$ needed for convergence has not been addressed. A related
issue is to estimate the error by the truncation of series~\eqref{eq:p_ss}, after a finite number of terms are summed. We expect that~\eqref{eq:p_ss} ceases to be meaningful as $M\to\infty$.

%%%%%%%%%%%%%%%%%%%%%%%%%%%%%%%%%%%%%%%%%%%%%%%%%%%%%%%
%%%%%%%%%%%%%%%%%%%%%%%%%%%%%%%%%%%%%%%%%%%%%%%%%%%%%%%
\section{Averaging: Discrete mesoscale model} 
\label{sec:DiscreteBCF}
%%%%%%%%%%%%%%%%%%%%%%%%%%%%%%%%%%%%%%%%%%%%%%%%%%%%%%%
%%%%%%%%%%%%%%%%%%%%%%%%%%%%%%%%%%%%%%%%%%%%%%%%%%%%%%%

In this section, we heuristically show how discrete variables that form averages of microscale quantities on the lattice are plausibly related to mesoscale BCF-type observables of physical interest. 
In particular, diffusion equation (\ref{eq:diffusion_eq_bcf}) and step velocity law (\ref{eq:step_velocity_bcf}) have clear discrete counterparts. A noteworthy finding of our approach is a set of discrete boundary conditions at the step which {\em partially} agree with linear kinetic relation~\eqref{eq:lkr_bcf}. In fact, our exact formulas for discrete fluxes at the left and right of the step edge manifest corrections to~\eqref{eq:lkr_bcf}. 

%%%%%%%%%%%%%%%%%%%%%%%%%%%%%%%%%%%%%%%%%%%%%%%%%%%%%%%
\subsection{Definitions of basic averages} 
\label{subsec:Averages}
%%%%%%%%%%%%%%%%%%%%%%%%%%%%%%%%%%%%%%%%%%%%%%%%%%%%%%%

We now define the average step position, $\varsigma(t)$, in terms of probabilities $p_{\boldsymbol{\alpha},m}(t)$.
\medskip

{\bf Definition 4.} The average step position is
\begin{equation}\label{eq:varsigma}
\varsigma(t) = a\sum\limits_{\boldsymbol{\alpha},m} s(\boldsymbol{\alpha},m) p_{\boldsymbol{\alpha},m}(t)~,
\end{equation}
where $s(\boldsymbol{\alpha},m)$, given in Definition 2, is in an integer that denotes the site directly to the right of the step edge in the fixed reference frame of the 1D lattice.

Note that, in Definition 4, $\big||\boldsymbol{\alpha}|-(m-m_0)\big|$ is the number of adatoms that are exchanged with the step edge and, thus, solely contribute to step motion. 

Next, we define the adatom number per lattice site, which plays the role of the adatom density in the mesoscale picture. We use the following two interrelated variables. (i) The density, $c_{\hat{\jmath}}(t)$, of adatoms {\em relative} to the step, where $\hat{\jmath}$ counts the lattice sites to the {\em right} of the step. This $c_{\hat{\jmath}}(t)$ is a Lagrangian variable. (ii) The Eulerian density, $\rho_j(t)$, at site $j$ of the fixed 1D lattice.
\medskip

\begin{subequations}\label{eq:density}
{\bf Definition 5.} (i) The Lagrangian-type adatom density is defined by
\begin{align} \label{eq:c_j}
c_{\hat{\jmath}}(t) &= \sum\limits_{\boldsymbol{\alpha},m} \nu_{\hat{\jmath}}(\boldsymbol{\alpha}) p_{\boldsymbol{\alpha},m}(t)/a \nonumber\\
          &= \sum\limits_{\boldsymbol{\alpha}} \nu_{\hat{\jmath}}(\boldsymbol{\alpha}) p_{\boldsymbol{\alpha}}(t)/a,
\end{align}
where $\nu_{\hat{\jmath}}(\boldsymbol{\alpha})$ is the number of adatoms at site $\hat{\jmath}$ for a system with adatom configuration $\boldsymbol{\alpha}$. \par
(ii) The Eulerian adatom density is
\begin{equation}\label{eq:rho_j}
\rho_j(t) = \sum\limits_{\boldsymbol{\alpha},m} \nu_{j-s(\boldsymbol{\alpha},m)}(\boldsymbol{\alpha}) p_{\boldsymbol{\alpha},m}(t)/a~,
\end{equation}
where $j-s(\boldsymbol{\alpha},m)$ is the Lagrangian coordinate corresponding to Eulerian $j$.
%\par
%(iii) The Eulerian ``dilute'' adatom density is
%
%\begin{equation}\label{eq:varrho_j}
%\varrho_j(t) = \sum\limits_{\boldsymbol{\alpha},m} \mathbbold{1}(\nu_{j-s(\boldsymbol{\alpha},m)}(\boldsymbol{\alpha})>0) p_{\boldsymbol{\alpha},m}(t)/a~,
%\end{equation}
%
%which, in contrast to (\ref{eq:rho_j}), measures the {\em presence} of adatoms at a lattice site via the indicator function $\mathbbold{1}(\cdot)$ as opposed to the number of particles at that site.  
\end{subequations}
\medskip

Regarding Definition~5, it is important to note that $\nu_{\hat{\jmath}}(\boldsymbol{\alpha})$ is a factor counting the number of instances of $\hat{\jmath}$ in multiset $\boldsymbol{\alpha}$. Since both $\hat{\jmath}$ and $\boldsymbol{\alpha}$ use the same coordinate system, $\nu_{\hat{\jmath}}(\boldsymbol{\alpha})$ is independent of $m$, allowing for $c_{\hat{\jmath}}(t)$ to be expressed in terms of the marginalized PDF $p_{\boldsymbol{\alpha}}(t)$. In contrast, the definition of $\rho_j(t)$ cannot be written in terms of $p_{\boldsymbol{\alpha}}(t)$ because of the mass dependence in the index of $\nu_{j-s(\boldsymbol{\alpha},m)}(\boldsymbol{\alpha})$. %required to change coordinates.

The variable $c_{\hat{\jmath}}(t)$ is most useful in discrete equations for fluxes and boundary conditions at the step (Section~\ref{subsec:gen_flux}). 
On the other hand, $\rho_{j}(t)$ is more convenient to use in the derivation of the discrete diffusion equation, at lattice sites sufficiently away from the step edge (Section~\ref{subsec:discr_eq}).
%Because of step motion, these two variables near the step edge are related through a transformation inducing advection to the mesoscale description  (Section~\ref{subsec:discr_eq}).
\medskip

{\bf Remark 3.} For $F=0$, the equilibrium adatom density, $c^{eq}$, at any lattice site can be computed by (\ref{eq:p_eq}) and (\ref{eq:c_j}); or, alternatively, directly from partition function (\ref{eq:partition_function}) since the number of adatoms is constant everywhere on the terrace:
\begin{equation}\label{eq:ceq_def}
c^{eq} = \frac{\langle n \rangle}{(N-1)a} = \frac{k/a}{1-k}~,
\end{equation}
where $\langle n \rangle$ denotes the average total number of adatoms. 
\medskip

We now proceed to define the adatom fluxes at the step edge by virtue of our rules for atomistic transitions (Section~\ref{sssec:MasterEqn}). 
\medskip

{\bf Definition 6.}  The flux $J_\pm$ on the right ($+$, lower terrace) or left ($-$, upper terrace) of the step edge is
\begin{align} \label{eq:J_pm_def}
J_\pm(t) =& \mp \sum_{\boldsymbol{\alpha},m} \mathbbold{1}(\nu_{-1}(\boldsymbol{\alpha})=0) \notag\\
&\times \left[ T_{(\boldsymbol{\alpha}_\pm,m),(\boldsymbol{\alpha},m)} p_{\boldsymbol{\alpha},m}(t) -
                T_{(\boldsymbol{\alpha},m),(\boldsymbol{\alpha}_\pm,m)} p_{\boldsymbol{\alpha}_\pm,m}(t) \right]~.
\end{align}
Here, indicator function $\mathbbold{1}(\cdot)$ is one if its argument is true and zero otherwise, and $\boldsymbol{\alpha}_\pm = \boldsymbol{\tilde\alpha} \cup \{\pm 1\}$ denotes the adatom configuration resulting from a rightward ($+$) or leftward ($-$) detachment. The factor  $\mathbbold{1}(\nu_{-1}(\boldsymbol{\alpha})=0)$ excludes configurations that involve adatoms on top of the edge atom, i.e., those configurations for which detachment is forbidden and are inaccessible via attachment events. Additionally, the external deposition of adatoms does not contribute to the mass flux $J_\pm$, and therefore only a single value of $m$ enters (\ref{eq:J_pm_def}); see (\ref{eq:transition_zero_rates}).
%\medskip

%It should be noted that~\eqref{eq:J_pm_def} measures the mass flux at the step edge in terms of probability currents. This kinetic definition is expected to include two contributions: diffusion due to adatom hopping on each terrace; and advection because of step motion. The latter contribution should be relatively small for a dilute adatom gas~\cite{PatroneMargetis14}.

%%%%%%%%%%%%%%%%%%%%%%%%%%%%%%%%%%%%%%%%%%%%%%%%%%%%%%%
\subsection{Generalized discrete flux at step edge}
\label{subsec:gen_flux}
%%%%%%%%%%%%%%%%%%%%%%%%%%%%%%%%%%%%%%%%%%%%%%%%%%%%%%%

In this section, we derive an exact expression for the discrete flux at the step edge. This expression forms the basis for characterizing {\em corrections} to linear kinetic law~\eqref{eq:lkr_bcf} in the discrete setting.

By manipulating~(\ref{eq:J_pm_def}), we directly find a formula for the flux in terms of atomistic parameters, the differences $c_{\pm 1}(t) - c^{eq}$, and other discrete averages, viz.,
\begin{align}\label{eq:J_+_algebra}
J_+(t) =& \sum_{\boldsymbol{\alpha},m} \mathbbold{1}(\nu_{-1}(\boldsymbol{\alpha})=0) 
          \left[ T_{(\boldsymbol{\alpha}_+,m),(\boldsymbol{\alpha},m)} p_{\boldsymbol{\alpha},m}(t) -
                T_{(\boldsymbol{\alpha},m),(\boldsymbol{\alpha}_+,m)} p_{\boldsymbol{\alpha}_+,m}(t) \right] \notag \\
       =& Dk\phi_+ \sum_{\boldsymbol{\alpha},m} \mathbbold{1}(\nu_{-1}(\boldsymbol{\alpha})=0) p_{\boldsymbol{\alpha},m}(t) - D\phi_+ \sum_{\boldsymbol{\alpha},m} \mathbbold{1}(\nu_{1}(\boldsymbol{\alpha})=1) p_{\boldsymbol{\alpha},m}(t) \notag \\
       =& Dk\phi_+ \sum_{\boldsymbol{\alpha}} \mathbbold{1}(\nu_{-1}(\boldsymbol{\alpha})=0) p_{\boldsymbol{\alpha}}(t) - D\phi_+ \sum_{\boldsymbol{\alpha}} \mathbbold{1}(\nu_{1}(\boldsymbol{\alpha})=1) p_{\boldsymbol{\alpha}}(t) \notag \\
       =& Dk\phi_+ \left[ 1 - \sum_{\boldsymbol{\alpha}} \mathbbold{1}(\nu_{-1}(\boldsymbol{\alpha})>0) p_{\boldsymbol{\alpha}}(t) \right] \notag \\
       &\quad- D\phi_+ \sum_{\boldsymbol{\alpha}} \mathbbold{1}(\nu_{1}(\boldsymbol{\alpha})=1) \nu_{1}(\boldsymbol{\alpha}) p_{\boldsymbol{\alpha}}(t) \notag \\
       =& D\phi_+a \left[ \frac{k/a}{1-k}(1-k) - k\sum_{\boldsymbol{\alpha}} \mathbbold{1}(\nu_{-1}(\boldsymbol{\alpha})>0) p_{\boldsymbol{\alpha}}(t)/a \right] \notag \\
       &\quad- D\phi_+a \left[ c_1(t) - \sum_{\boldsymbol{\alpha}} \mathbbold{1}(\nu_{1}(\boldsymbol{\alpha})>1) \nu_{1}(\boldsymbol{\alpha}) p_{\boldsymbol{\alpha}}(t)/a \right] \notag \\
       =& -D\phi_+a\left[ c_1(t) - c^{eq} \right] - D\phi_+a f_+(t)~.
\end{align}
In the above, the second equality results from substitution of the transition rates~\eqref{eq:transition_rates}; the third equality results from summing over the mass variable; the fourth equality makes use of the complement rule, $\mathbbold{1}(\nu_{-1}(\boldsymbol{\alpha})=0) = 1-\mathbbold{1}(\nu_{-1}(\boldsymbol{\alpha})>0)$; and the fifth equality involves adding and subtracting the last sum to make $c_1(t)$ appear. Similar steps can be used to derive the corresponding formula for $J_-(t)$. Together, these fluxes can be written as 
\begin{equation}\label{eq:J_pm}
J_\pm(t) = \mp D\phi_\pm a\left[ c_{\pm 1}(t) - c^{eq} \right] \mp D\phi_\pm a f_\pm(t)~.
\end{equation}
In~\eqref{eq:J_pm}, the first term on the right-hand side is the discrete analog of the linear kinetic relation of the BCF model. We invoke the definitions for $c_{\pm 1}(t)$ and $c^{eq}$ according to (\ref{eq:c_j}) and (\ref{eq:ceq_def}), respectively; and define 
\begin{subequations}\label{eq:f-flux}
\begin{align} \label{eq:f_+}
f_+(t) &= k\biggl[c^{eq} + \sum_{\boldsymbol{\alpha}} \mathbbold{1}(\nu_{-1}(\boldsymbol{\alpha})>0) p_{\boldsymbol{\alpha}}(t)/a \biggr] \notag \\
       &\quad - \sum_{\boldsymbol{\alpha}} \mathbbold{1}(\nu_{1}(\boldsymbol{\alpha})>1) \nu_{1}(\boldsymbol{\alpha}) p_{\boldsymbol{\alpha}}(t)/a
\end{align}
and
\begin{align} \label{eq:f_-}
f_-(t) &= k\left[c^{eq} + \sum_{\boldsymbol{\alpha}} \mathbbold{1}(\nu_{-1}(\boldsymbol{\alpha})>0) p_{\boldsymbol{\alpha}}(t)/a \right] \notag \\
       &\quad - \sum_{\boldsymbol{\alpha}} \mathbbold{1}(\nu_{1}(\boldsymbol{\alpha})>0) \nu_{-1}(\boldsymbol{\alpha}) p_{\boldsymbol{\alpha}}(t)/a \notag \\
       &\quad - \sum_{\boldsymbol{\alpha}} \mathbbold{1}(\nu_{1}(\boldsymbol{\alpha})=0) \mathbbold{1}(\nu_{-1}(\boldsymbol{\alpha})>1)\notag\\ &\qquad \times \left[\nu_{-1}(\boldsymbol{\alpha})-1\right] p_{\boldsymbol{\alpha}}(t)/a~.
\end{align}
\end{subequations}

Equations~\eqref{eq:f-flux} signify corrections to the discrete analog of the linear kinetic law of the BCF model; cf.~\eqref{eq:lkr_bcf} and~\eqref{eq:J_pm}. The corresponding {\em corrective} fluxes, $f_\pm$, measure the frequency by which the atomistic system visits configurations that forbid detachment of the edge atom or attachment of an adatom from the right ($+$) or left ($-$) of the step edge. We expect that the magnitudes of correction terms $f_\pm$ are negligibly small in the appropriate low-density regime for adatoms \cite{PatroneEinsteinMargetis14, PatroneMargetis14}. 

In Section~\ref{sec:BCFregime}, we find bounds for the above corrective fluxes, $f_\pm$. In particular, we describe systematically a parameter regime for $k$ and $F$ in which these corrections are negligible. In contrast, if $f_\pm(t)$ are important then the discrete fluxes $J_\pm(t)$ may no longer have a linear dependence on $c_{\pm1}(t)$, which in turn signifies the onset of high-supersaturation behavior; see Section~\ref{sec:Corrections}.

%%%%%%%%%%%%%%%%%%%%%%%%%%%%%%%%%%%%%%%%%%%%%%%%%%%%%%%
\subsection{Discrete diffusion and step velocity law}
\label{subsec:discr_eq}
%%%%%%%%%%%%%%%%%%%%%%%%%%%%%%%%%%%%%%%%%%%%%%%%%%%%%%%

In this section, we formally describe the change of the adatom density and average
step position with time. In addition, we complement~\eqref{eq:J_pm} with a mass-transport relation between discrete fluxes and densities near the step edge, which forms a discrete analog of Fick's law for diffusion. What we find are equations for discrete diffusion and discrete versions of Fick's law that have corrections terms resulting from atomistic configurations with multiple adatoms at the same lattice site. %The effect of advection, which comes from step motion, is also included. To address this effect, we follow a non-rigorous approach which warrants a physically consistent result in agreement with continuum theory, but leaves open questions in regard to details of derivation.

First, by Definition 5 and master equation (\ref{eq:master_eq}), the time evolution of $\rho_j(t)$ is described by
\begin{align}\label{eq:rho_j_dot_master_eqn}
\dot{\rho}_j(t) &= \sum\limits_{\boldsymbol{\alpha},m} \nu_{j-s(\boldsymbol{\alpha},m)}(\boldsymbol{\alpha})
\sum\limits_{(\boldsymbol{\alpha}',m') \atop \ne (\boldsymbol{\alpha},m)} \left[
T_{(\boldsymbol{\alpha},m),(\boldsymbol{\alpha}',m')} p_{\boldsymbol{\alpha}',m'}(t) -
T_{(\boldsymbol{\alpha}',m'),(\boldsymbol{\alpha},m)} p_{\boldsymbol{\alpha},m}(t) \right]/a \nonumber\\
  &= \sum\limits_{\boldsymbol{\alpha},m} \sum\limits_{\boldsymbol{\alpha}',m'} \left[\nu_{j-s(\boldsymbol{\alpha},m)}(\boldsymbol{\alpha}) - \nu_{j-s(\boldsymbol{\alpha}',m')}(\boldsymbol{\alpha}') \right] T_{(\boldsymbol{\alpha},m),(\boldsymbol{\alpha}',m')} p_{\boldsymbol{\alpha}',m'}(t)/a.
\end{align}
In (\ref{eq:rho_j_dot_master_eqn}) we invoke property (\ref{eq:transitions_probability_conservation}) of the transition rates to make the difference $\nu_{j-s(\boldsymbol{\alpha},m)}(\boldsymbol{\alpha})-\nu_{j-s(\boldsymbol{\alpha}',m')}(\boldsymbol{\alpha}')$ appear. By identifying values of this difference, most of which are zero, it is possible to simplify our formula for $\dot{\rho}_j(t)$ and express the right-hand side of (\ref{eq:rho_j_dot_master_eqn}) in terms of known averages. 
\medskip

{\bf Remark 4.} It should be noted that the derivation of equation (\ref{eq:rho_j_dot_master_eqn}) is independent of our definition of $\nu_{\hat{\jmath}}(\boldsymbol{\alpha})$. Equations similar to  (\ref{eq:rho_j_dot_master_eqn}) can be found using general properties of master equations describing Markov processes; namely that transitions must conserve probability. In particular, the above also applies if $\nu_{\hat{\jmath}}(\boldsymbol{\alpha})$ is replaced with $\mathbbold{1}(\nu_{\hat{\jmath}}(\boldsymbol{\alpha})>0)$, which amounts to using the definition of adatom density in \cite{PatroneMargetis14}.
\medskip

Now, by transition rates (\ref{eq:transition_rates}) and our definition of $\nu_{\hat{\jmath}}(\boldsymbol{\alpha})$, we write equation (\ref{eq:rho_j_dot_master_eqn}) as~\cite{Schneider2016}
\begin{align}\label{eq:rho_j_dot}
\dot{\rho}_j(t) 
    &= D\left[ \rho_{j-1}(t) -2\rho_{j}(t)+\rho_{j+1}(t) \right] + \frac{F}{(N-1)a}  \notag\\
    &- D\left[ R_{j-1}(t) -2R_{j}(t)+R_{j+1}(t) \right] \notag\\
    &- \sum\limits_{\boldsymbol{\alpha},m} \bigg\{ \delta_{j,s(\boldsymbol{\alpha},m)} \Big[ D\mathbbold{1}(\nu_{-1}(\boldsymbol{\alpha})>0) + D\mathbbold{1}(\nu_{1}(\boldsymbol{\alpha})>0) + \frac{F}{N-1} \Big] \notag\\
    &\quad\; -\delta_{j,s(\boldsymbol{\alpha},m)-1} \Big[ D\mathbbold{1}(\nu_{-1}(\boldsymbol{\alpha})>0) + Dk\phi_- \mathbbold{1}(\nu_{-1}(\boldsymbol{\alpha})=0) \notag\\
    &\qquad\qquad\qquad\qquad- D\phi_-\mathbbold{1}(\nu_{1}(\boldsymbol{\alpha})=0)\mathbbold{1}(\nu_{-1}(\boldsymbol{\alpha})>0) \Big] \notag \\
    &\quad\; - \delta_{j,s(\boldsymbol{\alpha},m)+1} \Big[ D\mathbbold{1}(\nu_{1}(\boldsymbol{\alpha})>0) + Dk\phi_+\mathbbold{1}(\nu_{-1}(\boldsymbol{\alpha})=0) \notag\\
    &\qquad\qquad\qquad\qquad - D\phi_+\mathbbold{1}(\nu_{1}(\boldsymbol{\alpha})=1) \Big] \bigg\} p_{\boldsymbol{\alpha},m}(t)/a~.
\end{align}
A few remarks on~\eqref{eq:rho_j_dot} are in order. The first two lines include a diffusion-type second-order difference scheme for $\rho_j(t)$ and the accompanying correction to discrete diffusion, respectively. The third line of (\ref{eq:rho_j_dot}) removes certain terms that do not contribute when $j=s(\boldsymbol{\alpha},m)$; no atoms are deposited to the step edge, for example. The remaining terms express boundary conditions at the right ($j=s(\boldsymbol{\alpha},m)+1$) or left ($j=s(\boldsymbol{\alpha},m)-1$) of the step; they contribute only when $j$ is in the vicinity of the step via the Kronecker delta functions. The high-occupation correction terms, $R_j(t)$, are defined as
\begin{equation}\label{eq:diffusion_corrections}
 R_j(t) = \sum\limits_{\boldsymbol{\alpha},m} \left[ \nu_{j-s(\boldsymbol{\alpha},m)}(\boldsymbol{\alpha}) - \mathbbold{1}(\nu_{j-s(\boldsymbol{\alpha},m)}(\boldsymbol{\alpha})>0) \right]p_{\boldsymbol{\alpha},m}(t)/a~.
\end{equation}
Equation (\ref{eq:diffusion_corrections}) measures discrete corrections related to certain correlated motion of adatoms. In particular, these corrections to discrete diffusion arise from the fact that the KRSOS model includes constant adatom hopping rates, regardless of the number of adatoms present at a given lattice site. In effect, atomistic configurations with multiple adatoms at the same lattice site introduce interactions between particles since only one is able to move. This high-occupancy effect can be seen in (\ref{eq:diffusion_corrections}) since $\nu_{j-s(\boldsymbol{\alpha},m)}(\boldsymbol{\alpha}) - \mathbbold{1}(\nu_{j-s(\boldsymbol{\alpha},m)}(\boldsymbol{\alpha})>0) \ne 0$ when two or more adatoms are at site $j$.

%In its present form, \eqref{eq:rho_j_dot} is slightly misleading, since it does not explicitly reveal advection terms coming from $\dot{\rho}_j(t)$, when $j$ is close enough to the step edge. Such terms should express the {\em jump} that the adatom density possibly exhibits across the edge, i.e., the property that $c_1(t)-c_{-1}(t)$ (in Lagrangian coordinates) is in principle an $\mathcal O(1)$ quantity. 

%To unveil this effect, we resort to Lagrangian coordinates on the lattice. If the Lagrangian index $\hat{\jmath}$ corresponds to site $j$ in the Eulerian reference frame, then the left-hand side of~\eqref{eq:rho_j_dot} is approximately written as 
%\begin{align}\label{eq:rho_j_dot_adv}
%\dot{\rho}_j(t)&\approx \frac{\rho_j(t+\Delta t)-\rho_j(t)}{\Delta t}\notag\\
%&\approx \frac{c_{\hat{\jmath}\mp 1}(t+\Delta t)-c_{\hat{\jmath}\mp 1}(t)}{\Delta t}+\frac{1}{\Delta t}[c_{\hat{\jmath}\mp 1}(t)-c_{\hat{\jmath}}(t)]~.
%\end{align}
%By this relation, the step edge is assumed to advance ($-$) or retreat ($+$) by one lattice spacing, $a$, within the appropriately chosen time $\Delta t$. Notice the somewhat inelegant use of a microscale timestep for $\rho_j(t)$ here. Now consider the second difference term on the right-hand side of~\eqref{eq:rho_j_dot_adv}, which yields the adatom density jump across the step edge. By Definition~2, no adatoms may occupy the site to the right of the step, therefore the adatom density at that site is zero, $c_0(t) \equiv 0$. Thus, for appropriately chosen $\hat{\jmath}$, (\ref{eq:rho_j_dot_adv}) corresponds to the desired boundary terms for advection.

The fourth and fifth lines of~\eqref{eq:rho_j_dot} suggest that
\begin{subequations}\label{eq:discrete_bc_rl_steps}
\begin{align}\label{eq:discrete_bc_l_steps}
 D\sum\limits_{\boldsymbol{\alpha},m} & \Big[ \mathbbold{1}(\nu_{-2}(\boldsymbol{\alpha})>0) - \mathbbold{1}(\nu_{-1} (\boldsymbol{\alpha})>0)  \notag \\
  &\quad + k\phi_- \mathbbold{1}(\nu_{-1}(\boldsymbol{\alpha})=0) - \phi_- \mathbbold{1}(\nu_{1}(\boldsymbol{\alpha})=0) \mathbbold{1}(\nu_{-1}(\boldsymbol{\alpha})>0)  \notag \\
  &= D\left[ c_{-2}(t) - c_{-1}(t) \right] - a^{-1}J_-(t) - D\left[ \hat{R}_{-2}(t) - \hat{R}_{-1}(t) \right]~,
\end{align}
where $J_-$ is defined by~(\ref{eq:J_pm_def}). The analogous result
can be reached for the last two lines of~(\ref{eq:rho_j_dot}) yielding the corresponding condition for $J_+$, at the left of the step edge:
\begin{align}\label{eq:discrete_bc_r_steps}
 D\sum\limits_{\boldsymbol{\alpha},m} & \Big[ \mathbbold{1}(\nu_{2}(\boldsymbol{\alpha})>0) - \mathbbold{1}(\nu_{1} (\boldsymbol{\alpha})>0)  \notag \\
  &\quad + k\phi_+ \mathbbold{1}(\nu_{-1}(\boldsymbol{\alpha})=0) - \phi_+ \mathbbold{1}(\nu_{1}(\boldsymbol{\alpha})=1)  \notag \\
  &= D\left[ c_{2}(t) - c_{1}(t) \right] + a^{-1}J_+(t) - D\left[ \hat{R}_{2}(t) - \hat{R}_{1}(t) \right]~.
\end{align}
\end{subequations}
It should be noted that the high-occupation terms $\hat{R}_{\hat{\jmath}}(t)$ found in equations (\ref{eq:discrete_bc_rl_steps}) are corrections of the same origin as (\ref{eq:diffusion_corrections}), but represented in Lagrangian coordinates. These are
\begin{equation}\label{eq:diffusion_corrections_lagrangian}
 \hat{R}_{\hat{\jmath}}(t) = \sum\limits_{\boldsymbol{\alpha}} \left[ \nu_{\hat{\jmath}}(\boldsymbol{\alpha}) - \mathbbold{1}(\nu_{\hat{\jmath}}(\boldsymbol{\alpha})>0) \right]p_{\boldsymbol{\alpha}}(t)/a~.
\end{equation}

In view of~\eqref{eq:discrete_bc_rl_steps}, we can now extract the discrete boundary terms at the step edge by setting $j=s(\boldsymbol\alpha,m)\pm1$ ($\hat{\jmath}=\pm 1$) in~\eqref{eq:rho_j_dot}. Hence, we find
\begin{equation}\label{eq:discrete_bc}
J_\pm(t) = \mp Da \left[ c_{\pm 2}(t) - c_{\pm 1}(t) \right] \mp Da \left[ \hat{R}_{\pm 2}(t) - \hat{R}_{\pm 1}(t) \right]~.
\end{equation}
%
%where the factor $\pm a/\Delta t$ entering this relation through~\eqref{eq:rho_j_dot_adv}   has been heuristically replaced by the average velocity, $\dot{\varsigma}(t)$, of the advancing ($+$) or retreating ($-$) step.
Equations~\eqref{eq:discrete_bc} are a semi-discrete version of Fick's law including corrections to diffusive fluxes at each side of the step edge.
\medskip

{\bf Remark 5.} In contrast to (\ref{eq:J-Fick_BCF}), the advection term at the step edge does not appear in~\eqref{eq:discrete_bc}. We have not been able to derive this term from the atomistic model. We attribute this inability to the fact that certain atomistic processes at the step edge are forbidden; see (\ref{eq:transition_zero_rates}a) and (\ref{eq:transition_zero_rates}b). In effect, these forbidden processes can cause adatoms to pile up in front of the step edge, thereby conserving mass at that atomistic scale. In Appendix \ref{app:Advection}, we develop a plausibility argument on the basis of the atomistic model for the appearance of the advection term, $\dot{\varsigma}~\partial \mathcal{C}/\partial \hat{x}$, away from the step edge; cf. (\ref{eq:J-Fick_BCF}).
\medskip

We now provide the anticipated mass conservation statement that involves the average step velocity, $\dot\varsigma(t)$; see Definition 4. This average is explicitly expressed by use of master equation~\eqref{eq:master_eq}, ensemble average~\eqref{eq:varsigma} for $\varsigma(t)$, and formula~\eqref{eq:J_pm_def} for fluxes $J_\pm$. It can be shown that~\cite{Schneider2016}
\begin{equation}\label{eq:step_velocity_discrete}
\dot{\varsigma}(t) = a\left[ J_-(t)-J_+(t) \right]~.
\end{equation}
We omit the details for derivation of this result.

In summary, motion laws~\eqref{eq:rho_j_dot} and~\eqref{eq:step_velocity_discrete} form discrete analogs to BCF-type equations \eqref{eq:step_velocity_bcf}-\eqref{eq:J-Fick_BCF}, notwithstanding advection.
%In particular, in our approach the advection term explicitly appearing in (\ref{eq:diffusion_eq_bcf}) in the step-continuum regime is reconciled with the present discrete setting in some asymptotic sense of sufficiently small lattice spacing.
The correction terms $R_j(t)$ and $\hat{R}_{\hat{\jmath}}(t)$, respectively appearing in equations~\eqref{eq:rho_j_dot} and~\eqref{eq:discrete_bc_r_steps}, are negligible if the system parameters are suitably scaled with the lattice spacing, $a$~\cite{PatroneMargetis14}. A systematic analysis of these corrections is given in Section~\ref{sec:BCFregime}.
%This discrepancy is due to the choice of coordinate systems: the former equation is %written in Lagrangian coordinates, while the latter is written in Eulerian form. 
%For reference, an equation for $\dot{c}_j(t)$, the Lagrangian corollary to %(\ref{eq:rho_j_dot)}), is included in \cite{PatroneEinsteinMargetis14}.

%%%%%%%%%%%%%%%%%%%%%%%%%%%%%%%%%%%%%%%%%%%%%%%%%%%%%%%
%%%%%%%%%%%%%%%%%%%%%%%%%%%%%%%%%%%%%%%%%%%%%%%%%%%%%%%
\section{Continuum step flow with estimates for discrete corrections} 
\label{sec:BCFregime}
%%%%%%%%%%%%%%%%%%%%%%%%%%%%%%%%%%%%%%%%%%%%%%%%%%%%%%%
%%%%%%%%%%%%%%%%%%%%%%%%%%%%%%%%%%%%%%%%%%%%%%%%%%%%%%%

In this section, we systematically derive the continuum step-flow equations of the discrete mesoscale model (Section~\ref{sec:DiscreteBCF}) in the limit where the lateral lattice spacing, $a$, approaches zero, and the microscale kinetic parameters scale properly with $a$. To investigate the behavior of discrete corrections with the kinetic parameters $k$ and $F$, we determine $L^\infty$-bounds for certain corrections via the ``maximum principle'', Proposition~1. Our estimates apply to (i) the correction terms $f_\pm(t)$  found in fluxes~\eqref{eq:J_pm}, and (ii) high-occupation corrections to discrete density, e.g. the terms $\hat{R}_{\hat{\jmath}}(t)$, described in Section~\ref{subsec:discr_eq}.

Our formal argument forms an extension of the deposition-free (with $F=0$) case~\cite{PatroneMargetis14} under the hypothesis that only a finite number of particle states contribute to system evolution; $|\boldsymbol\alpha|\le M$. Consequently, as $a\downarrow 0$ we extract a set of BCF-type equations for the moving step along with error estimates for the emerging diffusion equation and the linear kinetic law at the step edge. 

%%%%%%%%%%%%%%%%%%%%%%%%%%%%%%%%%%%%%
\subsection{Bounds for discrete corrections}
\label{subsec:BCFregime-bounds}
%%%%%%%%%%%%%%%%%%%%%%%%%%%%%%%%%%%%%

The first part of our program can be outlined as follows. First, in view of Remark~1 for the initial data $p_{\boldsymbol\alpha}(0)$, we estimate $f_\pm(t)$ and $\hat{R}_{\hat{\jmath}}(t)$ in terms of the marginal steady-state solution $p_{\boldsymbol\alpha}^{ss,\epsilon}$ of the master equation. Second, by invoking the $\epsilon$-series expansion of Section~\ref{subsec:SteadyStateSOS} for $p_{\boldsymbol\alpha}^{ss,\epsilon}$, we derive the desired estimates for $f_\pm(t)$ and $\hat{R}_{\hat{\jmath}}(t)$; these signify corrections to the linear kinetic law for the adatom fluxes, $J_\pm$, and high-occupation corrections to the discrete diffusion equation on the terrace, respectively. This second stage of our derivation of bounds on discrete corrections makes use of the equilibrium distribution $p_{\boldsymbol\alpha}^{eq}$, equation (\ref{eq:p_eq}), for $\epsilon^0$-terms in our formal expansion, and the $\epsilon^1$-terms found in (\ref{eq:p_ss}). In principle, higher order terms can also be computed, but are neglected in our analysis.

%This second stage of our derivation of bounds on discrete corrections uses the largest nonzero (negative) eigenvalue, $\lambda_2$, of matrix $\mathfrak A$ corresponding to transitions with $F=0$; $|\lambda_2|$, which expresses the rate by which the deposition-free system approaches equilibrium at long times, depends on system parameters $N,k$, and $\phi_\pm$. We comment on this dependence at the end of this section. %When introducing this prop, be sure to reference equations where f and R need estimating
\medskip

{\bf Proposition 2.} {\em The corrective fluxes (\ref{eq:f-flux}) at the step edge satisfy the estimate
\begin{equation} \label{eq:f_pm_estimate}
f_\pm(t) = \mathcal{O}\left(\frac{k}{1-k}\,\frac{k}{a},\, \frac{\epsilon N}{(1+\phi)a}\right)~,
\end{equation}
where $\phi = \phi_+$ or $\phi_-$. Similarly, the high-occupation corrections (\ref{eq:diffusion_corrections_lagrangian}) to densities satisfy
\begin{equation} \label{eq:R_j_estimate}
\hat{R}_{\hat{\jmath}}(t) = \mathcal{O}\left(\frac{k}{1-k}\,\frac{k}{a},\, \frac{\epsilon N}{a}\right)~.
\end{equation}
In these estimates, the constants entering the respective bounds are independent of time and parameters of the problem.}
\medskip

Thus, if we assume that $k/a=\mathcal{O}(1)$, $\phi \leq \mathcal{O}(1)$~\cite{PatroneMargetis14,LuLiuMargetis14} and $\epsilon < \mathcal{O}(a^2)$, we can assert that $f_\pm(t)$ can be neglected when compared to the linear-in-density part of the discrete flux $J_\pm(t)$. Furthermore, the corrections $\hat{R}_{\hat{\jmath}}(t)$ are small compared to density $c_{\hat{\jmath}}(t)$, and therefore $R_j(t) \ll \rho_j(t)$ as well. These controlled approximations reveal a kinetic regime in which equations~(\ref{eq:J_pm}), (\ref{eq:discrete_bc}), and the discrete diffusion equation in (\ref{eq:rho_j_dot}) reduce to discrete versions of linear kinetic law (\ref{eq:lkr_bcf}), Fick's law (\ref{eq:J-Fick_BCF}) and the continuum diffusion equation for density. %We place emphasis on the condition $\epsilon \ll a^2/L$, which implies that the deposition rate should be small compared to the equilibration rate of the deposition-free atomistic system.

The remainder of this subsection is devoted to proving Proposition 2. Note that the mass dependence of $f_\pm(t)$ and $\hat{R}_{\hat{\jmath}}(t)$ has already been summed out; see equations (\ref{eq:f-flux}) and (\ref{eq:diffusion_corrections_lagrangian}).

{\em Proof of Proposition 2}. We proceed to derive~\eqref{eq:f_pm_estimate} and (\ref{eq:R_j_estimate}) through heuristics. Define
\begin{subequations}\label{eq:f_ss}
\begin{align}
\tilde f_+^{ss,\epsilon} &:= k\left[c^{eq} + \sum_{\boldsymbol{\alpha}} \mathbbold{1}(\nu_{-1}(\boldsymbol{\alpha})>0) p_{\boldsymbol{\alpha}}^{ss,\epsilon}/a \right] \notag \\
         &\quad +\sum_{\boldsymbol{\alpha}} \mathbbold{1}(\nu_{1}(\boldsymbol{\alpha})>1) \nu_{1}(\boldsymbol{\alpha}) p_{\boldsymbol{\alpha}}^{ss,\epsilon}/a~, \label{eq:f_+_ss} \\
\tilde f_-^{ss,\epsilon} &:= k\left[c^{eq} + \sum_{\boldsymbol{\alpha}} \mathbbold{1}(\nu_{-1}(\boldsymbol{\alpha})>0) p_{\boldsymbol{\alpha}}^{ss,\epsilon}/a \right] \notag \\
         &\quad + \sum_{\boldsymbol{\alpha}} \mathbbold{1}(\nu_{1}(\boldsymbol{\alpha})>0) \nu_{-1}(\boldsymbol{\alpha}) p_{\boldsymbol{\alpha}}^{ss,\epsilon}/a \notag \\
         &\quad + \sum_{\boldsymbol{\alpha}} \mathbbold{1}(\nu_{1}(\boldsymbol{\alpha})=0) \mathbbold{1}(\nu_{-1}(\boldsymbol{\alpha})>1) \left[\nu_{-1}(\boldsymbol{\alpha})-1\right] p_{\boldsymbol{\alpha}}^{ss,\epsilon}/a~\mbox{, and }  \label{eq:f_-_ss} \\
\hat{R}_{\hat{\jmath}}^{ss,\epsilon} &:= \sum\limits_{\boldsymbol{\alpha}} \left[ \nu_{\hat{\jmath}}(\boldsymbol{\alpha}) - \mathbbold{1}(\nu_{\hat{\jmath}}(\boldsymbol{\alpha})>0) \right] p_{\boldsymbol{\alpha}}^{ss,\epsilon}/a~.
\end{align}
\end{subequations}
By Remark~1 and exact formulas~\eqref{eq:f-flux} and (\ref{eq:diffusion_corrections_lagrangian}) for $f_\pm(t)$ and $\hat{R}_{\hat{\jmath}}(t)$, respectively, we have
\begin{equation}\label{eq:f-ineq}
|f_\pm(t)|\lesssim \tilde f_\pm^{ss,\epsilon}~\mbox{, and }\ |\hat{R}_{\hat{\jmath}}(t)|\lesssim \hat{R}_{\hat{\jmath}}^{ss,\epsilon}~,~t>0~.
\end{equation}

Inequalities (\ref{eq:f-ineq}) are not particularly useful since they do not explicitly manifest the dependence on the kinetic parameters of interest. We need to use some results from Section \ref{subsec:SteadyStateSOS} in order to refine these estimates. 

In correspondence to~\eqref{eq:p_ss} for the truncated system, we write $p_{\boldsymbol{\alpha}}^{ss,\epsilon} = p_{\boldsymbol{\alpha}}^{ss,(0)} + \epsilon p_{\boldsymbol{\alpha}}^{ss,(1)} + \mathcal{O}(\epsilon^2)$, where $p_{\boldsymbol{\alpha}}^{ss,(l)}$ is the $l$-th order term of the underlying series expansion in $\epsilon=F/D$; in particular, $p_{\boldsymbol{\alpha}}^{ss,(0)} = p_{\boldsymbol{\alpha}}^{eq}$ is the zeroth-order contribution to the steady state. The constant entering the term $\mathcal O(\epsilon^2)$ may depend on parameters of the problem but is immaterial for our purposes. We will neglect terms with $l\ge 2$ in the $\epsilon$-expansion for $p_{\boldsymbol{\alpha}}^{ss,\epsilon}$.

By inspection of~\eqref{eq:f_ss}, we define the following sums.
\begin{subequations}\label{eq:S_n}
\begin{align}
 S_1^{(l)} &:= \sum_{\boldsymbol{\alpha}} \mathbbold{1}(\nu_{-1}(\boldsymbol{\alpha})>0) p_{\boldsymbol{\alpha}}^{ss,(l)}~, \label{eq:s1_n} \\
 S_2^{(l)} &:= \sum_{\boldsymbol{\alpha}} \mathbbold{1}(\nu_{1}(\boldsymbol{\alpha})>1) \nu_{1}(\boldsymbol{\alpha}) p_{\boldsymbol{\alpha}}^{ss,(l)}~, \label{eq:s2_n} \\
 S_3^{(l)} &:= \sum_{\boldsymbol{\alpha}} \mathbbold{1}(\nu_{1}(\boldsymbol{\alpha})>0) \nu_{-1}(\boldsymbol{\alpha}) p_{\boldsymbol{\alpha}}^{ss,(l)}~,  \label{eq:s3_n} \\
 S_4^{(l)} &:= \sum_{\boldsymbol{\alpha}} \mathbbold{1}(\nu_{1}(\boldsymbol{\alpha})=0) \mathbbold{1}(\nu_{-1}(\boldsymbol{\alpha})>1) \left[\nu_{-1}(\boldsymbol{\alpha})-1\right] p_{\boldsymbol{\alpha}}^{ss,(l)}~, \label{eq:s4_n} \\
 S_5^{(l)} &:= \sum\limits_{\boldsymbol{\alpha}} \left[ \nu_{\hat{\jmath}}(\boldsymbol{\alpha}) - \mathbbold{1}(\nu_{\hat{\jmath}}(\boldsymbol{\alpha})>0) \right] p_{\boldsymbol{\alpha}}^{ss,(l)}~. \label{eq:s5_n}
\end{align} 
\end{subequations}
Accordingly, formulas (\ref{eq:f_ss}) are recast to the forms
\begin{subequations}\label{eq:f_pm_ss_exp}
\begin{align} 
\tilde f_+^{ss,\epsilon} &= k c^{eq} + \frac{k}{a} \left[ S_1^{(0)} + \epsilon S_1^{(1)} \right] + \frac{1}{a} \left[ S_2^{(0)} + \epsilon S_2^{(1)} \right] + \mathcal{O}(\epsilon^2)~, \label{eq:f_pm_ss_exp+} \\
\tilde f_-^{ss,\epsilon} &= kc^{eq} + \frac{k}{a}\left[ S_1^{(0)} + \epsilon S_1^{(1)} \right] +\frac{1}{a} \left[ S_3^{(0)} + \epsilon S_3^{(1)} \right] \notag \\
         &\quad + \frac{1}{a} \left[ S_4^{(0)} + \epsilon S_4^{(1)} \right] + \mathcal{O}(\epsilon^2)~.\label{eq:f_pm_ss_exp-}
\end{align}
\end{subequations}

First, we compute $S_i^{(0)}$ ($i=1,\,2,\,3,\,4,\,5$), which amount to contributions from the equilibrium solution of the master equation, for $F=0$. By~(\ref{eq:p_eq}) and $0<k<1$, we write
\begin{align}\label{s1_0}
 S_1^{(0)} &= \frac{1}{Z} \sum_{\boldsymbol{\alpha}} \mathbbold{1}(\nu_{-1}(\boldsymbol{\alpha})>0) k^{|\boldsymbol{\alpha}|} \notag \\
	   &= \frac{1}{Z} \sum_{n=1}^\infty  \left( \begin{array}{c} n+N-3 \\ n-1 \end{array} \right) k^n = k~.
\end{align}
The binomial coefficient in~\eqref{s1_0} is the number of $n$-particle configurations with at least one adatom in the site immediately to the left of the step. Similarly, we have
\begin{align}\label{s2_0}
 S_2^{(0)} &= \frac{1}{Z} \sum_{\boldsymbol{\alpha}} \mathbbold{1}(\nu_{1}(\boldsymbol{\alpha})>1) \nu_{1}(\boldsymbol{\alpha}) k^{|\boldsymbol{\alpha}|} \notag \\
	   &= \frac{1}{Z} \sum_{l=2}^\infty  l k^l \sum_{n=l}^\infty \left( \begin{array}{c} n-l+N-3 \\ n-l \end{array} \right) k^{n-l} \notag \\
%       &= \frac{1}{Z} \sum_{l=2}^\infty lk^l \sum_{n=0}^\infty \left( \begin{array}{c} n+N-3 \\ n \end{array} \right) k^{n} \notag \\
	   &= (1-k)\sum_{l=2}^\infty lk^l = \frac{2k^2-k^3}{1-k}~,
\end{align}
\begin{align}\label{s3_0}
 S_3^{(0)} &= \frac{1}{Z} \sum_{\boldsymbol{\alpha}} \mathbbold{1}(\nu_{1}(\boldsymbol{\alpha})>0) \nu_{-1}(\boldsymbol{\alpha}) k^{|\boldsymbol{\alpha}|} \notag \\
	   &= \frac{k}{Z} \sum_{l=1}^\infty l k^l \sum_{n=l+1}^\infty  \left( \begin{array}{c} n-l-1+N-3 \\ n-l-1 \end{array} \right) k^{n-l-1} \notag \\
	   &= k (1-k) \sum_{l=1}^\infty l k^l  = \frac{k^2}{1-k}~,
\end{align}
\begin{align}\label{s4_0}
 S_4^{(0)} &= \frac{1}{Z} \sum_{\boldsymbol{\alpha}} \mathbbold{1}(\nu_{1}(\boldsymbol{\alpha})=0) \mathbbold{1}(\nu_{-1}(\boldsymbol{\alpha})>1) \left[\nu_{-1}(\boldsymbol{\alpha})-1\right] k^{|\boldsymbol{\alpha}|} \notag \\
	   &= \frac{k}{Z} \sum_{l=2}^\infty (l-1)k^{l-1} \sum_{n=l}^\infty \left( \begin{array}{c} n-l+N-4 \\ n-l \end{array} \right)k^{n-l} \notag \\
%           &= \frac{k}{Z} \sum_{l=1}^\infty lk^l \sum_{n=0}^\infty \left( \begin{array}{c} n+N-4 \\ n \end{array} \right)k^n \notag \\
           &= k(1-k)^2 \sum_{l=1}^\infty lk^l = k^2~,
\end{align}
and
\begin{align}\label{s5_0}
 S_5^{(0)} &= \frac{1}{Z} \sum\limits_{\boldsymbol{\alpha}} \left[ \nu_{\hat{\jmath}}(\boldsymbol{\alpha}) - \mathbbold{1}(\nu_{\hat{\jmath}}(\boldsymbol{\alpha})>0) \right] k^{|\boldsymbol{\alpha}|} \notag \\
	   &= \frac{1}{Z} \sum_{l=2}^\infty  (l-1) k^l \sum_{n=l}^\infty \left( \begin{array}{c} n-l+N-3 \\ n-l \end{array} \right) k^{n-l} \notag \\
	   &= k(1-k)\sum_{l=1}^\infty  l k^l = \frac{k^2}{1-k}~.
\end{align}
We have followed the convention that the index $l$ is used to account for restrictions on states coming from indicator functions and the index $n$ replaces the number of adatoms, $|\boldsymbol{\alpha}|$.

All that remains is to compute terms proportional
to $\epsilon$. This task calls for estimating the sums $S_i^{(l)}$, defined in~\eqref{eq:S_n}, for $l=1$. Hence, we would need to invoke the first-order steady-state solution, $p_{\boldsymbol{\alpha}}^{ss,(1)}$, of marginalized master equation~(\ref{eq:master_eq_marginal}). As alluded to in Section~\ref{subsec:SteadyStateSOS}, we do not have, strictly speaking, a simple closed-form solution. Instead, by restricting attention to a finite number of particle states ($|\boldsymbol\alpha|\le M$),
we provide approximations for the requisite (infinite) sums by finite sums via expansion~(\ref{eq:p_ss}). This approximation amounts to replacing the sums $S_i^{(1)}$ with quadratic forms of appropriately defined vectors, $\mathbf{y}_i\in \mathbbold{R}^{\Omega(M)}$; thus, we write $S_i^{(1)} \approx \mathbf{y}_i^T \mathfrak A^\dagger \mathbf{z}$, where the vectors $\mathbf{y}_i$ have entries that correspond to the indicator functions and/or $\nu_{\pm1}(\boldsymbol{\alpha})$ according to~(\ref{eq:S_n}) and $\mathbf{z} = -\mathfrak B \mathbf{p}^0$. 

Next, we obtain estimates for $S_i^{(1)}$ as follows (see Section~\ref{subsec:SteadyStateSOS}):
\begin{align}\label{eq:si_1}
|S_i^{(1)}| &\approx |\mathbf{y}_i^T \mathfrak A^\dagger \mathbf{z}| \notag\\
  &= \left|\left[\left(\frac{\mathbf{y}_i^T}{\mathbf{y}_i^T\mathbf{y}_i}\right)^T\right]^\dagger \mathfrak A^\dagger \left[\frac{\mathbf{z}^T}{\mathbf{z}^T\mathbf{z}}\right]^\dagger\right| \notag\\
  &= \left[ \frac{\left|\mathbf{z}^T \mathfrak A \mathbf{y}_i\right|}{\mathbf{y}_i^T\mathbf{y}_i \, \mathbf{z}^T\mathbf{z}} \right]^\dagger
  = \frac{\mathbf{y}_i^T\mathbf{y}_i \, \mathbf{z}^T\mathbf{z}}{\left|\mathbf{z}^T \mathfrak A \mathbf{y}_i\right|} \notag\\
  &\lesssim \left| \mathbf{z}^T \mathfrak A \mathbf{y}_i \right|^{-1}~,
\end{align}

whenever $\mathbf{z}^T \mathfrak A \mathbf{y}_i \ne 0$. The above calculation uses several properties of the Moore-Penrose pseudoinverse, i.e. $\mathbf{y}^\dagger = \mathbf{y}^T/(\mathbf{y}^T\mathbf{y})$ provided $\mathbf{y}\ne \mathbf{0}$ (second line), $(\mathfrak A\mathfrak B)^\dagger = \mathfrak B^\dagger\mathfrak A^\dagger$ (third line), and the pseudoinverse of a nonzero constant is its multiplicative inverse (fourth line). The last line of (\ref{eq:si_1}) results from observing that the numerator results in a fixed constant for each $i$.
%If $\mathbf{z}^T \mathfrak A \mathbf{y}_i = 0$, the right hand side of equation (\ref{eq:si_1}) should be replaced with zero, by definition of the Moore-Penrose pseudoinverse.
Since $\mathfrak A$, $\mathfrak B$, $\mathbf{p}^0$, and $\mathbf{y}_i$ are known, (\ref{eq:si_1}) is a computable estimate. After some linear algebra, we find
\begin{equation} \label{eq:si_1_values}
|S_i^{(1)}| \lesssim \left\{ \begin{array}{ll}
                     \frac{N}{1+\phi}~, & i=1,\,2,\,3,\,4, \\
                     N~, & i=5;
                     \end{array} \right.
\end{equation}
in the above, $\phi = \phi_+$ or $\phi_-$.  

In summary, if the initial data of the atomistic system is near the steady state in the sense of (\ref{eq:inital_data_near_eq}), then maximum principle (\ref{eq:maximum_principle}) implies~\eqref{eq:f-ineq}. Inequalities~\eqref{eq:f-ineq} along with~\eqref{eq:S_n}--\eqref{eq:si_1_values} 
yield estimates \eqref{eq:f_pm_estimate} and (\ref{eq:R_j_estimate}) for $0<k<1$ and sufficiently small $\epsilon$. 
This statement concludes our heuristic proof of Proposition 2. $\square$
\medskip

{\bf Remark 6.} The estimates in Proposition~2 are based upon several assumptions and approximations, including: (i) application of the maximum principle (Proposition~1); (ii) truncation of marginalized master equation for $\epsilon>0$; (iii) asymptotics for the Laplace transform and formal power-series expansion for steady-steady distribution, $p^{ss}_{\boldsymbol{\alpha}}$; and (iv) $L^\infty$-bounds for correction terms (\ref{eq:f-flux}) and (\ref{eq:diffusion_corrections_lagrangian}). Consequently, estimates (\ref{eq:f_pm_estimate}) and (\ref{eq:R_j_estimate}) are not expected to be optimal. In particular, the bounds involving $\epsilon$ can likely be improved.
\medskip

\subsection{BCF-type model as a scaling limit}
\label{subsec:BCF-lim}
%%%%%%%%%%%%%%%%%%%%%%%%%%%%%%%%%%%%%
 
By Proposition~2, we are now able to shed light on the scaling limit of the atomistic model,
which leads to the BCF-type description of Section~\ref{subsec:BCFmodel}. In view of (\ref{eq:R_j_estimate}), let us impose 
\begin{equation}\label{eq:suff-conds}
k= \mathcal O(a)~,\quad \epsilon = \mathcal O(a^3)~,
\end{equation}
which ensure that high-occupancy corrections $\hat{R}_{\hat{\jmath}}(t)$ and $R_j(t)$ are small compared to the corresponding densities. On the other hand, we assume $\rho_j(t) \to \rho(x,t)$ and $c_{\hat{\jmath}}(t) \to \mathcal{C}(\hat{x},t)$ as the lattice parameter approaches zero, $a \to 0$. The Taylor expansion of $\rho(x,t)$ about $x=ja$ yields
\begin{equation*}
 \frac{\rho_{j-1}(t) - 2\rho_j(t) + \rho_{j+1}(t)}{a^2}=\frac{\partial^2 \rho}{\partial x^2}(ja,t) + \mathcal{O}(a^2)~.
\end{equation*}
By the usual notion of macroscopic diffusion, it is natural to set
$\mathcal D:=Da^2=\mathcal{O}(1)$, the surface diffusivity; and $\mathcal F:=\frac{F}{(N-1)a}$, the deposition flux per unit length~\cite{PatroneMargetis14,LuLiuMargetis14}. Thus, the first two lines of~(\ref{eq:rho_j_dot}), with (\ref{eq:suff-conds}), imply
\begin{equation}\label{eq:diffusion_eq_eulerian}
 \frac{\partial \rho}{\partial t} = \mathcal{D} \frac{\partial^2 \rho}{\partial x^2} - \mathcal{D} \frac{\partial^2 \mathcal{R}}{\partial x^2} + \mathcal{F}~,
\end{equation}
as $a \to 0$, where $\mathcal{R}(x,t)$ is the continuum limit of $R_j(t)$. By (\ref{eq:suff-conds}), the terms $\mathcal{D}~\partial^2 \mathcal{R}/\partial x^2$ and $\mathcal{F}$ are $\mathcal{O}(a)$, and hence are negligible to leading order. With exception of the deposition term, the resulting leading-order equation for $\rho(x,t)$ is the anticipated diffusion equation in Eulerian coordinates; cf.~(\ref{eq:diffusion_eq_bcf}). By sharpening estimates for $\epsilon$ in Proposition~2, $\mathcal{F}$ may subsequently appear as a leading-order term.

Furthermore, the observation that $\hat{R}_{\hat{\jmath}}(t) = \mathcal{O}(a)$, enables us to obtain Fick's law of diffusion at the step edge by virtue
of~\eqref{eq:discrete_bc}. By Taylor expanding the Lagrangian adatom density, $\mathcal{C}$, about $\hat{x}=\hat{\jmath}a$ with $\hat{\jmath}=0$ for the right ($+$) side of the step and $\hat{\jmath}=N-1$ for the left ($-$) side of the step as $a\to 0$, we obtain the formulas
\begin{subequations}\label{eq:ficks_law_step}
\begin{align}
J_+(t) &= -\mathcal{D} \left.\frac{\partial \mathcal{C}}{\partial \hat{x}}\right|_{\hat{x}=0^+} + \mathcal O(a)~,\label{eq:ficks_law_r} \\
J_-(t) &= -\mathcal{D} \left.\frac{\partial \mathcal{C}}{\partial \hat{x}}\right|_{\hat{x}=L^-} + \mathcal O(a)~, \label{eq:ficks_law_l}
\end{align}
\end{subequations}
where $\hat{x}=\hat{\jmath}a$ and $L=Na=\mathcal O(1)$ by virtue of the screw-periodic boundary conditions. In the above, we used the expansions $c_2(t)-c_1(t)=a [(\partial\mathcal C/\partial \hat{x})|_{x=0^+}+\mathcal O(a)]$ and $c_{-2}(t)-c_{-1}(t)=a[-(\partial\mathcal C/\partial \hat{x})|_{x=L^-}+\mathcal O(a)]$. 

Equations (\ref{eq:ficks_law_step}) need to be complemented with kinetic boundary conditions at the step edge. Hence, we now apply~\eqref{eq:J_pm} with~\eqref{eq:f-flux} and~\eqref{eq:f_pm_estimate}. 
First, we set $\kappa_\pm:=D\phi_\pm a=\mathcal O(1)$~\cite{PatroneMargetis14,LuLiuMargetis14}. Second, by inspection of 
estimate~\eqref{eq:f_pm_estimate}, we assume $\phi_\pm \leq \mathcal{O}(1)$ along with (\ref{eq:suff-conds}). Consequently, we can assert that $|f_\pm(t)|= \mathcal O(a)$. Thus, by~\eqref{eq:J_pm} we obtain
\begin{subequations}\label{eq:lkr_coarse}
\begin{align}
 J_+(t)&= -\kappa_+ \left[\mathcal{C}(0^+,t) - c^{eq}\right]+\mathcal O(a)~,  \label{eq:lkr_coarse_r} \\
 J_-(t)&= \kappa_- \left[\mathcal{C}(L^-,t) - c^{eq}\right]+\mathcal O(a)~, \label{eq:lkr_coarse_lr}
\end{align}
\end{subequations}
as $a \to 0$; recall that $c^{eq}=(k/a)(1-k)^{-1}\approx k/a$ if $k=\mathcal O(a)$~\cite{PatroneMargetis14}.  

The last component of the BCF model that emerges from the discrete equations is the step velocity law. This law is provided by (\ref{eq:step_velocity_discrete}); in this equation, the factor multiplying the difference in the adatom flux across the step edge equals $\Omega/a'$. Thus, (\ref{eq:step_velocity_discrete}) is precisely (\ref{eq:step_velocity_bcf}) pertaining to the BCF model.

%%%%%%%%%%%%%%%%%%%%%%%%%%%%%%%%%%%
%%%%%%%%%%%%%%%%%%%%%%%%%%%%%%%%%%%
\section{Corrections to BCF linear kinetic relation: A numerical study} 
\label{sec:Corrections}
%%%%%%%%%%%%%%%%%%%%%%%%%%%%%%%%%%%
%%%%%%%%%%%%%%%%%%%%%%%%%%%%%%%%%%%

In this section, we carry out KMC simulations to illustrate the behavior of adatom fluxes at the step edge for high enough supersaturations, defined as $\sigma_\pm = c_{\pm1}/c^{eq}-1$. We demonstrate that in this regime of high detachment or deposition flux, described in more detail below, the boundary condition for the adatom fluxes at the step edge can deviate significantly from linear kinetic relation~\eqref{eq:lkr_coarse}; thus, in principle the corrective fluxes, $f_\pm(t)$, may {\em not} be negligible. Since we have been unable to express these contributions, $f_\pm$, in terms of mesoscale quantities such as the adatom density, KMC simulations remain our primary tool for describing the high-supersaturation effects. For a discussion of this point, see Section~\ref{sec:Discussion}.

First, we make an educated yet empirical attempt to outline plausible conditions by which the conventional BCF model, particularly the linear kinetic relation for the flux, becomes questionable. By estimate (\ref{eq:f_pm_estimate}) of Proposition~2, the microscale system may no longer be modeled in accord with the BCF theory if, for example: $k=O(1)$; or, $k=\mathcal O(a)$ and $\epsilon=\mathcal O(a)$.  These conditions indicate situations in which high supersaturation may occur, because of large enough detachment rate at the step edge or high enough deposition onto the surface from above. 

{\renewcommand*{\arraystretch}{0.5}
\begin{figure}[!h]
$\begin{array}{ccc}
    \includegraphics[width=0.34\textwidth]{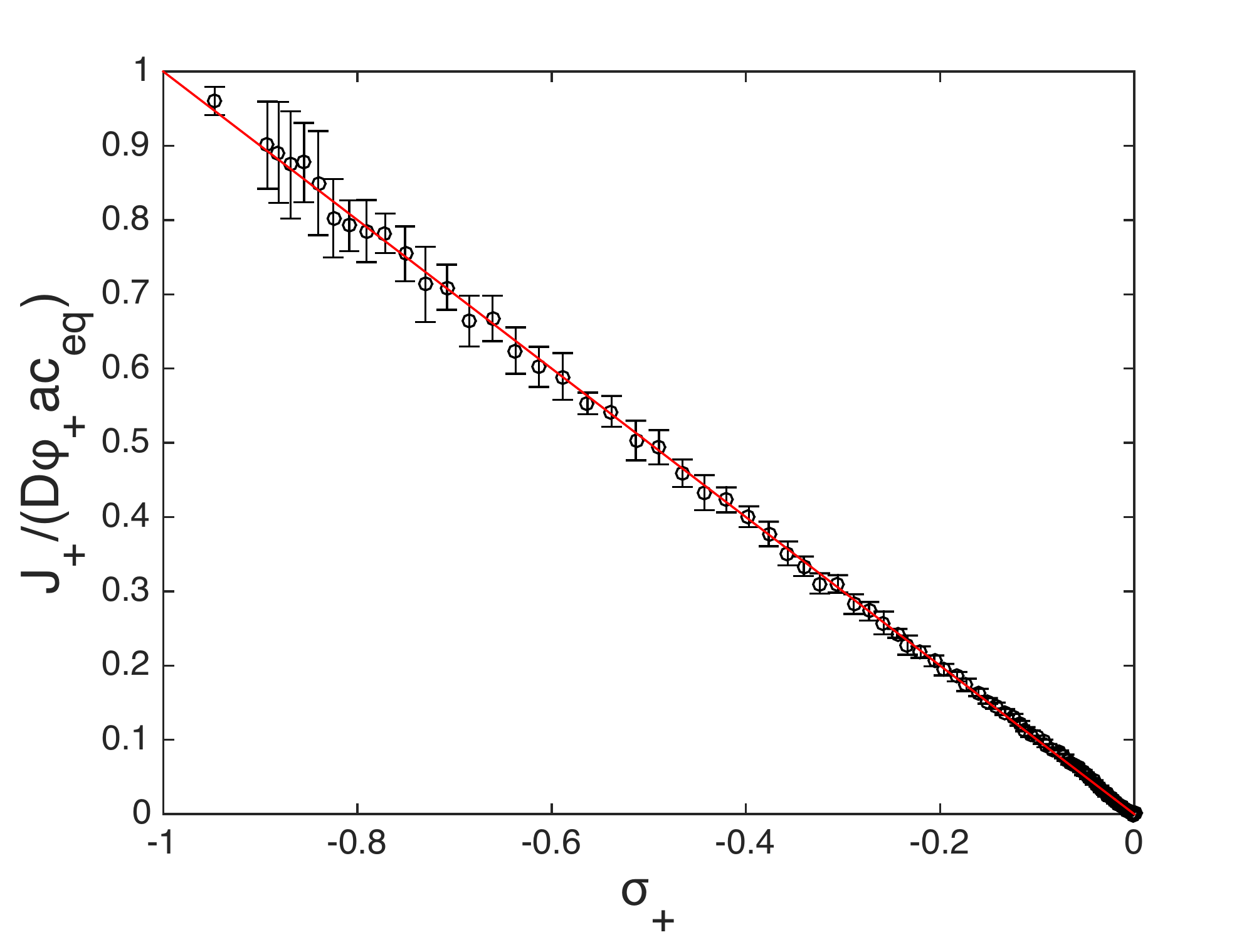} &
    \includegraphics[width=0.34\textwidth]{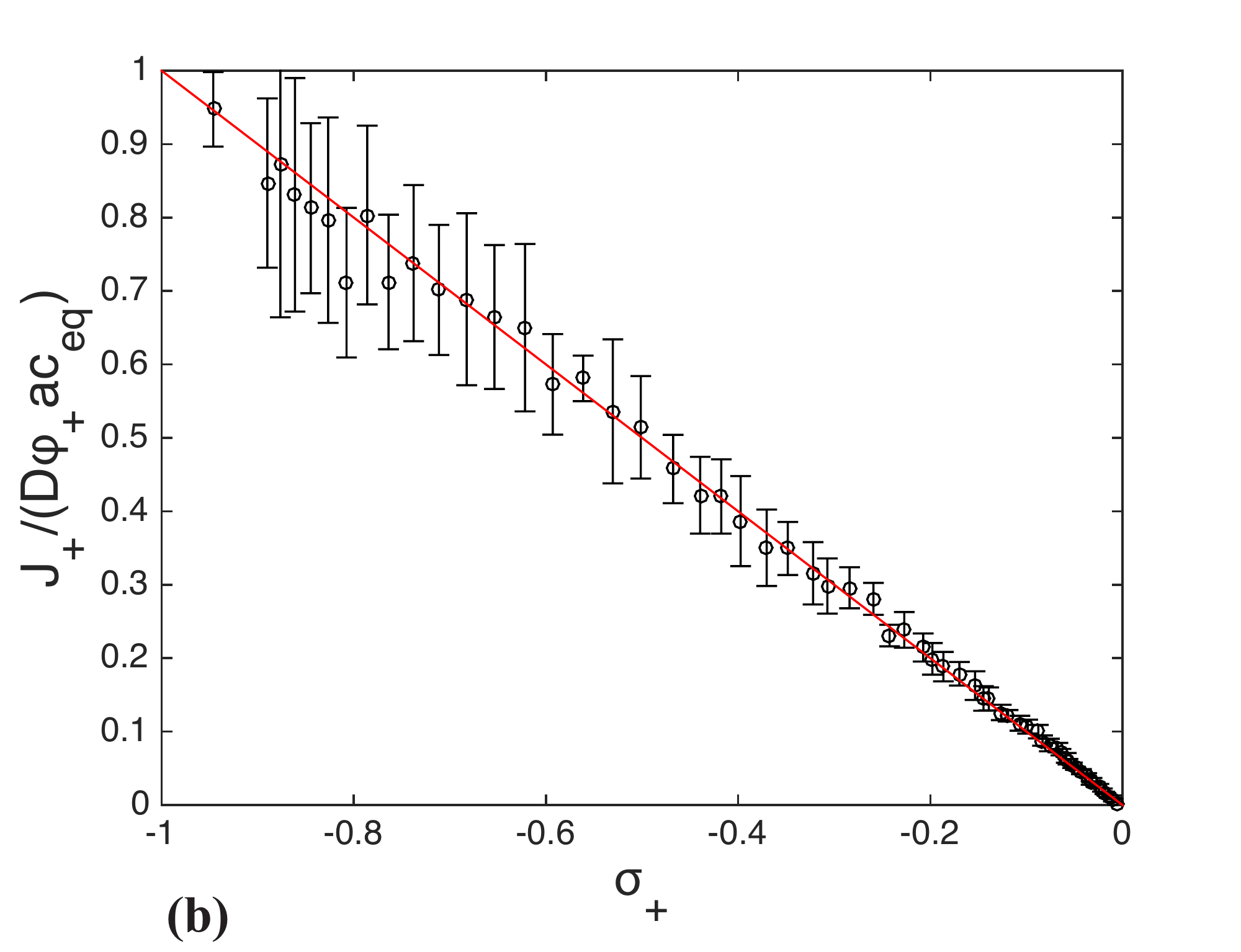} &
    \includegraphics[width=0.34\textwidth]{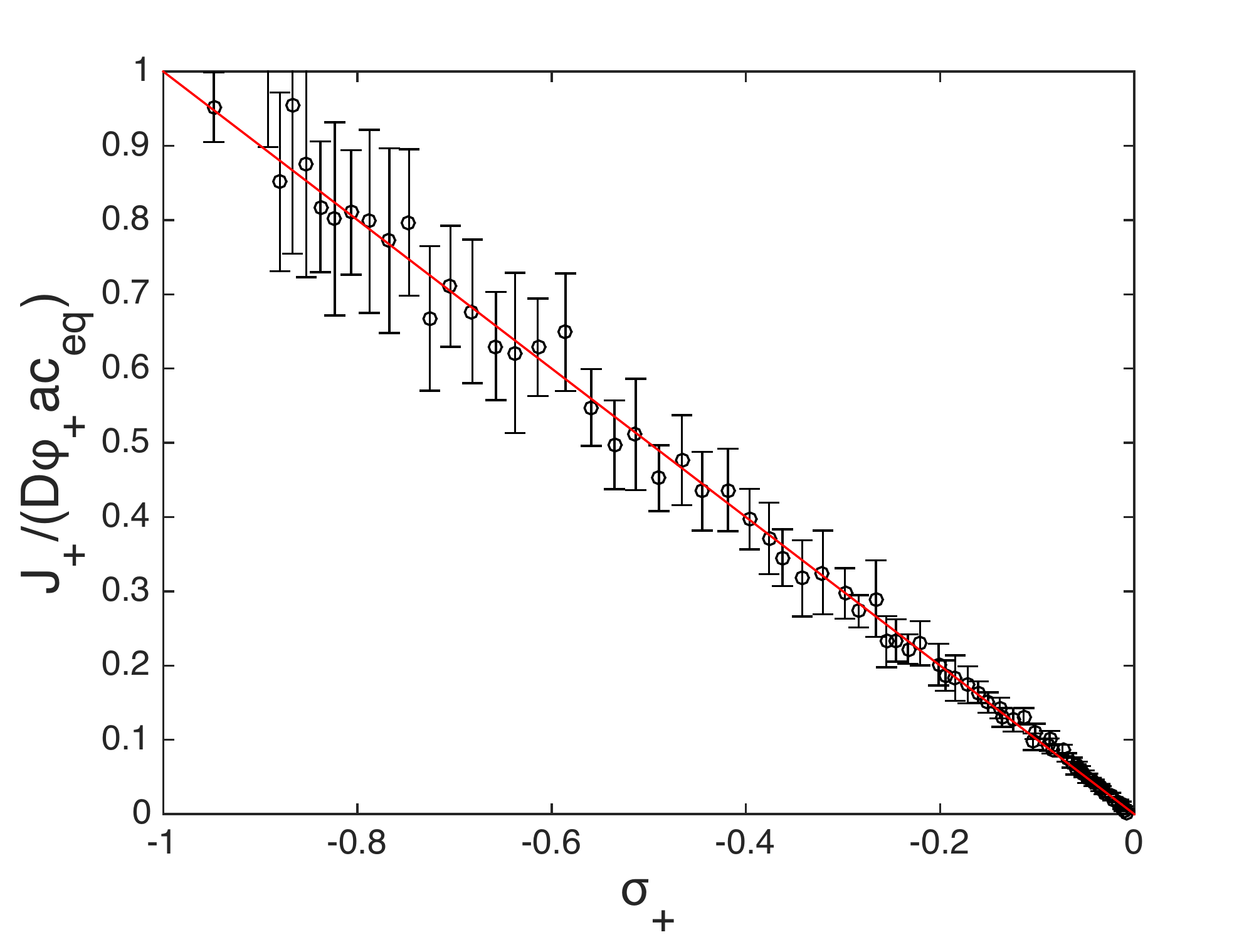}\\
    \mathbf{(a)} & \mathbf{(b)} & \mathbf{(c)}
\end{array}$
\caption{(Color online) Plots of KMC simulations (circles) and the linear kinetic law (red solid line) for adatom flux versus supersaturation on the right of step edge for different values of deposition rate, $F$, with $N=50$, $D=10^{10}$, $k\approx0.0025$, and $\phi_\pm=1$. From left to right: (a) $F=0$ ($\epsilon=0$); (b) $F=10^{6}$ ($\epsilon=10^{-4}$); and (c) $F=10^{7}$ ($\epsilon=10^{-3}$). In each plot, the solid line represents linear kinetic law~(\ref{eq:lkr_coarse}), by neglect of $f_+$. The error bars are determined by use of the standard deviation of flux in: (a) $10$ ensembles of $10^7$ simulations; and (b), (c) $10$ ensembles of $10^6$ simulations. Evidently, on average the system evolves close to thermodynamic equilibrium, in accord with the BCF model.}
\label{fig:lkr}
\end{figure}
}

Figure~\ref{fig:lkr} depicts the adatom flux on the right of the step edge under conditions that enable the system to remain close to thermodynamic equilibrium, i.e., for sufficiently small detachment rate or external deposition rate. In these cases, the supersaturation has small values. Linear kinetic law~\eqref{eq:lkr_coarse}, with neglect of the corrective flux, $f_+$,  is found on average to provide a reasonably accurate approximation for the adatom flux at the step edge.

The remaining plots of this section depict situations in which the adatom flux on the right of the step edge may deviate from linear kinetic law~\eqref{eq:lkr_coarse}, and thus $f_{+}$ may become significant. In particular, Figures~\ref{fig:no_lkr_k} and \ref{fig:no_lkr_F} reveal the behavior of the adatom flux versus supersaturation on the right of the step for large enough $k$ or $F$, respectively. The deviation from the conventional linear behavior predicted by~\eqref{eq:lkr_coarse} is manifested differently in each case. 

Let us consider the high-detachment rate cases with zero deposition, as these are depicted in Figure~\ref{fig:no_lkr_k}. If the supersaturation is sufficiently close to zero, when the flux is small enough, then the flux is approximately linear with supersaturation but with a slope that can be different from the value $D\phi_+ a$ predicted by kinetic law~\eqref{eq:lkr_coarse}. Farther away from equilibrium, the dependence of  adatom flux on supersaturation evidently becomes nonlinear. This nonlinear behavior becomes more pronounced for larger $k$.

{\renewcommand*{\arraystretch}{0.5}
\begin{figure}[!h]
$\begin{array}{ccc}
    \includegraphics[width=0.33\textwidth]{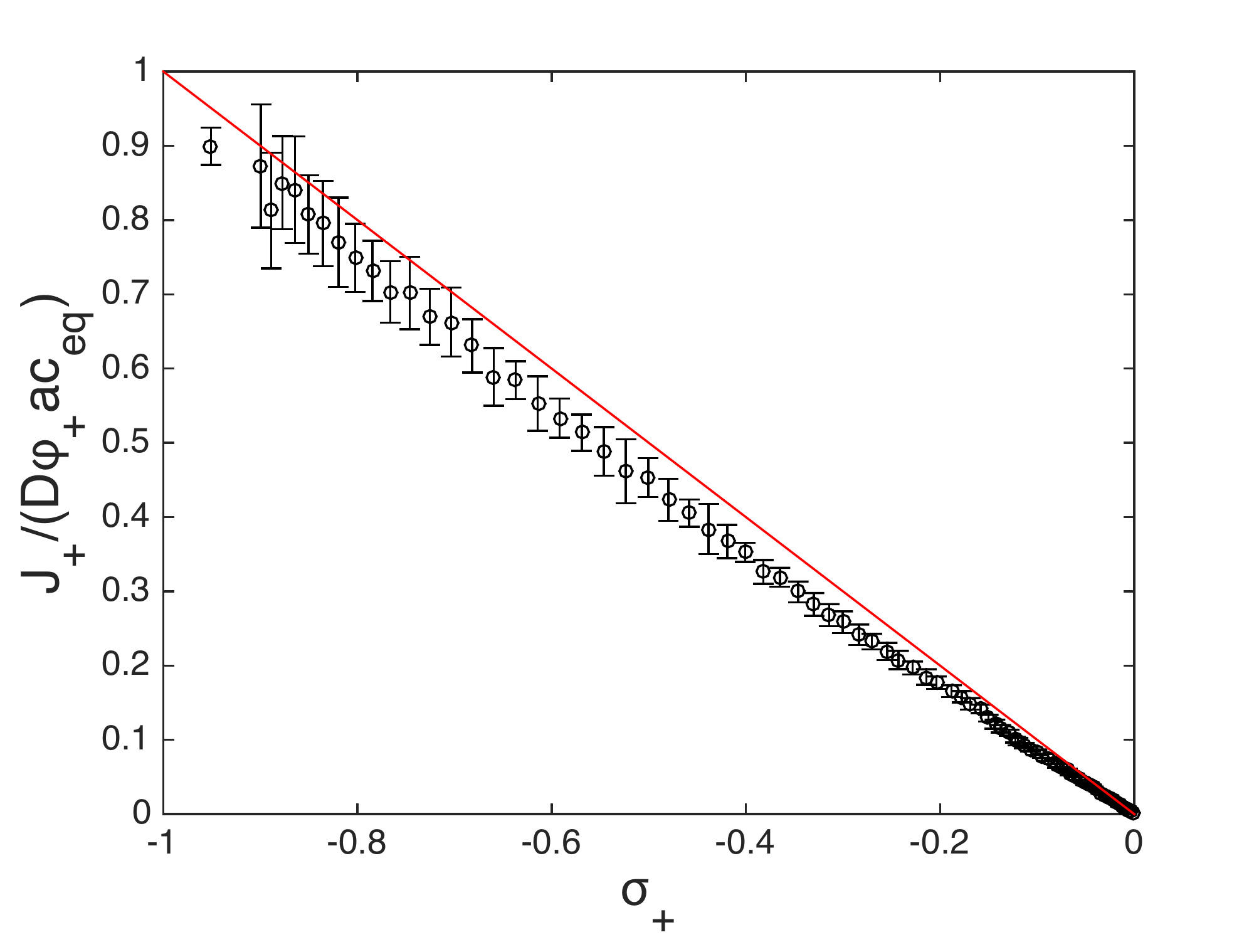} &
    \includegraphics[width=0.33\textwidth]{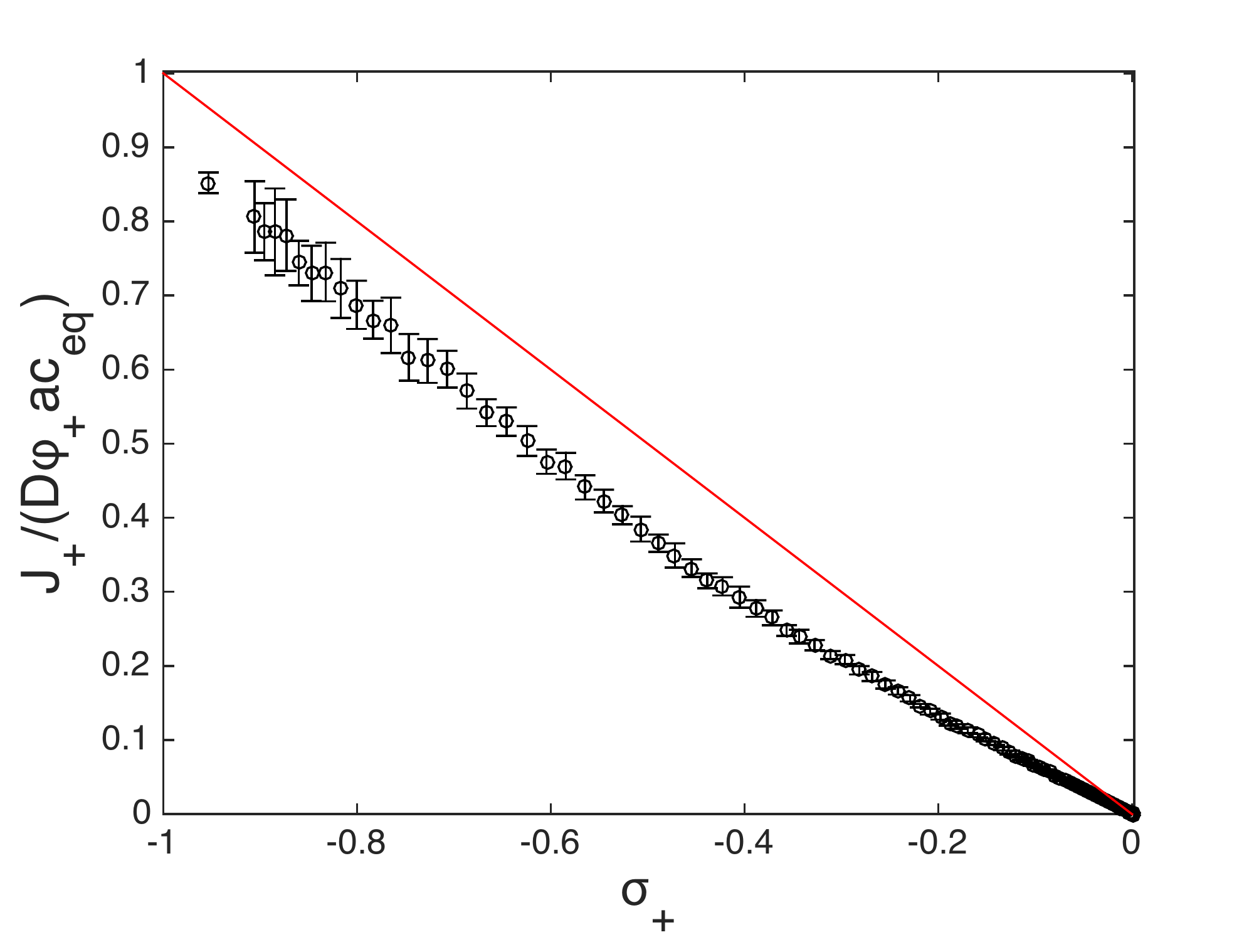} &
    \includegraphics[width=0.33\textwidth]{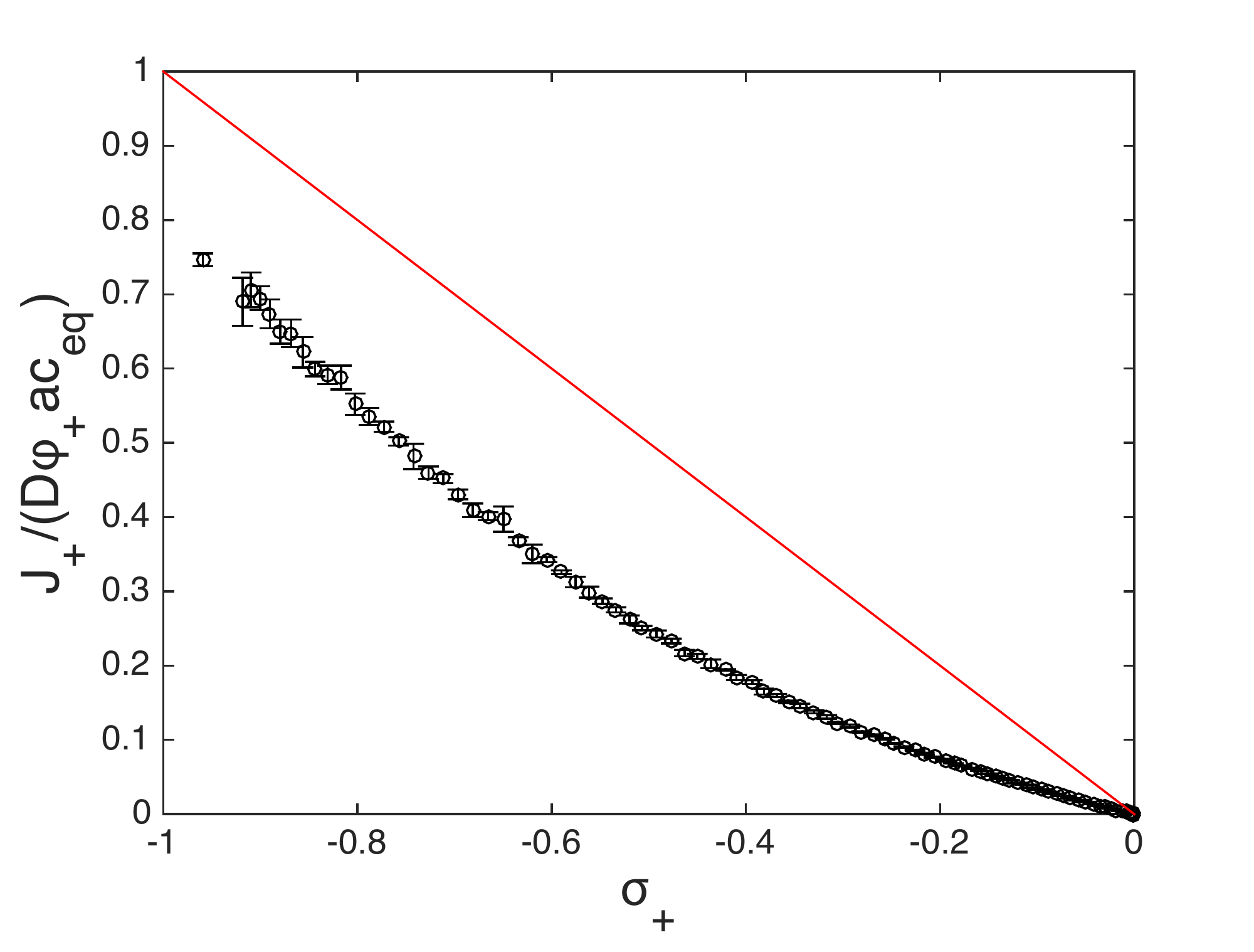}\\
    \mathbf{(a)} & \mathbf{(b)} & \mathbf{(c)}
\end{array}$
\caption{(Color online) Plots of KMC simulations (circles) and the linear kinetic law (red solid line) for adatom flux versus supersaturation on the right of step edge without external deposition ($F=0$) for different values of detachment factor, $k$, and fixed parameters $N=50$, $D=10^{10}$, and $\phi_\pm=1$. From left to right: (a) $k=0.04$; (b) $k=0.09$; and (c) $k=0.20$. In each plot, the solid line represents linear kinetic law~(\ref{eq:lkr_coarse}), by neglect of $f_+$. The error bars are determined by use of the standard deviation of flux in $10$ ensembles of $10^5$ simulations.}
\label{fig:no_lkr_k}
\end{figure}
}

Next, consider a small detachment factor, $k$, but large deposition rate $F$; see Figure~\ref{fig:no_lkr_F}.  We observe that for the smallest value of $F$ used in these plots [Figure~\ref{fig:no_lkr_F}(a)], the flux computed via KMC simulations agrees reasonably well with linear kinetic law~\eqref{eq:lkr_coarse} for a wide range of values for the supersaturation, $\sigma_+$.  For larger values of $F$ [Figures~\ref{fig:no_lkr_F}(b), (c)], the flux remains linear in the density with a slope equal to the predicted value, $D\phi_+ a$, as the density approaches its equilibrium value. However, as the supersaturation increases in magnitude, the nonlinear dependence of the flux is noticeable and becomes more pronounced with increasing varied parameter, $F$. It is worth noting that an increase in the deposition rate $F$ used in KMC simulations beyond the one used in Figure~\ref{fig:no_lkr_F}(c), even by a factor of two, drastically alters the long-time behavior of the system: apparently, no steady state can be established for sufficiently large $F$.

{\renewcommand*{\arraystretch}{0.5}
\begin{figure}[!h]
$\begin{array}{ccc}
    \includegraphics[width=0.33\textwidth]{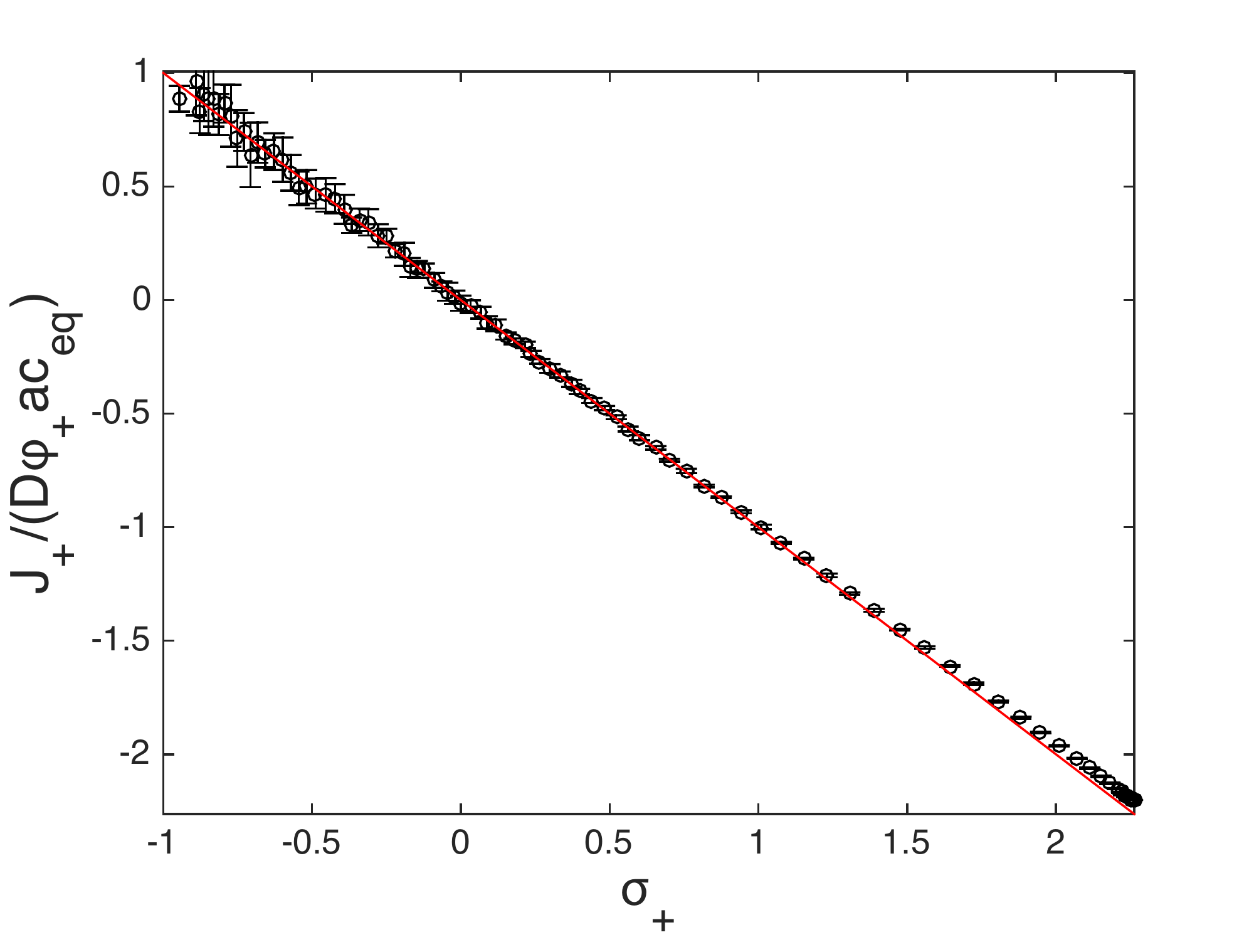} &
    \includegraphics[width=0.34\textwidth]{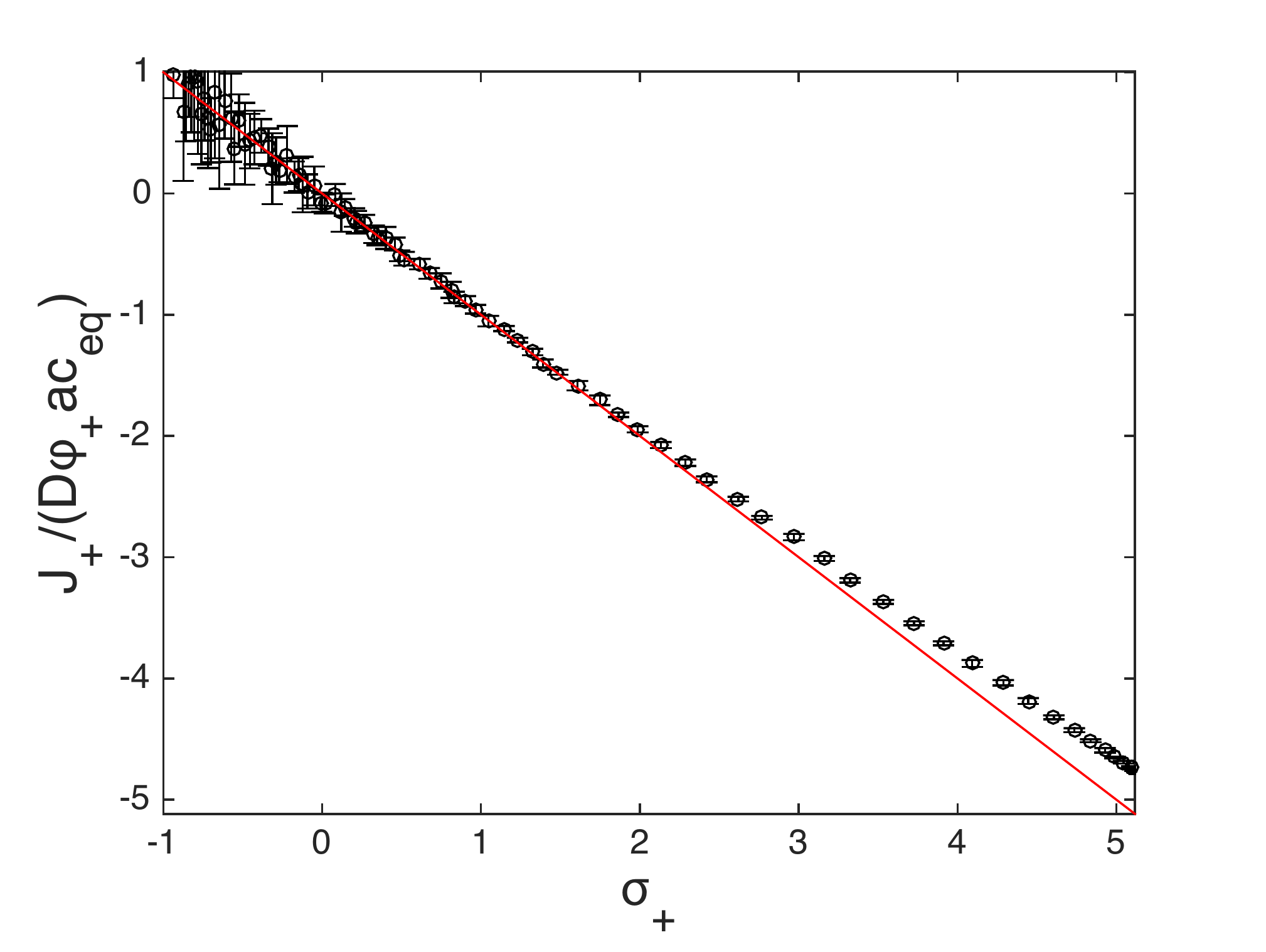} &
    \includegraphics[width=0.33\textwidth]{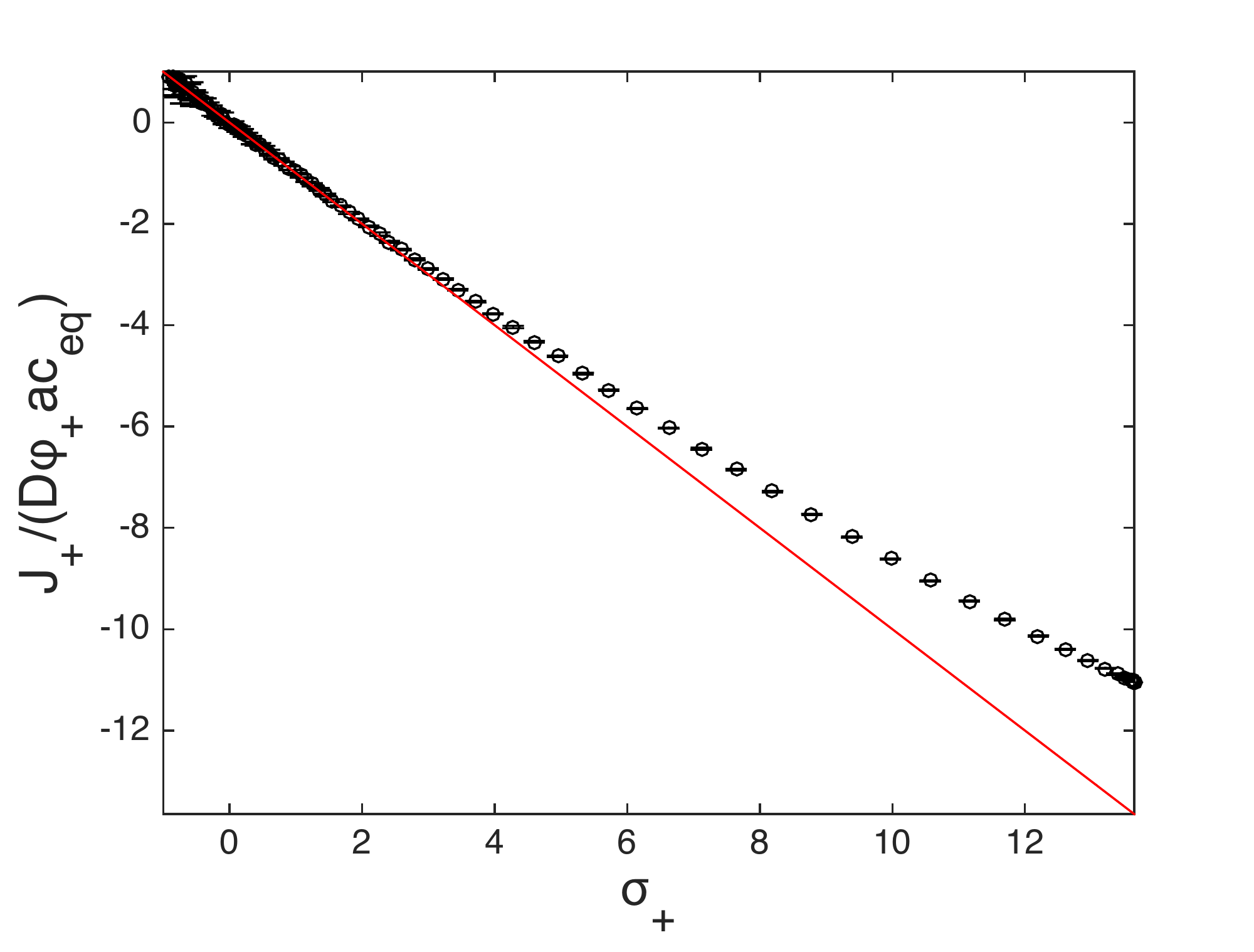}\\
    \mathbf{(a)} & \mathbf{(b)} & \mathbf{(c)}
\end{array}$
\caption{(Color online) Plots of KMC simulations (circles) and the linear kinetic law (red solid line) for adatom flux versus supersaturation on the right of step edge for small yet fixed detachment factor, $k=2.5\times 10^{-3}$,  and different values of deposition rate $F$, with fixed parameters $N=50$, $D=10^{10}$, and $\phi_\pm=1$. From left to right: (a) $F=10^8$ ($\epsilon=10^{-2}$); (b) $F=2\times 10^8$ ($\epsilon=2\times 10^{-2}$); and (c) $F=4\times 10^8$ ($\epsilon=4\times 10^{-2}$). The solid line represents linear kinetic law~(\ref{eq:lkr_coarse}), by neglect of $f_+$, in each plot. The error bars are determined using the standard deviation of flux in $10$ ensembles of: (a) $10^6$ simulations; and (b), (c) $10^5$ simulations.}
\label{fig:no_lkr_F}
\end{figure}
}

As described above, the high-$k$ and high-$F$ cases of Figures~\ref{fig:no_lkr_k} and~\ref{fig:no_lkr_F}, respectively, differ in the way that the effect of corrective flux $f_+$ is manifested in the observed value of the flux {\em near equilibrium}. Let us make an effort to discuss the origin of this behavior in the context of the atomistic model by resorting to formula (\ref{eq:f_+}). The first line in these formulas contains the prefactor $k$ along with a sum over states with one or more adatoms in the lattice site corresponding to the edge atom. This set of configurations does {\em not} allow for atom detachment; thus, according to this contribution to $f_+$, the change of the flux with supersaturation should be suppressed. This prediction should explain the behavior of the slope of the flux versus supersaturation as shown in Figure~\ref{fig:no_lkr_k}. The remaining terms in (\ref{eq:f_+}) come from two- or higher-particle states, which furnish significant contributions if $k$ or $F$ is sufficiently large. These remaining corrections account for configurations in which attachment is inhibited, thus causing an overall increase of the flux out of the step. This prediction is consistent with Figure \ref{fig:no_lkr_F}.

We have been unable to explicitly express the corrective fluxes, $f_\pm(t)$, as a function of adatom densities $c_{\pm 1}(t)$ on the basis of the
analytical model. In order to quantify
the nonlinear behavior of the flux near the step edge, we fit the fluxes computed by KMC simulations to polynomials of $\sigma_+ = c_{1}/c^{eq}-1$. Figure \ref{fig:fitted_flux} shows the fitted flux in two cases where deviations are significant: High $k$ with small $F$; and high $F$ with small $k$. In each case, a quadratic polynomial of supersaturation appears to capture adequately the behavior of the flux versus supersaturation.

{\renewcommand*{\arraystretch}{0.5}
\begin{figure}[!h]
\begin{center}
$\begin{array}{cc}
    \includegraphics[width=0.4\textwidth]{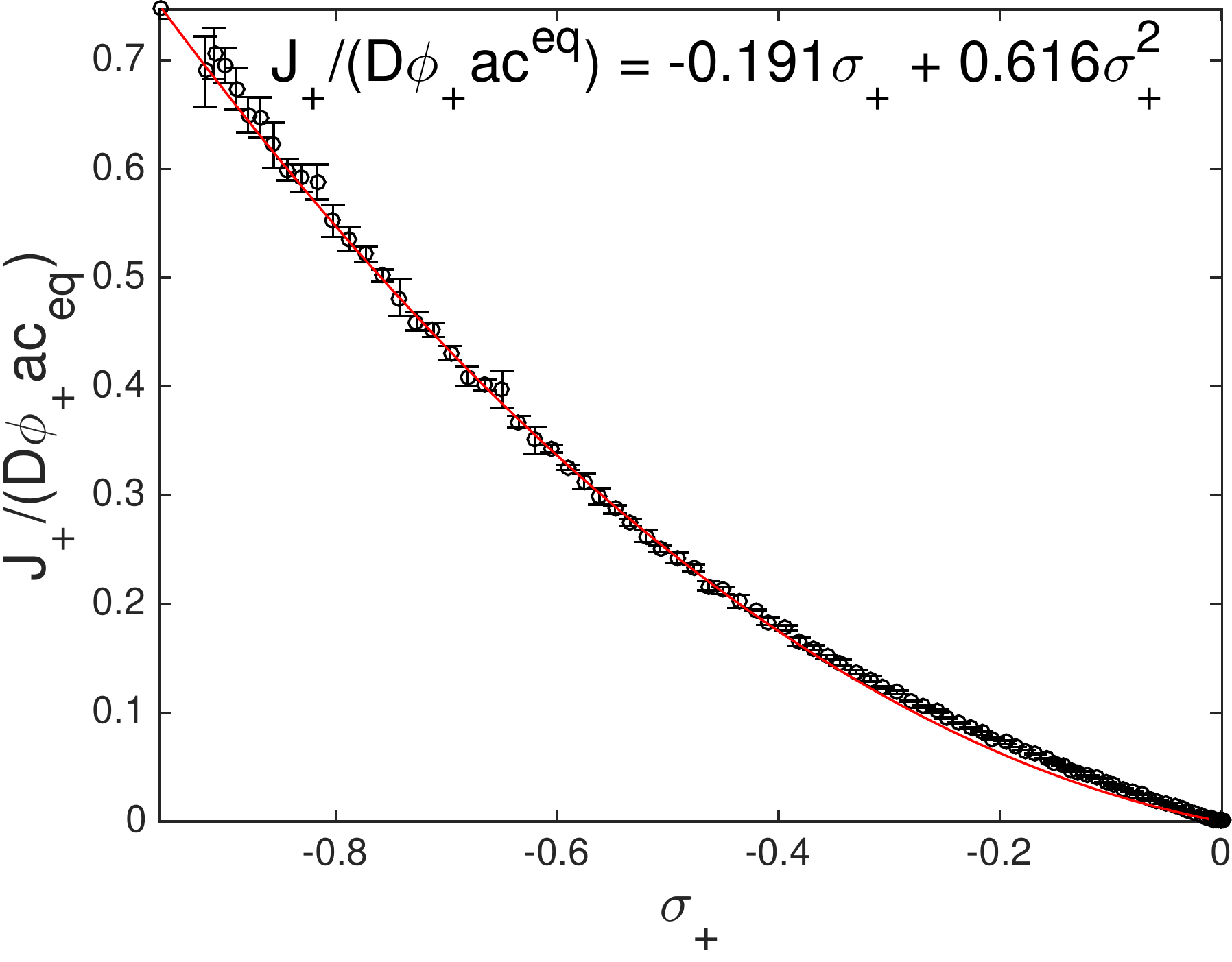} &
    \includegraphics[width=0.4\textwidth]{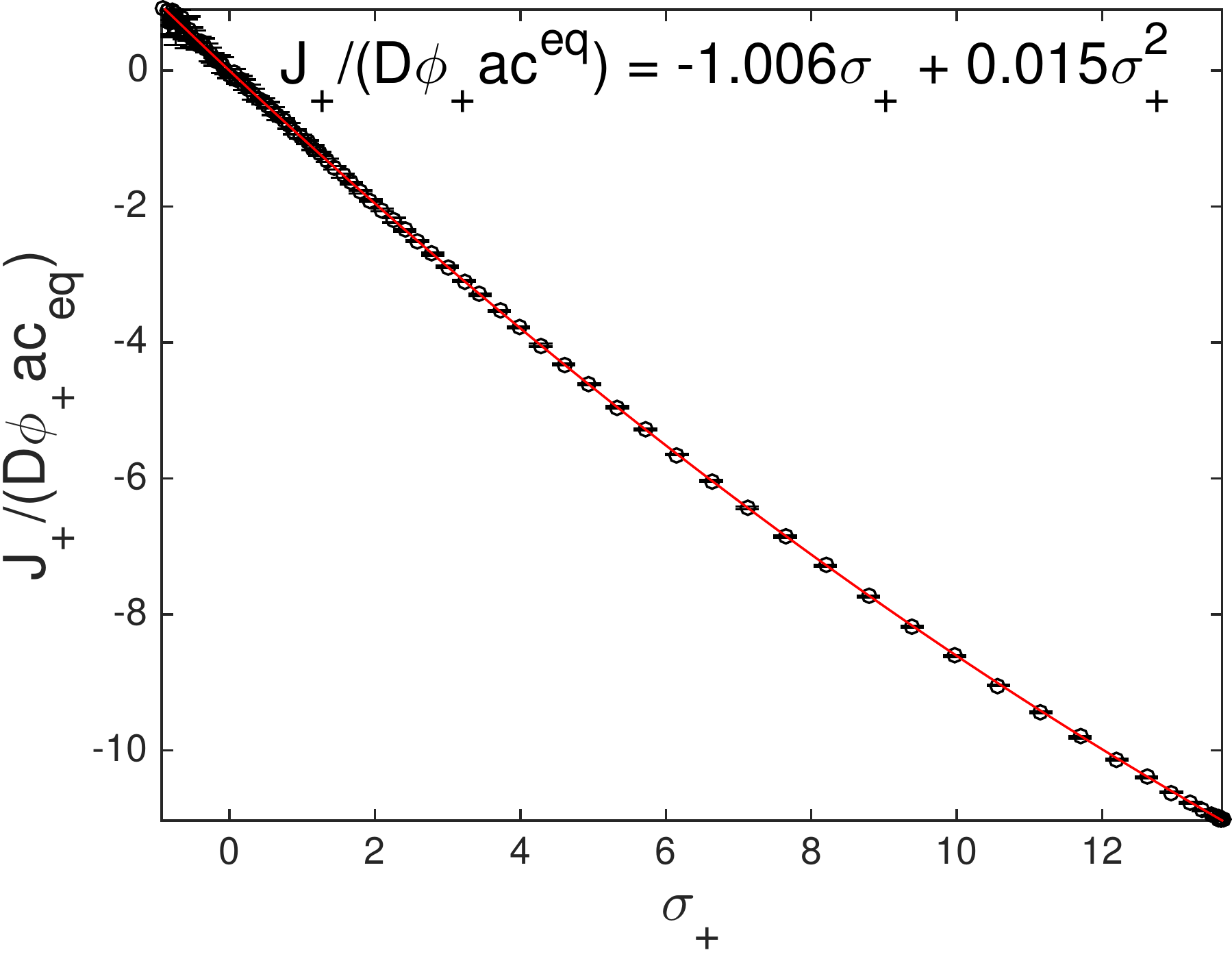}\\
    \mathbf{(a)} & \mathbf{(b)}
\end{array}$
\caption{(Color online) Plots of KMC simulations (circles) and fitted quadratic polynomials of $\sigma_+$ (red solid line) for adatom flux versus supersaturation on the right of step edge, for high detachment rate or high deposition rate. The fixed parameters are: $N=50$, $D=10^{10}$, and $\phi_\pm=1$. From left to right: (a) $k=0.20$ with $F=0$; and (b) $k= 2.5\times 10^{-3}$ with $F=4\times 10^8$. In each plot, the widths of the error bars are determined by use of the standard deviation of flux in $10$ ensembles of $10^5$ simulations.}
\label{fig:fitted_flux}
\end{center}
\end{figure}
}

We conclude that a linear kinetic relation for the adatom flux at the step edge in principle does {\em not} suffice to capture the full range of phenomena displayed by the atomistic solid-on-solid model. Instead, it is more reasonable to propose a discrete expansion of the form
\begin{equation} \label{eq:nlkr_discrete}
 \mp J_\pm \approx  c^{eq}\sum_{n=1}^{N_*} B_\pm^{(n)} \sigma_\pm^n~,
\end{equation}
where the number, $N_*$, would be speculated empirically.
At the mesoscale, the corresponding {\em nonlinear kinetic relation} for flux at the step edge is provided by (\ref{nlkr_bcf}). A systematic derivation of this relation from the atomistic model is still elusive. 
\medskip

{\bf Remark 7.} Based on our KMC results, we expect that~\eqref{eq:nlkr_discrete} reasonably reduces to conventional linear kinetic relation (\ref{eq:lkr_bcf}) of the BCF model if
\begin{equation}\label{eq:cond-micro}
\mathcal A:=k+\frac{\epsilon}{\phi_++\phi_-}\ll 1~.
\end{equation}
This empirical criterion appears less restrictive on $\epsilon$ than estimates (\ref{eq:f_pm_estimate}) and (\ref{eq:R_j_estimate}), suggesting that the bounds in Proposition~2 may be improved; see Remark~6.  Accordingly, if $\mathcal A$ is large enough, then (nonlinear) terms with $n\ge 2$ should become significant. %Note that~\eqref{eq:cond-micro} involves the parameters $k$, $\epsilon$, and $\phi_\pm$, whereas Proposition~2 asserts that the corrective fluxes $f_\pm(t)$ should be small when $\epsilon \ll |\lambda_2|$ provided $k=\mathcal{O}(a)$. These statements, equation~\eqref{eq:cond-micro} and Proposition 2, are mutually consistent; see Remark~6.
\medskip

In our KMC simulations, we observe that the linear kinetic relation for the adatom flux is reasonably accurate if the quantity $\mathcal A$ of (\ref{eq:cond-micro}) does not exceed $0.01$.

%%%%%%%%%%%%%%%
%%%%%%%%%%%%%%%
\section{Discussion and conclusion} 
\label{sec:Discussion}
%%%%%%%%%%%%%%%
%%%%%%%%%%%%%%%
Starting with an idealized atomistic solid-on-solid-type model in 1D, we studied
the mesoscale description of the kinetics of a single step with some emphasis on  the relation between the adatom flux and density at the step edge. Our approach relied on a combination of a master equation for adatom states, in the context of an analytical model, and KMC simulations in 1D.  A noteworthy result was our heuristic derivation from the master equation of exact formulas for the adatom flux; these indicate the physical origin, in terms of atomistic transitions, of corrections to the linear kinetic relation of the BCF model. Furthermore, by using a ``maximum principle'' inherent to the master equation, we estimated the aforementioned corrections for small lattice spacing. By KMC simulations, we observed deviations of the behavior of the flux from the conventional (linear-in-density) prediction of the BCF model.

The master equation approach in this paper forms a nontrivial extension of the analytical model invoked in~\cite{PatroneMargetis14,PatroneEinsteinMargetis14}. By including material deposition onto the surface from above, we accounted for adatom states that do not conserve the total mass of the system. We formally showed that the system has a steady state only for sufficiently small external deposition flux. 

Our atomistic model, despite its inability to include truly 2D effects, captures a few basic elements of diffusion processes on crystal surfaces below the roughening transition. In particular, our model describes hopping of adatoms; and, most importantly, attachment/detachment of atoms at the step in the absence of kinks.  The 1D character of the model, however, poses a few severe limitations. For example, nucleation cannot be included in our analysis. Because of such limitations, more work is needed in order to connect atomistic processes to the wealth of realistic phenomena accompanying crystal evolution below the roughening transition.

A possible criticism of our approach concerns our analysis about the structure of the corrective fluxes, $f_\pm$. This structure appears to be specific to the 1D character of our model. Admittedly, in our approach these correction terms are only associated with rules for attachment and detachment of atoms at the step from/to the terrace. Thus, we leave out the effect of kinks, which is expected to partially alter the mass flux since kink sites can act as local sources or sinks for atoms~\cite{Caflischetal_99}. Because of kinks, the local curvature of the step is expected to affect the equilibrium adatom density in a 2D mesoscale setting, giving rise to the step ``stiffness''~\cite{deGennes_68,Einstein_03}. To derive this effect from a fully atomistic model remains an open problem~\cite{MargetisCaflisch_08}. In spite of these complications, it is reasonable to expect that our 1D model captures features of 2D step motion if kinks are sufficiently far apart from each other, that is, if  the kink density is small.  

In this vein, it is natural to ask: Would it be possible to physically improve our one-step atomistic model by retaining its 1D character yet enriching it with effects that become significant for low {\em and} high supersaturations? A possibility is to account for pair correlations of adatoms due to their energetic interactions. The next stage in this direction would be to consider two steps by including entropic and elastic step-step interactions in the modeling. This task requires appropriate discretization of elastic effects on the lattice~\cite{Saitoetal_01,Baskaranetal_15}. Another scenario that could be examined in the 1D setting is the process of step permeability, which thus far is modeled phenomenologically at the mesoscale~\cite{OzdemirZangwill_92}. Since our model does not describe processes by which adatoms may pass from one side of the step to the other without attaching to the step edge, it seems that permeability is not inherent in our treatment. It is possible that permeability emerges from elastic effects since these induce long-range correlations.

In this work, we focused on averages for the adatom density, step position, and adatom flux, motivated by the known structure of the BCF model. Hence, we have not addressed the stochastic fluctuations arising in step motion. These fluctuations are often significant~\cite{JeongWilliams99}. We believe that this stochastic effect should be more pronounced at high enough supersaturations. The step fluctuations are known to be intimately connected to step stiffness~\cite{Einstein_03}; thus, a 2D atomistic model would be a natural starting point for their systematic in-depth study.

%%%%%%%%%%%%%%%%%%%%%%%%%%%%%%%%%%%%%%%%%%%%%%%%%%
%%%%%%%%%%%%%%%%%%%%%%%%%%%%%%%%%%%%%%%%%%%%%%%%%%

\section*{Acknowledgments}
The authors wish to thank Professors T.~L. Einstein, R.~V. Kohn, E. Lubetzky and J.~D. Weeks for illuminating discussions, as well as T.~J. Burns and W.~F. Mitchell for helpful reviews of this manuscript. The research of the first author (JPS) and third author (DM) was supported by NSF DMS-1412769 at the University of Maryland.
%The second author's (PNP) research was supported by ....

%%%%%%%%%%%%%%%%%%%%%%%%%%%%%%%%%%%%%%%%%%%%%%%%%%
%%%%%%%%%%%%%%%%%%%%%%%%%%%%%%%%%%%%%%%%%%%%%%%%%%

\appendix

%%%%%%%%%%%%%%%%%%%%%%%%%%%%%%%%%%%%%%%%%%%%%%%%%%
%%%%%%%%%%%%%%%%%%%%%%%%%%%%%%%%%%%%%%%%%%%%%%%%%%

%%%%%%%%%%%%%%%%%%%%%%%%%%%%%%%%%%%%%%%%%%%%%%%%%%%%%%%
\section{On birth-death processes} 
\label{app:BirthDeath}
%%%%%%%%%%%%%%%%%%%%%%%%%%%%%%%%%%%%%%%%%%%%%%%%%%%%%%%
In this appendix, we use a toy model for the birth-death Markov process (see, for example, \cite{Kelly2011}) to indicate that no steady state of the master equation exists for large enough external deposition flux, $F$.

A birth-death process is a Markov process with infinitely many states, labeled by $n=0,\,1,\,\dots$, for which transitions are only allowed between the $n$-th and $(n+1)$-th states. Our discrete KRSOS model has this structure if we classify configurations $\boldsymbol{\alpha}$ by the number of adatoms on the terrace, i.e., by the cardinality $|\boldsymbol{\alpha}|=n$ for $n=0,\,1,\,\dots$. For our purposes, the detachment and deposition-from-above events correspond to births while attachment events correspond to deaths.
\begin{figure}[!h]
$\begin{array}{c}
    \includegraphics[width=\textwidth]{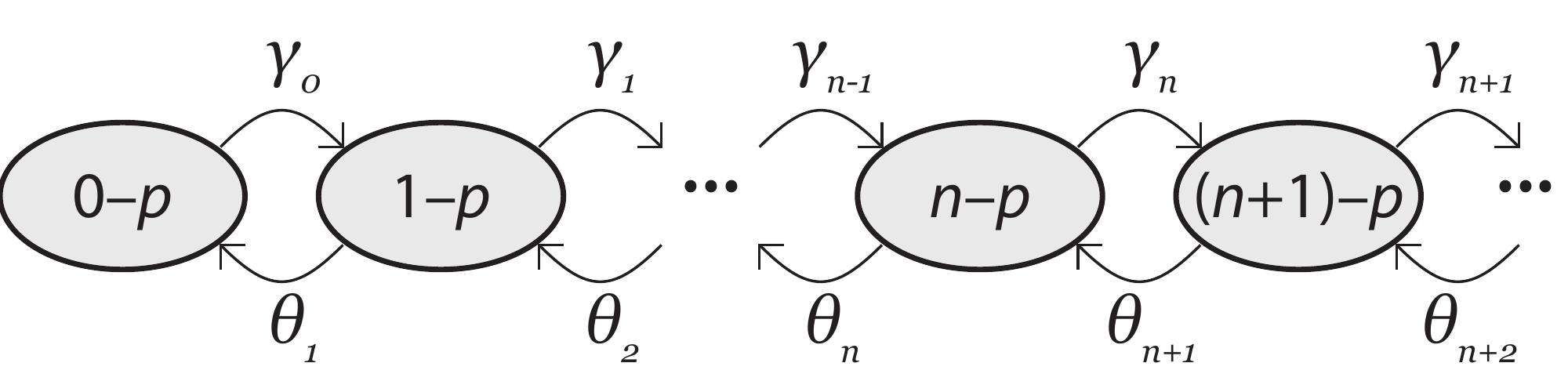}
\end{array}$
\caption{Schematic of the birth-death process showing the hierarchy of particle states and respective transitions. The arrows connecting successive states depict the underlying Markovian character of  transitions. The symbol ``n-p'' stands for the $n$-particle state; $\gamma_n$ denotes the transition from the $n$-particle to the $(n+1)$-particle state; and $\theta_n$ is the rate for the transition from the $n$-particle state to the $(n-1)$-particle state.}
\label{fig:birth_death}
\end{figure}

Let $\gamma_n$ denote the rate for the transition from the $n$-particle state to the $(n+1)$-particle state (the process of ``birth'') and $\theta_n$ denote the rate for the transition from the $n$-particle state to the $(n-1)$-particle state (``death''); see Figure~\ref{fig:birth_death}. If $p_n^{ss}$ is the steady-state probability of the $n$-particle configuration, we have the following balance equations.
\begin{align}\label{eq:birth_death}
\gamma_n p_n^{ss} &= \theta_{n+1} p_{n+1}^{ss}~, \notag \\
\gamma_0 p_0^{ss} &= \theta_{1} p_{1}^{ss}~.
\end{align}
Solving (\ref{eq:birth_death}) for $p_n^{ss}$, we find the formula
\begin{equation}\label{eq:p_n_eq}
p_n^{ss} = \frac{\gamma_{n}\gamma_{n-1}\cdots\gamma_{0}}{\theta_{n+1}\theta_{n}\cdots\theta_{1}} p_{0}^{ss},
\end{equation}
which of course must satisfy the normalization constraint $\sum\limits_{n=0}^\infty p_n^{ss} = 1$. This constraint  can be written as
\begin{equation}\label{eq:norm_condition}
p_0^{ss}\sum_{n=0}^\infty  \frac{\gamma_{n}\gamma_{n-1}\cdots\gamma_{0}}{\theta_{n+1}\theta_{n}\cdots\theta_{1}} < \infty~.
\end{equation}
In other words, if condition (\ref{eq:norm_condition}) does not hold, the probabilities in \eqref{eq:p_n_eq} are not normalizable and, thus, no steady state exists.

In our setting of epitaxial growth in 1D, the rates $\gamma_n$ and $\theta_n$ are determined by the transition rates for attachment, detachment and deposition. These rates satisfy the following equations.
\begin{align}\label{eq:births}
\gamma_n &= \frac{1}{p_n^{ss}}\sum\limits_{\boldsymbol{\alpha} \atop |\boldsymbol{\alpha}|=n} \left\{ F + Dk(\phi_+ + \phi_-)\mathbbold{1}(\nu_{-1}(\boldsymbol{\alpha})=0) \right\}p_{\boldsymbol{\alpha}}^{ss} \notag \\
   &= F + Dk(\phi_++\phi_-)a_n~,
\end{align}
\begin{align}\label{eq:deaths}
\theta_n &= \frac{1}{p_n^{ss}}\sum\limits_{\boldsymbol{\alpha} \atop |\boldsymbol{\alpha}|=n} \left\{ D\phi_+\mathbbold{1}\left(\nu_1(\boldsymbol{\alpha})=1\right) + D\phi_-\mathbbold{1}\left(\nu_1(\boldsymbol{\alpha})=0\right) \right.\notag\\
&\quad \times \left. \mathbbold{1}\left(\nu_1(\boldsymbol{\alpha})>0\right) \right\}p_{\boldsymbol{\alpha}}^{ss} \notag \\
   &= D\phi_+b_n + D\phi_-d_n~,
\end{align}
where $a_n$, $b_n$ and $d_n$ are the probabilities of $n$-particle configurations that forbid detachment, attachment from the right, and attachment from the left, respectively. By (\ref{eq:births}) and (\ref{eq:deaths}), the ratio of birth and death rates is bounded below, i.e., we have the inequality 
\begin{equation}\label{eq:bd_lower_bound}
\frac{\gamma_n}{\theta_{n+1}} \geq \frac{F}{D(\phi_+ + \phi_-)}\quad (n=0,\,1,\,\ldots)~.
\end{equation}
We deduce that the condition $F>D(\phi_+ + \phi_-)$ implies that no steady state may exist because normalization condition (\ref{eq:norm_condition}) is violated. 
In the context of the 1D epitaxial system, having  $F>D(\phi_+ + \phi_-)$ means that the deposition rate is faster than the attachment rate. In this case, the number of particles on the terrace constantly grows and, hence, no steady state can be established.

%%%%%%%%%%%%%%%%%%%%%%%%%%%%%%%%%%%%%%%%%%%%%%%%%%%%%%%
\section{On steady-state solution: Asymptotics of inverse Laplace transform} 
\label{app:Asymptotics}
%%%%%%%%%%%%%%%%%%%%%%%%%%%%%%%%%%%%%%%%%%%%%%%%%%%%%%%
In this appendix, we derive formula (\ref{eq:p_ss}) for the steady-state probability density.
Our derivation relies on the following ideas. (i) We assume that $\epsilon=F/D$ is sufficiently small so that $[I-\epsilon V\mathfrak D(s)V^{-1}\mathfrak B]^{-1}$ exists and we can write this matrix as $\sum_{n=0}^\infty [\epsilon V\mathfrak D(s)V^{-1}\mathfrak B]^n$; and (ii) the quantity $\mathbf{p}^{ss}$, which is the limit of $\mathbf{p}(t)$ as $t \to\infty$, comes from contributions to the inverse Laplace transform of (\ref{eq:laplace_trans_p}) corresponding to the pole at $s=\lambda_1=0$ in the Laplace complex ($s$-) domain. 
Recall that $\mathfrak D(s) := \mbox{diag}\{(s-D\lambda_j)^{-1}\}_{j=1}^{\Omega(M)}$, where $\lambda_j$ are the eigenvalues of matrix $\mathfrak A$ of the deposition-free problem. For a review of basic techniques in computing inverse Laplace transforms, which we do not elaborate on here, see \cite{Schiff2013}.

Let us now elaborate on these ideas. From (i) we may write $\mathbf{p}(t)$ as
\begin{align}\label{eq:p_laplace_integral}
 \mathbf{p}(t) &= \frac{1}{2\pi i} \int_{\gamma-i\infty}^{\gamma+i\infty} \sum\limits_{n=0}^\infty \left(\epsilon V\mathfrak D(s)V^{-1}\mathfrak B\right)^n V\mathfrak D(s)V^{-1}\mathbf{p}(0)e^{st}ds\notag \\
      &= \frac{1}{2\pi i}  \sum\limits_{n=0}^\infty \epsilon^n V \mathcal{I}_n(t) V^{-1}\mathbf{p}(0)~,
\end{align}
where $\gamma$ is a positive constant and the integrals $\mathcal{I}_n(t)$ are defined by
\begin{equation}\label{eq:complex_integral}
 \mathcal{I}_n(t) = \int_{\gamma-i\infty}^{\gamma+i\infty} \left(\mathfrak D(s)\tilde{\mathfrak B}\right)^n \mathfrak D(s)\,e^{st}ds~;\quad \tilde{\mathfrak B}:= V^{-1}\mathfrak B V~.
\end{equation}

The integrand in (\ref{eq:complex_integral}) is a matrix resulting from the product $(\mathfrak D\tilde{\mathfrak B})^n \mathfrak D$, whose entries have the form
\begin{equation}\label{eq:complex_integrand}
 \left[(\mathfrak D\tilde{\mathfrak B})^n \mathfrak D\right]_{ij} = \frac{ \tilde{b}_{ik_1}\tilde{b}_{k_1k_2}\cdots\tilde{b}_{k_{n-2}k_{n-1}}\tilde{b}_{k_{n-1}j} }{ (s-\lambda_i)(s-\lambda_{k_1})\cdots(s-\lambda_{k_{n-1}})(s-\lambda_j) }~;
\end{equation}
$\tilde b_{kl}$ are appropriate coefficients independent of the Laplace variable, $s$.

By (ii) above, since we seek the steady-state solution $\mathbf{p}^{ss}$, the main contribution to the integral~\eqref{eq:complex_integral} comes from the pole at $s=\lambda_1=0$. The only terms which include $(s-\lambda_1)^{-1}$ are those with $j=1$ in (\ref{eq:complex_integrand}). This conclusion can be reached after simplification in the algebra, which can be described as follows. The rows of the matrix $V^{-1}$ contain the left-eigenvectors of matrix $\mathfrak A$. Because of conservation of probability, by which the column sums of $\mathfrak A$ and $\mathfrak B$ are zero, we must have $\left(V^{-1}\right)_{1i} = \left(V^{-1}\right)_{1j}$ for all $i,j$. The matrix $V^{-1}\mathfrak B$, which forms the left matrix product in the $\tilde{\mathfrak B}$ defined in~\eqref{eq:complex_integral}, must obey
\begin{equation}\label{eq:b_property}
 \sum_i \left(V^{-1}\right)_{1i} \left(\mathfrak B\right)_{ij} = 0 \quad \mbox{for all } j.
\end{equation}
Hence, the right-hand side of (\ref{eq:complex_integrand}) vanishes whenever any of the indices $i,k_1,k_2,\dots k_{n-1}$ is equal to unity. Consequently, the steady-state contributions to integral (\ref{eq:complex_integral}) come only from simple poles at $s=\lambda_1$, specifically the terms in (\ref{eq:complex_integrand}) for which $j=1$.

If $n=0$,  integral (\ref{eq:complex_integral}) equals $e^{Dt\Lambda}$, the inverse transform of $\mathfrak D(s)$. If $n>0$ an asymptotic expansion for $\mathcal{I}_n(t)$ as $t\to\infty$ may be computed using the residue theorem, viz.,
\begin{align}\label{eq:complex_integral_computation}
 \left[\mathcal{I}_n(t)\right]_{ij} 
    &= \int_{\gamma-i\infty}^{\gamma+i\infty} \left[(\mathfrak D\tilde{\mathfrak B})^n \mathfrak D\right]_{ij}e^{st}ds \notag \\
    &= \int_{\gamma-i\infty}^{\gamma+i\infty} \frac{ \tilde{b}_{ik_1}\tilde{b}_{k_1k_2}\cdots\tilde{b}_{k_{n-2}k_{n-1}}\tilde{b}_{k_{n-1}j} }{ (s-\lambda_i)(s-\lambda_{k_1})\cdots(s-\lambda_{k_{n-1}})(s-\lambda_j) }e^{st}ds \notag \\
    &\approx 2\pi i \sum_{i,\,k_1,\,k_2,\,\dots,\, k_{n-1}\neq1} \frac{ \tilde{b}_{ik_1}\tilde{b}_{k_1k_2}\cdots\tilde{b}_{k_{n-2}k_{n-1}}\tilde{b}_{k_{n-1}1} }{ (-\lambda_i)(-\lambda_{k_1})\cdots(-\lambda_{k_{n-1}}) } \notag \\
    &= \left\{ \begin{array}{ll}
               2\pi i \left[\left( -\Lambda^\dagger\tilde{\mathfrak B} \right)^n\right]_{i1}~, & j=1~, \\
               0~,      & j>1~, 
               \end{array} \right.                            
\end{align}
as $t\to\infty$. In the above calculation, the symbol $\approx$ implies that the respective result of contour integration leaves out contributions from poles other than $s=\lambda_1$ in the limit $t\to\infty$. For the same reason, all entries in the matrix $\mathcal{I}_n(t)$ other than the first column are asymptotically small and are neglected.

Finally, by substitution of asymptotic formula~\eqref{eq:complex_integral_computation} into~(\ref{eq:p_laplace_integral}), we compute the steady-state probability distribution as
\begin{align}\label{eq:p_ss_asymp}
  \mathbf{p}^{ss,\epsilon} &= \sum_{n=0}^\infty \epsilon^n V\left( -\Lambda^\dagger\tilde{\mathfrak B} \right)^n [\mathbf{e}_1,\mathbf{0},\mathbf{0},\dots,\mathbf{0}] V^{-1}\mathbf{p}(0) \notag \\
  &= \sum_{n=0}^\infty \left( -\epsilon \mathfrak A^\dagger \mathfrak B \right)^n \mathbf{p}^0~,
\end{align}
which leads to~\eqref{eq:p_ss}. In the above, we used the definition $\tilde{\mathfrak B} = V^{-1}\mathfrak BV$ along with $\mathfrak A^\dagger = V\Lambda^\dagger V^{-1}$. Equation (\ref{eq:p_ss_asymp}) is written in terms of the equilibrium distribution, $\mathbf{p}^0$, which satisfies (\ref{eq:truncated_master_equation}) when $\epsilon = 0$. This distribution can be derived in the same way as (\ref{eq:p_eq}), or computed as $\mathbf{p}^0 = \lim_{t\to\infty} \exp(D\mathfrak{A}t)\mathbf{p}(0) = V[\mathbf{e}_1,\mathbf{0},\mathbf{0},\dots,\mathbf{0}]V^{-1}\mathbf{p}(0)$, where $\mathbf{e}_1$ is the $\Omega(M)$-dimensional vector $\mathbf{e}_1 = (1,0,0,\dots,0)^{T}$. 

% Do I need this? %% provided $\epsilon \| \mathfrak A^\dagger \mathfrak B\|< 1$ where $\|\cdot \|$ is any reasonable matrix norm, e.g. an operator norm corresponding to the vector $p$-norm.

%%%%%%%%%%%%%%%%%%%%%%%%%%%%%%%%%%%%%%%%%%%%%%%%%%%%%%%
\section{On the extraction of advection from a microscopic average}
\label{app:Advection}
%%%%%%%%%%%%%%%%%%%%%%%%%%%%%%%%%%%%%%%%%%%%%%%%%%%%%%%

In this appendix, we develop a plausibility argument for the extraction of the continuum-scale advection term, $-v \partial_{\hat{x}}\mathcal C(\hat{x},t)$,  which enters diffusion equation~(\ref{eq:diffusion_eq_bcf}), from the atomistic model ($\partial_x=\partial/\partial x$). The derivation of estimates for corrections entering our formula lie beyond our scope. Our argument provides a heuristic reconciliation of continuum-scale advection with the atomistic and probabilistic perspectives of the master equation approach followed in our work. We will invoke the notation $v=\dot \varsigma$ for the average step velocity; recall that $\varsigma=\varsigma(t)$ is the average step position.

Consider the Eulerian adatom density of (\ref{eq:rho_j}), Definition 5. First, note that the corresponding sum can be conveniently rewritten as 
\begin{equation}\label{eq:rho-j-alt}
\rho_j(t)=\sum_{n\in \mathbb{Z}}\;\sum_{(\boldsymbol{\alpha},m)\in \mathfrak{S}(n)}\nu_{j-s_0+n}({\boldsymbol\alpha})\,p(\boldsymbol{\alpha},m; t)/a~,
\end{equation}
where $|n|$ counts the total number of adatoms detached from ($n>0$) or attached to ($n<0$) the step edge and, thus, determines the microscopic position, $s_0-n$, of the step on the lattice; %$\underline{{\boldsymbol\alpha}}=(\boldsymbol{\alpha},m)$ represents the state of the system;
$p(\boldsymbol{\alpha},m; t):=p_{\boldsymbol{\alpha},m}(t)$ for notational convenience; and $\mathfrak{S}(n):=\{(\boldsymbol{\alpha},m)\,\big|\,|\boldsymbol{\alpha}|=n+(m-m_0)\}$, the set of all allowed values of $(\boldsymbol{\alpha},m)$ for fixed $n$. 

In order to extract the advection term sufficiently away from the step edge, we take into account the decomposition of $p(\boldsymbol{\alpha},m; t)$ into products of the form $p(\boldsymbol{\alpha},m \big| n ;t)\,\wp(n;t)$. In this product, $\wp(n;t)$ is the probability that the microscopic step lies at the lattice site $s_0-n$ at time $t$, and $p(\boldsymbol{\alpha},m\big| n; t)$ is the conditional probability for state $(\boldsymbol{\alpha},m)$ to occur {\it given} that the step edge is at site $s_0-n$. Hence, \eqref{eq:rho-j-alt} is recast to the formula
\begin{equation}\label{eq:rho-j-cond}
\rho_j(t)=\sum_n \wp(n;t)\sum_{(\boldsymbol{\alpha},m)\in\mathfrak{S}(n)}\nu_{j-s_0+n}({\boldsymbol\alpha})\,p(\boldsymbol{\alpha},m\big| n;t)/a~,
\end{equation}
for fixed $j$. Clearly, the right-hand side of~\eqref{eq:rho-j-cond} becomes the discrete Lagrangian density $c_{\hat{\jmath}}(t)$ if $j-s_0+n$ under the summation sign is replaced by the index $\hat{\jmath}$; cf. (26a) in Definition 5.

At this stage, by inspection of~\eqref{eq:rho-j-cond}, we define 
\begin{equation}
c(\hat{\jmath} \big| \aleph;t):=\sum_{(\boldsymbol{\alpha},m)\in\mathfrak{S}(\aleph)}\nu_{j-s_0+\aleph}({\boldsymbol\alpha})\,p(\boldsymbol{\alpha},m\big| \aleph;t)/a~.
\end{equation}
This formula expresses the (conditional) Lagrangian adatom density at fixed site $\hat{\jmath}$ {\it given} that the step position is at site $s_0-\aleph$. Here, $\aleph$ is the discrete {\it random variable} with values $n\in\mathbb{Z}$ that represents the number of adatoms detached from the step edge. Accordingly, we compute 
\begin{equation}\label{eq:Lagran-time-deriv}
\frac{dc_{\hat{\jmath}}(t)}{dt}=\sum_n \dot{\wp}(n;t) c(\hat{\jmath}\big| n;t)+\langle \partial_t{c}(\hat{\jmath}\big|\aleph;t)\rangle,
\end{equation}
where $\langle f(\aleph;t)\rangle$ is the expectation of the random variable $f(\aleph;t)$ under the probability distribution $\wp(n;t)$, viz., $\langle f(\aleph;t)\rangle:=\sum_n\wp(n;t) f(n;t)$, with $f(~\cdot~;t)=\partial_tc(\hat{\jmath}\big| \,\cdot\, ;t)$.

Next, we show that~\eqref{eq:Lagran-time-deriv} plausibly generates a discrete version of the anticipated advection term at long times. For this purpose, we assume that the density of adatoms is sufficiently low, and thereby hypothesize that $\wp(n;t)$ is well approximated by the Poisson distribution with parameter $\varsigma(t)/a$ and $t\gg F^{-1}$; cf.~(17) with $\varsigma(t)=aFt$. Hence, we write $\dot{\wp}(n;t) \approx [\dot{\varsigma}(t)/a]\{\wp(n-1;t)-\wp(n;t)\}$, bearing in mind that correction terms neglected in this formula should account for finite times and the effect of higher adatom numbers per site, controlled by $k$ and $F$. By applying summation by parts in the screw-periodic setting of our system we obtain
\begin{align}\label{eq:adv-discr}
\sum_n \dot{\wp}(n;t) c(\hat{\jmath}\big| n;t)&\approx [\dot{\varsigma}(t)/a]\sum_n \wp(n;t)\{c(\hat{\jmath}\big| n+1;t)-c(\hat{\jmath}\big| n;t)\}\notag\\
&= \dot{\varsigma}(t) a^{-1} \left\{\langle c(\hat{\jmath}\big| \aleph+1;t)\rangle-\langle c(\hat{\jmath}\big| \aleph;t)\rangle \right\}~.
\end{align}
%It remains to make an additional approximation: 
Were it true that $\langle c(\hat{\jmath}\big| \aleph+\ell;t) \rangle \approx \langle c(\hat{\jmath}-\ell\big| \aleph;t) \rangle$ for any integer $\ell$, expressing the translation invariance of the adatom system relative to the step edge, \eqref{eq:adv-discr} would imply
\begin{equation}
\sum_n \dot{\wp}(n;t) c(\hat{\jmath}\big| n;t) \approx -\dot{\varsigma}(t) a^{-1} \left\{\langle c(\hat{\jmath}\big| \aleph;t)\rangle-\langle c(\hat{\jmath}-1\big| \aleph;t)\rangle \right\}~,
\end{equation}
which approaches $-\dot{\varsigma}~\partial_{\hat{x}}\mathcal C(\hat{x},t)$ as $a\downarrow 0$. We leave it as an open question to what extent this approximation is true.

\end{document}